\documentclass[aps,preprint,superscriptaddress]{revtex4-1}

\usepackage{graphicx}
\usepackage{amsmath}
\usepackage{amssymb}
\usepackage{slashed}
\usepackage{braket}
\def\Dslash{D \hspace{-2.7mm}/ \;}
\newcommand{\tr}{{\rm tr \,}}

\usepackage{color}

\usepackage{feynmp}
\makeatletter
\def\endfmffile{%
  \fmfcmd{\p@rcent\space the end.^^J%
          end.^^J%
          endinput;}%
  \if@fmfio
    \immediate\closeout\@outfmf
  \fi
  \ifnum\pdfshellescape=\@ne
    \immediate\write18{mpost \thefmffile}%
  \fi}
\makeatother

\begin{document}

\title{Axial-vector form factors of the baryon octet and chiral symmetry}

\author{Ulrich Sauerwein}
\affiliation{GSI Helmholtzzentrum f\"ur Schwerionenforschung GmbH, \\Planckstra\ss e 1, 64291 Darmstadt, Germany}
\affiliation{Technische Universit\"at Darmstadt, D-64289 Darmstadt, Germany}
\affiliation{Van Swinderen Institute for Particle Physics and Gravity, University of Groningen, 9747 AG Groningen, The Netherlands}

\author{Matthias F.M. Lutz}
\affiliation{GSI Helmholtzzentrum f\"ur Schwerionenforschung GmbH, \\Planckstra\ss e 1, 64291 Darmstadt, Germany}
\affiliation{Technische Universit\"at Darmstadt, D-64289 Darmstadt, Germany}

\author{Rob G.E. Timmermans}
\affiliation{Van Swinderen Institute for Particle Physics and Gravity, University of Groningen, 9747 AG Groningen, The Netherlands}

\date{\today}

\begin{abstract}
We consider the axial-vector form factors of the baryon octet in flavor-SU(3) chiral perturbation theory. The baryon octet and decuplet and the pseudoscalar-meson octet are included as explicit degrees of freedom. We explore the use of on-shell meson and baryon masses in the one-loop contributions to the axial-vector form factors and focus on a consistent treatment in terms of chiral power counting. The convergence properties of such an approach are scrutinized. We discuss the potential for comparison to upcoming QCD lattice data.
\end{abstract}



\maketitle
\tableofcontents

\newpage
\section{Introduction}
\label{sec:Intro}
The axial-vector form factors of the octet baryons are, next to the baryon masses, important testing grounds for our understanding of the flavor and chiral structure of low-energy, non-perturbative QCD. Cabibbo's flavor-SU(3) symmetric model \cite{Cabibbo:1963wer, Cabibbo:2003cu} for neutron $\beta$ decay and the semileptonic hyperon decays was remarkably successful, but nowadays one requires a more fundamental effective field theory description of these processes in the framework of chiral perturbation theory ($\chi$PT). Unfortunately, previous work within flavor-SU(3) heavy-baryon $\chi$PT has shown serious convergence problems of the chiral expansion \cite{Jenkins:1990jv, Jenkins:1991es}.  The expected progress in flavor-SU(3) lattice studies motivates us to revisit the axial-vector form factors of the nucleon and hyperons in $\chi$PT.

In Ref. \cite{Lutz:2020dfi} we investigated the axial-vector form factor of the nucleon in flavor-SU(2) $\chi$PT with nucleons and isobars. Calculations of masses and form factors in $\chi$PT face the problem that in the chiral expansion terms occur that violate standard power-counting rules that assume $m_{\pi}\sim\Delta\sim$ small momenta, where $\Delta$ is the isobar-nucleon mass difference. We explored the use of on-shell hadron masses in the loop diagrams to improve the convergence of the chiral expansion when the isobars are included. We showed that it was possible in our scheme to renormalize the one-loop amplitudes in terms of subtracted Passarino-Veltman integrals.
We performed a successful global fit of the LECs to the available flavor-SU(2) QCD lattice data for the nucleon and isobar masses and the nucleon axial-vector form factor \cite{Rajan:2017lxk, Capitani:2017qpc, Alexandrou:2017hac, Bali:2018qus}. An interesting prediction of our approach that results from the use of on-shell masses in the loops is a non-analytic behavior \cite{Semke:2006hd,Guo:2019nyp} of the masses and the form factor as function of the pion mass, which should become prominent for larger lattice volumes than were used so far.

In this work we extend this approach to calculate the axial-vector form factors of the baryon octet in flavor-SU(3) $\chi$PT. The baryon decuplet is included as an explicit degree of freedom and consistently incorporated via a suitable subtraction and renormalization scheme similar to Ref. \cite{Lutz:2020dfi}. We investigate whether the use of on-shell masses in loop contributions leads to better convergence properties of the flavor-SU(3) chiral expansion. Unfortunately, a full lattice description of all axial-vector form factors in flavor-SU(3) is not available yet. Refs. \cite{Ishikawa:2018rew, Alexand:2018pln, Bali:2019yiy} are restricted to the axial-vector form factor of the nucleon. In Ref. \cite{Bali:2019svt} first results for axial-vector form factors of the $\Sigma$ and $\Xi$ hyperons are presented and more results, also for strangeness-changing axial-vector currents, are announced.

We have organized our work as follows. In Section II we discuss the flavor-SU(3) chiral Lagrangian with the pseudoscalar-meson Goldstone-boson octet, the baryon octet, and the baryon decuplet as degrees of freedom. Section III discusses the axial-vector currents and form factors in QCD, the prime observables covered in our approach. In Section IV we calculate the axial-vector form factors at the one-loop level from the chiral Lagrangian. Next, in Section V we discuss our power-counting scheme and the implication for renormalization of the one-loop diagrams. In Section VI we investigate the convergence properties of our scheme at the physical point of the form factors and in the flavor-SU(3) limit. Finally, Section VII contains our summary and an outlook for further work. Several appendices are devoted to definitions of amplitudes, Clebsch-Gordan coefficients and recoupling constants, and kinematical constants.

\section{The chiral Lagrangian}

The starting point of all calculations in $\chi$PT is the effective chiral Lagrangian, which contains fields of all considered particles: the Goldstone-boson octet $\Phi$, the baryon octet $B$, and the baryon decuplet $\Delta_\mu$ \cite{Lutz:2001yb}. The fields are given,
respectively,  by
\begin{eqnarray}
\Phi&=&
\begin{pmatrix}
\frac{\eta}{\sqrt{3}}+\pi^0 &\sqrt{2}\pi^{+} &\sqrt{2} K^+ \\
\sqrt{2}\pi^{-}&\frac{\eta}{\sqrt{3}}-\pi^0&\sqrt{2} K^0 \\
\sqrt{2}K^{-}&\sqrt{2}\bar{K}^{0} &-\frac{2}{\sqrt{3}}\eta
\end{pmatrix}\,, 
\end{eqnarray}
\begin{eqnarray}
B&=& 
\begin{pmatrix}
\frac{\Lambda}{\sqrt{6}}+\frac{\Sigma^0}{\sqrt{2}} &\Sigma^{+} & p \\
\Sigma^{-}&\frac{\Lambda}{\sqrt{6}}-\frac{\Sigma^0}{\sqrt{2}}  & n \\
-\Xi^{-}&\Xi^{0} &-\sqrt{\frac{2}{3}}\Lambda
\end{pmatrix}\,, 
\end{eqnarray}
and
\begin{eqnarray}
 &&\Delta_{\mu}^{111}=\Delta_{\mu}^{++}\,, \hspace{0.6cm}	
   \Delta_{\mu}^{113}=\Sigma_{\mu}^{+}/\sqrt{3}\,, \hspace{0.6cm}
   \Delta_{\mu}^{133}=\Xi_{\mu}^{0}/\sqrt{3}\,, \hspace{0.6cm} 
   \Delta_{\mu}^{333}=\Omega_{\mu}^{-}\,, 
\nonumber\\
 && \hspace{1.0cm} \Delta_{\mu}^{112}=\Delta_{\mu}^{+}/\sqrt{3}\,, \hspace{0.6cm}
   \Delta_{\mu}^{123}=\Sigma_{\mu}^{0}/\sqrt{6}\,, \hspace{0.6cm}
   \Delta_{\mu}^{233}=\Xi_{\mu}^{-}/\sqrt{3}\,,
\nonumber\\
 && \hspace{2.4cm} \Delta_{\mu}^{122}=\Delta_{\mu}^{0}/\sqrt{3}\,, \hspace{0.6cm}
   \Delta_{\mu}^{223}=\Sigma_{\mu}^{-}/\sqrt{3}\,,
\nonumber\\
 && \hspace{4.2cm} \Delta_{\mu}^{222}=\Delta_{\mu}^{-}\,.
\label{Eq:Def-Fields}
\end{eqnarray}

We list the chiral flavor-SU(3) Lagrangian \cite{Lutz:2001yb, Lutz:2010se, Frink:2006hx}, but we neglect the terms without explicit impact on the axial-vector form factors. The terms that are relevant for this work are divided into chiral orders,
 \begin{eqnarray}
  \mathcal{L} = \mathcal{L}_\text{kin} + \mathcal{L}^{(1)} + \mathcal{L}^{(2)} + \mathcal{L}^{(3)}\,,
  \label{Eq:Def-chiralLagrangian}
 \end{eqnarray}
where $i$ denotes the chiral order of $\mathcal{L}^{(i)}$ and $\mathcal{L}_\text{kin}$ contains the kinetic part of Goldstone bosons $\Phi$, spin-1/2 baryons $B$, and spin-3/2 baryons $\Delta_\mu$, 
\begin{eqnarray}
 &&\mathcal{L}_\text{kin} =  -\, f^2\, \tr \Big[ U_\mu \, U^\mu \Big]
+\tr\, \Big[ \bar B\, (i\, \Dslash\, - M)\, B \Big]  
\nonumber\\
&& \hspace{0.3cm} - \,\tr\, \Big[ \bar \Delta_\mu \cdot \big((i\,\Dslash\, - (M+\Delta))\,g^{\mu\nu} 
 -\,i\,(\gamma^\mu D^\nu + \gamma^\nu D^\mu) + \gamma^\mu(i\,\Dslash + (M+\Delta))\gamma^\nu \big)\, \Delta_\nu \Big]\,.
\label{Eq:Def-Lkin}
\end{eqnarray}
We introduced the covariant derivative $D_\mu$, which is constructed such that it transforms in the same way as the corresponding field by incorporating the chiral connection $\Gamma$. The mass of the baryon octet in the chiral limit is given by $M$ and the mass of the baryon decuplet in the chiral limit by $M+\Delta$. The Goldstone bosons are massless in the chiral limit. 
The covariant derivative $D_\mu$ and the chiral connection $\Gamma_\mu$ are given by
\begin{eqnarray}
&&  D_\mu \, B \;\,= \partial_\mu B +  \Gamma_{\mu}\, B - B\,\Gamma_{\mu} \,,
 \nonumber\\
 &&\big(D_{\mu} \Delta_{\nu}\big)^{abc}=\partial_{\mu} \Delta^{abc}_{\nu}
 +\Gamma^a_{\mu,n}\Delta_{\nu}^{nbc}
 +\Gamma^b_{\mu,n}\Delta_{\nu}^{anc}
+\Gamma^c_{\mu,n}\Delta_{\nu}^{abn} , 
 \nonumber\\
 &&\Gamma_\mu =
\genfrac{}{}{1pt}{1}{1}{2} \,u^\dagger \,\big[\partial_\mu -i\,(v_\mu + a_\mu) \big] \,u
+\genfrac{}{}{1pt}{1}{1}{2}\, u \,
\big[\partial_\mu -i\,(v_\mu - a_\mu)\big] \,u^\dagger\,, \qquad
 u = e^{i\,\frac{\Phi}{2\,f}}  \,,
\nonumber\\
 &&  U_\mu = \genfrac{}{}{1pt}{1}{1}{2} \,u^\dagger \, \big(\partial_\mu \,e^{i\,\frac{\Phi}{f}} \big)\, u^\dagger
-\genfrac{}{}{1pt}{1}{i}{2}\,u^\dagger \,(v_\mu+ a_\mu)\, u
+\genfrac{}{}{1pt}{1}{i}{2}\,u \,(v_\mu-a_\mu)\, u^\dagger\;, 
\label{Eq:Def-CovariantDerivative}
\end{eqnarray}
and $U_\mu$ contains the Goldstone-boson field $\Phi$ and possible external currents with vector $v_\mu$ or axial-vector structure $a_\mu$. The pion-decay constant in the chiral limit is denoted by $f \simeq 87\text{ MeV}$ \cite{Aoki:2019cca}.
Furthermore, we specify
 \begin{eqnarray}
&& \mathcal{L}^{(1)} =
\,F \, \tr \Big[ \bar{B}\, \gamma^\mu \,\gamma_5\, [i\,U_\mu,B]\, \Big] + D\, \tr\Big[ \bar{B}\, \gamma^\mu \,\gamma_5\, \{i\,U_\mu,\,B\}\, \Big]
\nonumber \\
&& \hspace{0.83cm} + \, C\, \tr \Big[ (\bar{\Delta}_\mu \cdot i\, U^\mu)\, B + \textrm{h.c.} \Big] 
+ H\, \tr \Big[ (\bar{\Delta}^\mu\cdot \gamma_5\gamma_\nu   \Delta_\mu)\, i\,U^\nu \Big]\,.
\label{Eq:Def-L1}
\end{eqnarray}
In the lowest-order Lagrangian $\mathcal{L}^{(1)}$ we introduced for the decuplet fields the short-hand notation \cite{Lutz:2001yb}
\begin{eqnarray}
 &&\big(\Phi\cdot\Delta_{\mu}\big)^a_b=\epsilon_{klb} \, \Phi_n^l \, \Delta_{\mu}^{kna}\,, \hspace{1cm}
 \big(\bar{\Delta}^{\mu}\cdot\Phi\big)^a_b=\epsilon^{kla} \, \bar{\Delta}^{\mu}_{knb} \, \Phi^n_l\,,
 \nonumber\\
 &&\big(\bar{\Delta}^{\mu}\cdot\Delta_{\mu}\big)^a_b=\bar{\Delta}^{\mu}_{bcd} \, \Delta_{\mu}^{acd}\,,
 \label{Eq:Def-PhiDeltaInteractions}
\end{eqnarray}
and the dimensionless LECs $F$, $D$, $C$, and $H$. 
For convenience the Lagrangians $\mathcal{L}^{(2)}$ and $\mathcal{L}^{(3)}$ are split as
 \begin{eqnarray}
  \mathcal{L}^{(2)} &=& \mathcal{L}^{(S)} + \mathcal{L}^{(V)} + \mathcal{L}^{(T)} + \mathcal{L}^{(A,a)} + \mathcal{L}^{(A,b)} + \mathcal{L}^{(F)}\,,
  \label{Eq:Def-L2split}
  \\
 \mathcal{L}^{(3)} &=& \mathcal{L}^{(\chi)} + \mathcal{L}^{(R)}\,,
  \label{Eq:Def-L3split}
 \end{eqnarray}
with
 \allowdisplaybreaks[1] 
  \begin{eqnarray}
   &&\mathcal{L}^{(S)}=-\frac12\,g_0^{(S)}\,\tr\big[\bar{B}\,B\big]
  \,\tr\big[U_{\mu}\,U^{\mu}\big]   
  -\frac12\,g_1^{(S)}\,\tr\big[\bar{B}\,U^{\mu}\big]
  \,\tr\big[U_{\mu}\,B\big]  
 \nonumber \\
  &&\quad\; -\frac14\,g_F^{(S)}\,\tr\Big[\bar{B}\,
  \big[\big\{U^{\mu}\,,\,U_{\mu}\big\}\,,\,B\big]\Big]  
   -\frac14\,g_D^{(S)}\,\tr\Big[\bar{B}\,
   \big\{\big\{U^{\mu}\,,\,U_{\mu}\big\}\,,\,B\big\}\Big] \,,
 \\
\nonumber\\
 &&\mathcal{L}^{(V)}=-\frac14\, g_0^{(V)}\,\tr\big[\bar{B}\,i\,\gamma^{\mu} \, D^{\nu} B\big]
 \,\tr\big[U_{\nu}\,U_{\mu}\big]+\textrm{h.c.}
 \nonumber\\
 &&\quad\; -\frac18\, g_1^{(V)}\Big(\tr\big[\bar{B}\,U_{\mu}\big]\,i\,\gamma^{\mu}
 \,\tr\big[U_{\nu}\, D^{\nu} B\big]+
 \tr\big[\bar{B}\, U_{\nu}\big]\,i\,\gamma^{\mu} \,\tr\big[U_{\mu}\,  D^{\nu} B\big]\Big)\hspace{-0.1cm}+\textrm{h.c.}
 \nonumber\\&&\quad\;
 -\frac18\, g_F^{(V)}\,\tr\Big[\bar{B}\,i\,\gamma^{\mu}
 \big[\big\{U_{\mu}\,,\,U_{\nu}\big\}\, , \, D^{\nu} B\big]\Big]+\textrm{h.c.}
 \nonumber\\&&\quad\;
 -\frac18\, g_D^{(V)}\,\tr\Big[\bar{B}\,i\,\gamma^{\mu}
 \big\{\big\{U_{\mu}\,,\,U_{\nu}\big\}\, , \, D^{\nu} B\big\}\Big]+\textrm{h.c.}\,,
     \\
  \nonumber\\
&&\mathcal{L}^{(T)}= 
-\frac12\,g_{1}^{(T)}\,\tr\big[\bar{B}\,U_{\mu}\big]\,i\,\sigma^{\mu\nu} \,
\tr\big[U_{\nu}\, B\big]
\nonumber\\ 
&&\quad \;
-\frac14\,g_{F}^{(T)}\,\tr\big[\bar{B}\,i\,\sigma^{\mu\nu} \,
\big[\big[U_{\mu}\,,\,U_{\nu}\big]\,, B\big]\big]
-\frac14\,g_{D}^{(T)}\,\tr\big[\bar{B}\,i\,\sigma^{\mu\nu} \,
\big\{\big[U_{\mu}\,,\,U_{\nu}\big]\,, B\big\}\big],
\\
\nonumber\\
&&\mathcal{L}^{(A,a)}= -\frac14 \, f_{1}^{(A)}\,\tr\Big[\big(\bar{\Delta}^{\mu}\cdot \,\gamma^{\nu}\,\gamma_5\,B\big)
 \,\big\{U_{\mu}\,,\,U_{\nu}\big\}\Big]+\textrm{h.c.}
\nonumber \\
 &&\quad \;
 -\frac14 \, f_{3}^{(A)} \, \tr\Big[\big(\bar{\Delta}^{\mu}\cdot U_{\nu}\big)
 \,\gamma^{\nu}\,\gamma_5 \,U_{\mu}\, B
 +\big(\bar{\Delta}^{\mu}\cdot U_{\mu}\big)
 \,\gamma^{\nu}\,\gamma_5 \,U_{\nu}\,B\Big]+\textrm{h.c.}\,,
\\
 \nonumber\\
 &&\mathcal{L}^{(A,b)}= 
 -\frac14 \, f_{2}^{(A)}\,\tr\Big[\big(\bar{\Delta}^{\mu}\cdot \,\gamma^{\nu}\,\gamma_5\,B\big)
 \,\big[U_{\mu}\,,\,U_{\nu}\big]\Big]+\textrm{h.c.}
 \nonumber\\
&&\quad\,  -\frac14 \, f_{4}^{(A)} \, \tr\Big[\big(\bar{\Delta}^{\mu}\cdot U_{\nu}\big)
 \,\gamma^{\nu}\,\gamma_5 \,U_{\mu}\, B
 -\big(\bar{\Delta}^{\mu}\cdot U_{\mu}\big)
 \,\gamma^{\nu}\,\gamma_5 \,U_{\nu}\,B\Big]+\textrm{h.c.} \, ,
 \nonumber\\
 \nonumber\\
&& \mathcal{L}^{(F)}  =	  g_{F}^{(F)} \, \tr\Big[\bar{B} \, [ F_{\mu\nu}^+ \, , \, \sigma^{\mu\nu} \, B]\Big] +
 g_{D}^{(F)} \, \tr\Big[\bar{B}\, \{ F_{\mu\nu}^+ \, , \, \sigma^{\mu\nu} \, B\}\Big] 
\nonumber \\
 && \hspace{0.86cm} + \,  f_M^{(F)} \, \tr \Big[ \bar{B} \, \gamma^\mu \, \gamma_5 \, \big(i \,F_{\mu\nu}^+\cdot \Delta^\nu\big) \Big] + \textrm{h.c.}
 - \,  f_E^{(F)} \, \tr \Big[ \bar{B} \, \gamma^\mu \, \big(i \,F_{\mu\nu}^-\cdot \Delta^\nu\big) \Big] + \textrm{h.c.} \,. 
 \label{Eq:Def-L2}
 \end{eqnarray}
 The impact of $\mathcal{L}^{(F)}$ is not considered in this work. 
Furthermore we specify
  \begin{eqnarray}
  &&\mathcal{L}^{(\chi)}=
  -\frac12\,g_1^{(\chi)}\,\tr\,\big[\bar{B}\,\gamma_5\,\gamma^{\mu}\,\{i\,U_{\mu}\,,\,
\{\chi_0\,,\,B\}\}\big]+\textrm{h.c.}
-\frac12\,g_2^{(\chi)}\,\tr\,\big[\bar{B}\,\gamma_5\,\gamma^{\mu}\,\{i\,U_{\mu}\,,\,
[\chi_0\,,\,B]\}\big]+\textrm{h.c.}
\nonumber\\
  &&\hspace{1.29cm}
  -\frac12\,g_3^{(\chi)}\,\tr\,\big[\bar{B}\,\gamma_5\,\gamma^{\mu}\,[i\,U_{\mu}\,,\,
\{\chi_0\,,\,B\}]\big]+\textrm{h.c.}
  -\frac12\,g_4^{(\chi)}\,\tr\,\big[\bar{B}\,\gamma_5\,\gamma^{\mu}\,[i\,U_{\mu}\,,\,
[\chi_0\,,\,B]]\big]+\textrm{h.c.}
\nonumber\\
   &&\hspace{1.29cm}
   -\frac12\,g_5^{(\chi)}\,\tr\big[\bar{B}\,\gamma_5\,\gamma^{\mu}\,B\big]\,
   \tr[\chi_0\, i\,U_{\mu}]+\textrm{h.c.}
     -\frac12\,g_6^{(\chi)}\,\tr\big[\bar{B}\,\gamma_5\,\gamma^{\mu}\,i\,U_{\mu}\big]\,
   \tr[\chi_0\, B]+\textrm{h.c.}
   \nonumber\\
&&\hspace{1.29cm}
 -\,g_7^{(\chi)}\,\tr\big[\bar{B}\,\gamma_5\,\gamma^{\mu}\,[i\,U_{\mu}\,,\,B]\big]\, \tr[\chi_0]\,,
 \end{eqnarray}
 and
 \begin{eqnarray}
 \mathcal{L}^{(R)}=g_F^{(R)}\,\tr\Big[\bar{B}\,\gamma^{\mu}\,\gamma_5\,\big[[D^\nu\,,\,F_{\mu\nu}^-]\,,\,B\big]\Big]
+\,g_D^{(R)}\,\tr\Big[\bar{B}\,\gamma^{\mu}\,\gamma_5\,\big\{[D^\nu\,,\,F_{\mu\nu}^-]\,,\,B\big\}\Big]\,.
 \label{Eq:Def-L3}
\end{eqnarray}
Here two new operators appear, $\chi_0$ and $F_{\mu\nu}^\pm$, where
\begin{eqnarray}
\chi_0=2\,B_0
\begin{pmatrix}
m & 0 & 0 \\
0 & m & 0 \\
0&0&m_s
\end{pmatrix}
.
\label{Eq:Def-chi}
\end{eqnarray}
In order to implement explicit chiral symmetry breaking, 
$\chi_0$ introduces the finite quark masses $m$ and $m_s$, where we assume the strict isospin limit $m_u=m_d\equiv m$. 
Furthermore, a derivative acting on the external axial-vector current $a_\mu$ is needed,
\begin{eqnarray}
F_{\mu\nu}^\pm&=&u^\dagger\,F^R_{\mu\nu}\,u\, \pm \,u \,F^L_{\mu\nu}\,u^\dagger\,, 
  \label{Eq:Def-Fmunu}
\end{eqnarray}
with
\begin{eqnarray}
 F^R_{\mu\nu}&=&\partial_\mu\,a_\nu-\partial_\nu\,a_\mu-i\,[a_\mu\,,\,a_\nu]\,, 
\;\quad  \\
F^L_{\mu\nu}&=&-(\partial_\mu\,a_\nu-\partial_\nu\,a_\mu)-i\,[a_\mu\,,\,a_\nu]\,. 
 \end{eqnarray}
 All coefficients $g_{...}^{(...)}$ and $f_{...}^{(...)}$ are a priori unknown LECs. 
 Three LECs in Ref. \cite{Lutz:2001yb}, corresponding to $\mathcal{L}^{(\chi)}$, are redundant.
In Table \ref{Tab:LEC_Comparison} we link our Lagrangian $\mathcal{L}$ to other conventions found in the literature. Terms including exclusively the baryon octet have been derived in Refs. \cite{Frink:2006hx, Oller:2006yh, Oller:2007qd}, whereas the terms also including the decuplet are compared to Ref. \cite{Holmberg:2018dtv}. 
Our LECs $g_{...}^{(...)}$ and $f_{...}^{(...)}$ are normalized in such a way that we expect their values to be in the range $\sim[-10,10]$. 
Refs. \cite{Frink:2006hx, Oller:2007qd, Holmberg:2018dtv} use in part a  different normalization. The comparison between our LECs $f^{(A)}_{1,2,3,4}$ and the LECs $c_{F,D,(10),(27)}$, performed in Ref. \cite{Holmberg:2018dtv}, agrees with our translation in Table \ref{Tab:LEC_Comparison}. 

\begin{table}[ht]
\centering
\renewcommand{\arraystretch}{1.0}
\begin{tabular}{|c c|c c|}
\hline
this work & \cite{Frink:2006hx}, \cite{Oller:2007qd}, \cite{Holmberg:2018dtv}  & 
this work & \cite{Frink:2006hx}, \cite{Oller:2007qd}, \cite{Holmberg:2018dtv} 
\\ \hline
$D$ & $D$ & $g_0^{(S)}$ & $8\,(-b_1+b_2+b_4)$ \\
$F$ & $F$ & $g_1^{(S)}$ & $16\,(-b_1+b_2)$ \\
$C$ & $\frac{1}{\sqrt{2}}h_A$ & $g_F^{(S)}$ & $8\,b_3$ \\
$H$ & $H_A$ & $g_D^{(S)}$ & $8\,(3\,b_1-b_2)$  
\\ \hline
$g_0^{(V)}$ & $16\,(-b_5+b_7+b_8)$ & $g_1^{(T)}$ & $8\,d_3$ \\
$g_1^{(V)}$ & $32\,(-b_5+b_7)$ & $g_F^{(T)}$ & $16\,d_2$ \\
$g_F^{(V)}$ & $16\,b_6$ & $g_D^{(T)}$ & $16\,d_1$ \\
$g_D^{(V)}$ & $16\,(3\,b_5-b_7)$ & &
\\ \hline 
$g_F^{(F)}$ & $b_{M,F}$ & $f_1^{(A)}$ & $\frac{4}{5}\,c_{(27)}-4\, c_D$ \\
$g_D^{(F)}$ & $b_{M,D}$ & $f_2^{(A)}$ & $-\frac{8}{3}\,c_{(10)}-4 \, c_F$ \\
$f_M^{(F)}$ & $c_M$     & $f_3^{(A)}$ & $4\,c_{(27)}$ \\
$f_E^{(F)}$ & $c_E$     & $f_4^{(A)}$ & $-8\,c_{(10)}$  
\\ \hline 
 $g_F^{(R)}$ & $d_{64}$ & $g_D^{(R)}$ & $d_{65}$ 
\\ \hline
$g_1^{(\chi)}$ & $2\,(4\,d_{44}+3\,d_{46})$ & $g_5^{(\chi)}$ & $4\,(-d_{46}+d_{47})$ \\
 $g_2^{(\chi)}$ & $8\,d_{43}$ & $g_6^{(\chi)}$ & $-8\,d_{46}$ \\
 $g_3^{(\chi)}$ & $8\,d_{42}$ & $g_7^{(\chi)}$ & $4\,d_{45}$ \\
$g_4^{(\chi)}$& $2\,(4\,d_{41}+d_{46})$ &&
\\
\hline
\end{tabular}
\caption{Link of the LECs, used in this work, to previous publications \cite{Frink:2006hx},  \cite{Oller:2007qd}, and \cite{Holmberg:2018dtv}. }
\label{Tab:LEC_Comparison}
\end{table}

\section{Axial-Vector Currents} 
\subsection{Motivation}
The matrix element $\mathcal{M}$ of the semileptonic decay process of a nucleon or hyperon ($b\rightarrow \bar{b}+e^-+\bar{\nu}$) can be parameterized \cite{Cabibbo:2003cu} as
\begin{eqnarray}
&& \hspace{-0.3cm} \mathcal{M}=\frac{G_F\, V_{u\,d/s}}{\sqrt{2}} \, \bar{u}_{\bar{b}} \, \big(O_\alpha^V + O_\alpha^A\big) \, u_b \, \bar{u}_e \, \gamma^\alpha \, (1+\gamma_5) \, v_\nu \,,
 \nonumber\\
&& \hspace{-0.3cm} O_\alpha^V = f_1^{(\bar{b}b)}(q^2) \, \gamma_\alpha +  \frac{q^\beta}{M_B}\, f_2^{(\bar{b}b)}(q^2)\, \sigma_{\alpha\beta} + \frac{q_\alpha}{M_B}\, f_3^{(\bar{b}b)}(q^2)\,,
 \nonumber\\
&& \hspace{-0.3cm} O_\alpha^A = g_1^{(\bar{b}b)}(q^2) \, \gamma_\alpha \, \gamma_5+  \frac{q^\beta}{M_B}\, g_2^{(\bar{b}b)}(q^2)\, \sigma_{\alpha\beta}\,  \gamma_5 + \frac{q_\alpha}{M_B}\, g_3^{(\bar{b}b)}(q^2)\, \gamma_5\,,
\label{Eq:Cabibbo-Model}
\end{eqnarray}
with the Fermi constant $G_F$ and the Cabibbo-Kobayashi-Maskawa (CKM) matrix elements 
$V_{ud}\simeq 0.97$ for strangeness-conserving and $V_{us}\simeq 0.22$ for strangeness-changing decays \cite{Aoki:2019cca}. 
The form factors $g^{(\bar{b}b)}_{1,2,3}(q^2)$ and $f^{(\bar{b}b)}_{1,2,3}(q^2)$ depend on the momentum transfer $q^2$. 
We focus on the axial-vector form factor $g^{\bar{b}b}_1(q^2)$. Based on group-theoretical considerations and assuming exact flavor-SU(3) symmetry, the Cabibbo model is able to describe $g^{(\bar{b}b)}_1(q^2)$ of 12 processes with only two reduced matrix elements. These refer, translated to our framework, to the LECs $D$ and $F$ from the chiral Lagrangian $\mathcal{L}^{(1)}$ from Eq. \eqref{Eq:Def-L1}. 
We show Cabibbo's results for $g^{\bar{b}b}_1(q^2)$ in terms of  $D$ and $F$ in Table \ref{Tab:Cabibbo-Model}.

\begin{table}[ht]
\centering
\renewcommand{\arraystretch}{1.0}
\begin{tabular}{| c | c | c | c | c |}\hline
\rule{0pt}{3ex}
process $ \, b\rightarrow \bar{b} \, $&
\, CKM \,& 
$g_1^{(\bar{b}b)}(0)$ &
\, estimate \, &
experiment \cite{Tanabashi:2018oca}
\\ \hline
$n\rightarrow p \, e^- \, \bar{\nu}$ & $V_{ud}$ & $D+F$ & 1.25 & 1.2724 $\pm$ 0.0023
\\ 
$\Xi^-\rightarrow \Xi^0 \, e^- \, \bar{\nu}$ & $V_{ud}$ & $D-F$ & 0.35 & -
\\ 
$\Sigma^{-}\rightarrow \Lambda \, e^{-} \, \bar{\nu}$ & $V_{ud}$ & $\sqrt{\frac{2}{3}}\, D$ & 0.65 & 0.601 $\pm$ 0.015 \cite{Lutz:2001yb}
\\ 
$\Sigma^-\rightarrow \Sigma^0 \, e^- \, \bar{\nu}$ & $V_{ud}$ & $\sqrt{2} \, F$ & 0.64 & -
\\  \hline
$\Xi^0\rightarrow \Sigma^+ \, e^- \, \bar{\nu}$ & $V_{us}$ & $D+F$ & 1.25 & 1.22 $\pm$ 0.05
\\ 
$\Sigma^- \rightarrow n \, e^- \, \bar{\nu}$ & $V_{us}$ & $D-F$ & 0.35 & 0.340 $\pm$ 0.017
\\ 
$\Lambda \rightarrow p \, e^- \, \bar{\nu}$ & $V_{us}$ & $-\frac{1}{\sqrt{6}}\,(D+3\,F)$ & -0.88 & -0.879 $\pm$ 0.018 
\\ 
$\Xi^- \rightarrow \Lambda \, e^- \, \bar{\nu}$ & $V_{us}$ & $-\frac{1}{\sqrt{6}}\,(D-3\,F)$ & 0.22 & 0.31 $\pm$ 0.06
\\ \hline 
$n\rightarrow n$ & - & $-\frac{1}{\sqrt{6}}\, (D-3\, F)$ & 0.22 & -
\\ 
$\Sigma^- \rightarrow \Sigma^-$ & - & $\sqrt{\frac{2}{3}}\, D$ & 0.65 & -
\\ 
$\Lambda \rightarrow \Lambda$ & - & $-\sqrt{\frac{2}{3}}\, D$ & -0.65 & -
\\ 
$\Xi^- \rightarrow \Xi^-$ & - & $-\frac{1}{\sqrt{6}}\,(D+3\,F)$ & -0.88 & -
\\ \hline
\end{tabular}
\caption{Cabibbo theory for the form factor $g_1^{(\bar{b}b)}(0)$ \cite{Cabibbo:2003cu}. We use $F=0.45$ and $D=0.80$ \cite{Semke:2005sn} for the numerical estimate. }
\label{Tab:Cabibbo-Model}
\end{table}

Roughly ten years ago first results from non-perturbative calculations of axial-vector form factors on the lattice became available (e.g. Refs. \cite{Lin:2007ap, Gockeler:2011ze}). The quark mass is an input parameter in these kinds of calculations. Therefore additional data points can be produced, also in unphysical regions \cite{Ishikawa:2018rew, Alexand:2018pln, Bali:2019yiy, Bali:2019svt}. 
Since for larger quark masses flavor-SU(3)-breaking effects play an important role, the Cabibbo model is not suitable for more detailed investigations. $\chi$PT offers the opportunity to consistently add higher-order contributions, such as finite quark masses. Therefore we scrutinize axial-vector form factors in $\chi$PT to one-loop order and derive their dependence on the LECs of the chiral Lagrangian of Eq. \eqref{Eq:Def-chiralLagrangian}.

\subsection{Definition of Axial-Vector Form Factors}

In QCD the axial-vector currents take the form \cite{Gasser:1984gg}
\begin{eqnarray}
 A_i^\mu(x) = \bar{q}(x) \gamma^\mu \gamma_5 \frac{\lambda_i}{2} q(x)\,, 
 \label{Eq:QCD-Axial-Vector-Current}
\end{eqnarray}
with the 3-component quark field $q(x)$ and the Gell-Mann matrices $\lambda_i$ for flavor-SU(3).
The axial-vector couplings can be incorporated as external fields $a_\mu(x)= \sum_{i} a_\mu^i(x) \frac{\lambda_i}{2}$ in the QCD Lagrangian. They enter the chiral Lagrangian as a part of $U_\mu$ in Eq. \eqref{Eq:Def-CovariantDerivative}.

We consider an axial-vector current $A_{i}^{\mu}(x=0)$ coupling to an incoming baryon $B$ (with mass $M_p$ and momentum $p$) and an outgoing baryon $\bar{B}$ (with mass $M_{\bar{p}}$ and momentum $\bar{p}$). In contrast to the baryons $\bar{b}$ and $b$ in the previous Section, we present the baryons here in isospin multiplets $\bar{B}$ and $B$, shown in Table \ref{Tab:Baryon-Multiplets}.  Isospin-breaking effects are neglected. We restrict the external baryon fields to the octet.
The matrix element for the axial-vector current can be decomposed into three parts as
\begin{eqnarray}
 &&\bra{\bar{B}(\bar{p})}A_{i}^{\mu}(0)\ket{B(p)}=
 \bar{u}_{\bar{B}}(\bar{p})\,\Big(\gamma^{\mu}\,G_{A,i}^{(\bar{B}B)}(q^2)+
 \frac{q^{\mu}}{M_{\bar{p}}+M_p}\,G_{P,i}^{(\bar{B}B)}(q^2)  \nonumber\\ 
 &&\hspace{1,7cm} +\frac{(\slashed{\bar{p}}+\slashed{p})\,(\bar{p}^\mu+p^\mu)}{(M_{\bar{p}}+M_p)^2} \,G_{I,i}^{(\bar{B}B)}(q^2)\Big)\,\gamma_5\, T_{I(\bar{B})I(B)}^{(i)}\,u_B(p)\,,
 \label{Eq:Def-Axial-Form-Factor-SU3}
\end{eqnarray}
with $q^\mu=(\bar{p}-p)^\mu$ and the index $i=1,...,8$ running over the different types of strangeness-conserving ($\Delta S=0$) and strangeness-changing ($\Delta S=\pm 1$) axial-vector currents.
\begin{table}[ht]
\centering
\renewcommand{\arraystretch}{1.0}
\begin{tabular}{|c| c | c |}\hline
singlet&
doublets&
triplet
\\ \hline
$\Lambda=\big(\Lambda^0\big)$
&$N=\begin{pmatrix}p\\n\end{pmatrix}$\hspace{0.4cm}
$\Xi=\begin{pmatrix}\Xi^{0}\\ \Xi^{-}\end{pmatrix}$&
$\Sigma=\begin{pmatrix}\frac{\Sigma^{+}+\Sigma^{-}}{\sqrt{2}}\\
\frac{i\Sigma^{+}-i\Sigma^{-}}{\sqrt{2}}\\
\Sigma^{0}\end{pmatrix}$
 \\ \hline
\end{tabular}
\caption{Baryon-octet multiplets $B$.}
\label{Tab:Baryon-Multiplets}
\end{table}
The coefficients $G_{A,i}^{(\bar{B}B)}(q^2)$ and $G_{P,i}^{(\bar{B}B)}(q^2)$ are known as the axial-vector and pseudoscalar form factors. The third form factor  $G_{I,i}^{(\bar{B}B)}(q^2)$ is, by construction, only non-zero for processes with $\bar{B}\neq B$ for some specific Feynman diagrams.
\\

Strict isospin symmetry forces the axial-vector form factor $G_{A,i}^{(\bar{B}B)}(q^2)$ to be equal within isospin multiplets. We express the form factor in the isospin basis and use the isospin equivalent  axial-vector current $a_\mu^X$ with $X\in\{\pi, K, \bar{K}, \eta\}$, {\it i.e.}
\begin{eqnarray}
 && G_{A,1}^{(\bar{B}B)}=G_{A,2}^{(\bar{B}B)}=G_{A,3}^{(\bar{B}B)} \rightarrow G_{A,a_\mu^\pi}^{(\bar{B}B)}\,, \hspace{0.4cm}
 G_{A,8}^{(\bar{B}B)} \rightarrow G_{A,a_\mu^\eta}^{(\bar{B}B)}\,,
 \nonumber\\ 
 && G_{A,4}^{(\bar{B}B)}=G_{A,5}^{(\bar{B}B)}\rightarrow G_{A,a_\mu^K}^{(\bar{B}B)}\,, \hspace{0.4cm}
 G_{A,6}^{(\bar{B}B)}=G_{A,7}^{(\bar{B}B)}\rightarrow G_{A,a_\mu^{\bar{K}}}^{(\bar{B}B)}\,. 
 \label{Eq:Form-Factor-Multiplets}
\end{eqnarray}
The isospin transition matrices $T_{I(\bar{B})I(B)}^{(i)}$ depend on the isospin of the outgoing/incoming baryon $I(\bar{B})/I(B)$ and the isospin of the axial-vector current $I(i)$,
\begin{eqnarray}
 T_{I(\bar{B})I(B)}^{(i)}=
 \begin{cases}\vspace{0.2cm}
 T_{I(\bar{B})I(B),i}^{(1)}&\text{if} \hspace{0.2cm} i=1,2,3 \, , 
 \\ \vspace{0.2cm}
 T_{I(\bar{B})I(B),i-3}^{(1/2)}&\text{if} \hspace{0.2cm} i=4,5 \, ,
 \\ \vspace{0.2cm}
 \tilde{T}_{I(\bar{B})I(B),i-5}^{(1/2)}&\text{if} \hspace{0.2cm} i=6,7 \, ,
 \\ 
 T_{I(\bar{B})I(B),i-7}^{(0)}&\text{if} \hspace{0.2cm} i=8 \, .
 \end{cases}
 \label{Eq:Def-Iso-Transition-Matrices}
\end{eqnarray}
An isospin transition matrix $T_{I(\bar{B})I(B),i}^{I(i)}$ has $2\,I(i)+1$ components, 
each of them being a $\big(2\, I(\bar{B})+1\big)\times \big(2\, I(B)+1\big)$ matrix. Specifically,
\begin{eqnarray}
&& \Big(T_{00,1}^{(0)}\Big)=\frac{1}{\sqrt{2}}\,, \hspace{0.6cm}
\Big(T_{\frac{1}{2}\frac{1}{2},1}^{(0)}\Big)_{jk}=\frac{1}{\sqrt{2}} \, (\delta)_{jk}\,, \hspace{0.6cm}
 \Big(T_{11,1}^{(0)}\Big)_{jk}=\frac{1}{\sqrt{2}} \, (\delta)_{jk}\,,
 \nonumber\\
 && \Big(T_{11,i}^{(1)}\Big)_{jk}=-\frac{i}{\sqrt{2}} \, \epsilon_{ijk}\, , \hspace{0.6cm} 
 \Big(T_{0  1,i}^{(1)}\Big)_j=\Big(T_{10,i}^{(1)}\Big)_j=\frac{1}{\sqrt{2}} \, (\delta^i)_j\,,
\Big(T_{\frac{1}{2}\frac{1}{2},i}^{(1)}\Big)_{jk}=(-1)^{m_{1/2}}\frac{1}{2} \, (\tau^i)_{jk}\,, 
 \nonumber\\
&& \Big(T_{\frac{1}{2}1,i}^{(1/2)}\Big)_{jk}= \frac{1}{2} \, (\tau^k)_{ji}\,,\hspace{0.6cm}
\Big(T_{\frac{1}{2}0,i}^{(1/2)}\Big)_{j}= \frac{1}{\sqrt{2}} \, (\delta^i)_{j}\,,
 \nonumber\\
&&\Big(T_{1\frac{1}{2},i}^{(1/2)}\Big)_{jk}= \frac{i}{2}  \, (\tau_2)_{il} \, \big(\tau^j\big)_{lk}\,, \hspace{0.6cm}
 \Big(T_{0\frac{1}{2},i}^{(1/2)}\Big)_{j}= -\frac{i}{\sqrt{2}} \, (\tau_2)_{ij}\,,  
  \nonumber\\
  && \Big(\tilde{T}_{1\frac{1}{2},i}^{(1/2)}\Big)_{jk}= \frac{1}{2} \, (\tau^j)_{ki}\,,\hspace{0.6cm}
\Big(\tilde{T}_{0\frac{1}{2},i}^{(1/2)}\Big)_{j}= \frac{1}{\sqrt{2}} \, (\delta^i)_{j}\,,
 \nonumber\\
&&\Big(\tilde{T}_{\frac{1}{2}1,i}^{(1/2)}\Big)_{jk}= -\frac{i}{2} \, \big(\tau^k\big)_{jl} \, (\tau_2)_{li}\,, \hspace{0.6cm}
 \Big(\tilde{T}_{\frac{1}{2}0,i}^{(1/2)}\Big)_{j}= \frac{i}{\sqrt{2}} \, (\tau_2)_{ij}\,,  
 \label{Eq:Def-Transition-Matrices}
\end{eqnarray}
with $ m_{1/2} = 1 $ for $ \bar{B}=B=\Xi $  and  $ m_{1/2} = 0 $ for $ \bar{B}=B=N$. 

The transition matrices are normalized in such a way that we recover the results of the Cabibbo model at tree level, 
\begin{eqnarray}
 G_{A,i\neq 8}^{(\bar{B}B)}\big(q^2\big) = g_1^{(\bar{b}b)}\big(q^2\big)\,.
\end{eqnarray}
We work in the isospin basis with the external baryons $\bar{B}$ and $B$, whereas the form factors in Cabibbo's model are given in particle basis with $\bar{b}$ and $b$. 
The singlet form factors $G_{A,i=8}^{(\bar{B}B)}\big(q^2\big)$ are not related to physical decays. 

We will determine the form factors $G_{A,i}^{(\bar{B}B)}(q^2)$ to one-loop level. Divergent loop integrals need to be treated carefully. The technique of dimensional regularization \cite{tHooft:1972tcz, Leibbrandt:1975are} introduces the renormalization scale $\mu$, which allows us to determine the integrals in $d$ space-time dimensions. Subtracting poles of the form $1/(d-4)$ leads to a finite result, which only depends logarithmically on $\mu$. 
The dimensionful renormalization scale $\mu$ does not interfere with the chiral power counting. Nevertheless, the existence of four different scales, namely the Goldstone-boson mass $m_{\text{GB}}$, the baryon octet(decuplet) mass in the chiral limit $M(M+\Delta)$, and the break-down scale $\Lambda_{\chi\text{PT}}$, require consistent power counting. 
There are different ways to achieve this. Jenkins and Manohar \cite{Jenkins:1990jv} introduced a scheme which uses an expansion in inverse baryon masses (heavy-baryon formalism). The method of infrared regularization was applied by Becher and Leutwyler \cite{Becher:1999he}. The so-called small-scale expansion \cite{Hemmert:1997ye} uses $\Delta$, the decuplet-octet mass difference in the chiral limit, as an expansion parameter. 
In this work we will follow the strategy of Refs. \cite{Lutz:2018cqo, Lutz:2020dfi}, where suitable subtractions assure a consistent power counting, but no expansion in $\Delta$ is performed. The details of our power-counting scheme are given in Chapter \ref{Section:Renormalization}.

\section{Determination of Axial-Vector Form Factors}

\subsection{Analytical results} 

\begin{figure}[ht]
\begin{center}
\includegraphics[scale=1.0]{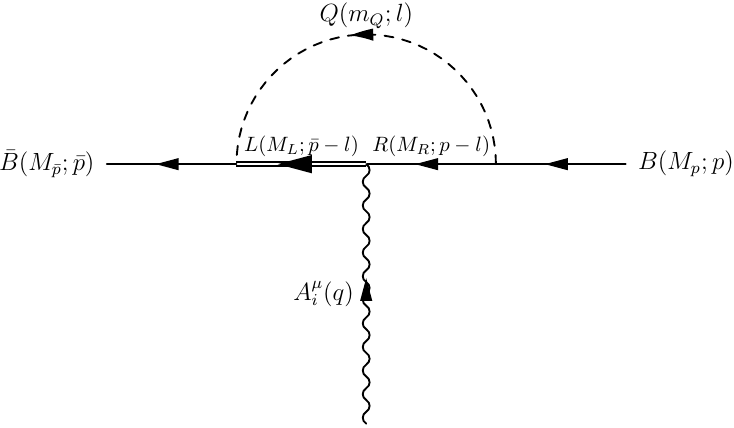}
\end{center}
\caption{Example of a one-loop diagram contributing to $G_{A,i}^{\bar{B}B}(t)$. The Goldstone boson $Q$ is represented by a dashed line, whereas the wiggly line is associated with the axial-vector current $A_i^\mu$. Baryons ($\bar{B}, L, R, B$) are shown with straight lines, a single line is connected to a octet particle and a double line represents a decuplet particle. We use the notation \textit{`label (mass ; momentum)'}.  }
\label{Fig:FeynBBJExampleDiagram}
\end{figure}

In this Section we investigate all axial-vector form factors $G_{A,i}^{\bar{B}B}(t)$, with $t=q^2$, to N$^2$LO. We use the baryon multiplets (see Table \ref{Tab:Baryon-Multiplets}) and the axial-vector currents $i=\{a_\mu^\pi, a_\mu^K, a_\mu^{\bar{K}}, a_\mu^\eta\}$ in the isospin basis.  
A typical one-loop diagram is shown in Figure \ref{Fig:FeynBBJExampleDiagram}.
In addition to the tree-level diagram there are three different types of one-loop diagrams contributing to $G_{A,i}^{\bar{B}B}(t)$: tadpole (Table \ref{Tab:FeynmanDiagrams1}), bubble (Table \ref{Tab:FeynmanDiagrams2}), and triangle diagrams (Table \ref{Tab:FeynmanDiagrams3}). 
All these contributions can be summarized in one condensed formula:
\begin{eqnarray}
&& G_{A,i}^{\bar{B}B}(t)=\,\sqrt{Z_{\bar{B}}Z_B}\, J^A\, C_{i}^{(\bar{B}B)}
 + t\, J^A\,T_{i}^{(\bar{B}B)}+  
 2\,B_0\,J^A\,X_{i}^{(\bar{B}B)}
 + \sum_{Q\in [8]}J_{Q}^{A}\,\frac{C_{Q,i}^{(\bar{B}B)}}{(2f)^2}
\nonumber\\
&& \hspace{1.5cm} +  \sum_{n=1}^5\sum_{Q\in [8]}\ \sum_{L\in [8],[10]} J_{LQ}^{A,n}\,
\frac{G^{(\bar{B})}_{QL}}{2f}\,\frac{C_{LQ,i}^{(\bar{B}B), n}}{2f}
+  \sum_{n=1}^5\sum_{Q\in [8]}\ \sum_{R\in [8],[10]} J_{QR}^{A,n}\,
\frac{C^{(\bar{B}B),n}_{QR,i}}{2f}\,\frac{G^{(B)}_{QR}}{2f}
\nonumber\\
&& \hspace{1.5cm} +  \sum_{n=1}^2 \sum_{Q\in [8]} \sum_{L,R\in [8],[10]}J_{LQR}^{A,n}\,
\frac{G^{(\bar{B})}_{QL}}{2f}\,
C_{LQR,i}^{(\bar{B}B),n}\,\frac{G^{(B)}_{QR}}{2f}\,,
\label{Eq:GA-all-contributions-of-diagrams}
\end{eqnarray}
where we separate the flavor-independent part $J_{...}^A$ from the Clebsch-Gordan coefficients (CGCs) $C$, $T$, $X$, and $G$, which carry the full flavor information. Sums over the multiplets $[8]$ and $[10]$ of the internal particles $L$, $R$, and $Q$ (see Figure \ref{Fig:FeynBBJExampleDiagram}) indicate loop contributions. 
The parameter $n$ distinguishes bubble diagrams with contributions from different vertices (see Table \ref{Tab:FeynmanDiagrams2}).
We also introduced the wave-function renormalization $\sqrt{Z_{\bar{B}}Z_B}$ \cite{Jenkins:1990jv, Ledwig:2014rfa} modifying the contribution of the tree-level diagram. 

The expressions $J^A_{...}$ in Eq. \eqref{Eq:GA-all-contributions-of-diagrams} represent the axial-vector component of the amplitudes $J_{...}^\mu$ of all considered Feynman diagrams. They are given by
\begin{eqnarray}
&&J_{...}^A=
-\frac{\sqrt{2} \, M_{\bar{p}} \, M_p}
{(d-2)\,(M_{\bar{p}} \, M_p+\bar{p}\cdot p)}\,
\tr \Bigg[\gamma_{\mu} \, \gamma_5\;
\frac{\bar{\slashed{p}}+M_{\bar{p}}}{2 \, M_{\bar{p}}} \, J^{\mu}_{...} \,\frac{\slashed{p}+M_p}{2 \, M_p}\Bigg]
\nonumber\\
&& \hspace{0.5cm} -\frac{M_{\bar{p}} \, M_p\, (M_{\bar{p}}+M_p)^2}
{\sqrt{2} \, (d-2) \, (M_{\bar{p}} \, M_p-\bar{p}\cdot p) \, (M_{\bar{p}} \, M_p+\bar{p}\cdot p)}
\, \tr\Bigg[\frac{\bar{p}_{\mu}-p_{\mu}}{M_{\bar{p}}+M_p} \, \gamma_5\;
\frac{\bar{\slashed{p}}+M_{\bar{p}}}{2 \, M_{\bar{p}}} \, J^{\mu}_{...} \,\frac{\slashed{p}+M_p}{2 \, M_p}\Bigg]
\label{Eq:Projection-JA}
\\ \nonumber
&& \hspace{0.5cm} -\frac{M_{\bar{p}} \, M_p \, (M_{\bar{p}}+M_p)^2}
{\sqrt{2} \, (d-2) \, (M_{\bar{p}} \, M_p-\bar{p}\cdot p) \, (M_{\bar{p}} \, M_p+\bar{p}\cdot p)}
\, \tr\Bigg[\frac{(\slashed{\bar{p}}+\slashed{p}) \, (\bar{p}_\mu + p_{\mu})}{(M_{\bar{p}}+M_p)^2} \, \gamma_5\;
\frac{\bar{\slashed{p}}+M_{\bar{p}}}{2 \, M_{\bar{p}}}\, J^{\mu}_{...} \,\frac{\slashed{p}+M_p}{2 \, M_p}\Bigg]\,, 
\end{eqnarray}
where the projector $\frac{\bar{\slashed{p}}+M_{\bar{p}}}{2M_{\bar{p}}}\Big(\frac{\slashed{p}+M_p}{2M_p}\Big)$ sets the outgoing (incoming) baryon $\bar{B}(B)$ on shell. The trace is taken in Dirac space. For completeness we also give the expressions $J_{...}^P$ and $J_{...}^I$, which project onto the form factors $G_P$ and $G_I$ in Eqs. \eqref{JPallcontributionsnew} and \eqref{JIallcontributionsnew} in Appendix \ref{Chapter:Appendix-Amplitudes}.
Using the method of dimensional regularization \cite{Leibbrandt:1975are}, we work in $d$ space-time dimensions.

In Tables \ref{Tab:FeynmanDiagrams1}-\ref{Tab:FeynmanDiagrams3} we show all considered Feynman diagrams with the link to the amplitude $J_{...}^\mu$, defined in Appendix \ref{Chapter:Appendix-Amplitudes}. The resulting projection $J_{...}^A$ enters the formula for the axial-vector form factor \eqref{Eq:GA-all-contributions-of-diagrams}.

\begin{table}[ht]
\centering
\renewcommand{\arraystretch}{1.0}
\begin{tabular}{|c|c|c|c|c|c|}\hline
type & Feynman diagram & projection & amplitude $J_{...}^\mu$ & CGCs & Lagrangian $\mathcal{L}$ \\ \hline

tree&
\includegraphics[scale=1.0]{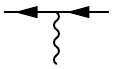}
&$J^A$
&\eqref{Eq:Amplitude-Tree} \,
& $C_{i}^{(\bar{B}B)}$ \, 
&$\mathcal{L}^{(1)}$ \,
\\ \hline

counterterms &
\includegraphics[scale=1.0]{ZFeynTree}
&$J^A$
&\parbox[c]{2.4cm}{\vspace{-0.6cm} \eqref{Eq:Amplitude-Tree}\,,\\ \eqref{Eq:Amplitude-Counter-Term} \, }
&\parbox[c]{2.4cm}{\vspace{-0.8cm} $X_{i}^{(\bar{B}B)}$\,,\\ $T_{i}^{(\bar{B}B)}$ \, }
&\parbox[c]{2.4cm}{\vspace{-0.8cm} $\mathcal{L}^{(\chi)}$\,,\\ $\mathcal{L}^{(R)}$ \, }
\\ \hline

tadpole&
\includegraphics[scale=1.0]{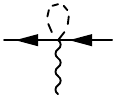}
&$J_Q^A$
& \eqref{Eq:Amplitude-Tadpole} \,
& $C_{Q,i}^{(\bar{B}B)}$ \, 
&$\mathcal{L}^{(1)}$ \,
\\ \hline

\end{tabular}
\caption{Contribution of all tree-level and tadpole diagrams entering Eq. \eqref{Eq:GA-all-contributions-of-diagrams}. The meaning of different lines is explained in Figure \ref{Fig:FeynBBJExampleDiagram}.}
\label{Tab:FeynmanDiagrams1}
\end{table}

\begin{table}[ht]
\centering
\renewcommand{\arraystretch}{1.0}
\begin{tabular}{|c|c|c|c|c|c|}\hline
type & Feynman diagram & projection & amplitude $J^\mu_{...}$ & Lagrangian $\mathcal{L}$ \\ \hline

\parbox[b]{1.3cm}{ bubble \\ $L\in[8]$ \\ }&
\includegraphics[scale=1.0]{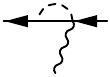}
&$J_{LQ}^{A,n}$
& \parbox[c]{1.2cm}{
\eqref{Eq:Amplitude-BubbleL8-1}\,, \\
\eqref{Eq:Amplitude-BubbleL8-2}\,, \\
\eqref{Eq:Amplitude-BubbleL8-3}\,, \\
\eqref{Eq:Amplitude-BubbleL8-4}\,, \\
\eqref{Eq:Amplitude-BubbleL8-5}\,\,}
&\parbox[c]{2.6cm}{
$n=1 \leftrightarrow \mathcal{L}_{\text{kin}}$\,,   \\
$n=2 \leftrightarrow \mathcal{L}^{(S)}$\,, \\
$n=3 \leftrightarrow \mathcal{L}^{(V)}$\,, \\
$n=4 \leftrightarrow \mathcal{L}^{(T)}$\,, \\
$n=5 \leftrightarrow \mathcal{L}^{(F)}$\,\, }
\\  \hline 

\parbox[b]{1.3cm}{bubble \\ $R\in[8]$  \\}&
\includegraphics[scale=1.0]{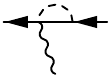}
&$J_{QR}^{A,n}$
&\parbox[c]{1.2cm}{
\eqref{Eq:Amplitude-BubbleR8-1}\,, \\
\eqref{Eq:Amplitude-BubbleR8-2}\,, \\
\eqref{Eq:Amplitude-BubbleR8-3}\,,\\
\eqref{Eq:Amplitude-BubbleR8-4}\,, \\
\eqref{Eq:Amplitude-BubbleR8-5}\,\,\,}
&\parbox[c]{2.6cm}{ 
$n=1 \leftrightarrow \mathcal{L}_{\text{kin}}$\,,   \\
$n=2 \leftrightarrow \mathcal{L}^{(S)}$\,, \\
$n=3 \leftrightarrow \mathcal{L}^{(V)}$\,, \\
$n=4 \leftrightarrow \mathcal{L}^{(T)}$\,, \\
$n=5 \leftrightarrow \mathcal{L}^{(F)}$\,\, }
\\ \hline

\parbox[b]{1.5cm}{ bubble \\ $L\in[10]$ \\ }&
\includegraphics[scale=1.0]{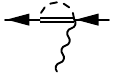}
&$J_{LQ}^{A,n}$
& \parbox[c]{1.8cm}{
\eqref{Eq:Amplitude-BubbleL10-1}\,,  \\
\eqref{Eq:Amplitude-BubbleL10-2}\,,  \\
\eqref{Eq:Amplitude-BubbleL10-3}\,\,\,\;}
&\parbox[c]{2.9cm}{ 
$n=1 \leftrightarrow \mathcal{L}^{(A,a)}$\,,  \\
$n=2 \leftrightarrow \mathcal{L}^{(A,b)}$\,, \\
$n=3 \leftrightarrow \mathcal{L}^{(F)}$\phantom{x,}\,\,\, 
}
\\ \hline

\parbox[b]{1.5cm}{ bubble \\ $R\in[10]$ \\}&
\includegraphics[scale=1.0]{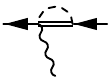}
&$J_{QR}^{A,n}$
& \parbox[c]{1.2cm}{ 
\eqref{Eq:Amplitude-BubbleR10-1}\,,  \\
\eqref{Eq:Amplitude-BubbleR10-2}\,,  \\
\eqref{Eq:Amplitude-BubbleR10-3}\,\,\,\; }
&\parbox[c]{2.9cm}{
$n=1 \leftrightarrow \mathcal{L}^{(A,a)}$\,,  \\
$n=2 \leftrightarrow \mathcal{L}^{(A,b)}$\,, \\
$n=3 \leftrightarrow \mathcal{L}^{(F)}$\phantom{x,}\,\,\, 
}
\\ \hline

\end{tabular}
\caption{Contribution of all bubble diagrams entering Eq. \eqref{Eq:GA-all-contributions-of-diagrams}. The meaning of different lines is explained in Figure \ref{Fig:FeynBBJExampleDiagram}. The parameter $n$ specifies the origin of the 4-point vertices in bubble diagrams.}
\label{Tab:FeynmanDiagrams2}
\end{table}

\begin{table}[ht]
\centering
\renewcommand{\arraystretch}{1.0}
\begin{tabular}{|c|c|c|c|c|c|}\hline
type & diagram & projection & amplitude $J^\mu_{...}$   \\ \hline

\parbox[b]{1.5cm}{triangle \\ $L\in[8]$ \\$R\in[8]$ \\}&
\includegraphics[scale=1.0]{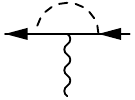}
&$J_{LQR}^{A,1}$
&\eqref{Eq:Amplitude-TriangleL8R8}
\\ \hline 

\parbox[b]{1.5cm}{triangle \\ $L\in[10]$ \\$R\in[8]$ \\}&
\includegraphics[scale=1.0]{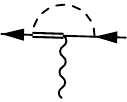}
&$J_{LQR}^{A,1,2}$
&\eqref{Eq:Amplitude-TriangleL10R8}, \eqref{Eq:Amplitude-TriangleL10R8-2}
\\ \hline 

\parbox[b]{1.5cm}{triangle \\ $L\in[8]$ \\$R\in[10]$ \\}&
\includegraphics[scale=1.0]{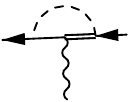}
&$J_{LQR}^{A,1,2}$
&\eqref{Eq:Amplitude-TriangleL8R10}, \eqref{Eq:Amplitude-TriangleL8R10-2} 
\\ \hline 

\parbox[b]{1.5cm}{triangle \\ $L\in[10]$ \\$R\in[10]$ \\}&
\includegraphics[scale=1.0]{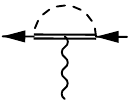}
& $J_{LQR}^{A,1}$
&\eqref{Eq:Amplitude-TriangleL10R10}
\\ \hline
\end{tabular}
\caption{Contribution of all non-vanishing triangle diagrams entering Eq. \eqref{Eq:GA-all-contributions-of-diagrams}. The meaning of different lines is explained in Figure \ref{Fig:FeynBBJExampleDiagram}.}
\label{Tab:FeynmanDiagrams3}
\end{table}

We now turn to the flavor structure and take a closer look at the CGCs $C$, $T$, $X$, and $G$. 
The 3-point CGCs $G^{(B)}_{QR}$ enter the calculation of the baryon masses and are given in Ref. \cite{Semke:2005sn}, but for completeness we give them in Tables \ref{Tab:Clebsches3pointG88} - \ref{Tab:Clebsches3pointG1010}. 
The counterterm CGCs, which originate from the Lagrangian $\mathcal{L}_3$ \eqref{Eq:Def-L3}, $T_{i}^{(\bar{B}B)}$ and $X_{i}^{(\bar{B}B)}$, are used to absorb the divergences of the one-loop contributions. 
Explicit symmetry breaking gives raise to the coefficients $X_{i}^{(\bar{B}B)}$, which are proportional to the quark masses $m$ and $m_s$. They are shown in Table \ref{Tab:Clebsches-Counterterms-X}.
The coefficients $T_{i}^{(\bar{B}B)}\propto t=q^2$ are important to describe the momentum transfer dependence of the axial-vector current. Their explicit contributions are shown in Table \ref{Tab:Clebsches-Counterterms-T}. 

The CGCs $C^{(\bar{B}B)}_{...,i}$ can further be decomposed: 
\begin{eqnarray}
 && C_i^{(\bar{B}B)}=A_{\bar{B}B}^{(i)}\,,
 \nonumber \\
 && C_{Q,i}^{(\bar{B}B)}=A_{Q,i}^{(\bar{B}B)}\,,
 \nonumber \\
 && C_{LQ,i}^{(\bar{B}B),n}=A_{QL,(\bar{B})}^{(i B),n} \, \tilde{C}_{i}^{\bar{B}B}\,,
 \nonumber \\
 && C_{QR,i}^{(\bar{B}B),n}=A_{QR,(B)}^{(i\bar{B}),n} \, \tilde{C}_{\bar{i}}^{B\bar{B}}\,,
 \nonumber \\
 && C_{LQR,i}^{(\bar{B}B),1}=A_{LR}^{(i)} \, \tilde{C}_{\bar{B}B,LR}^{iQ}\,,
 \label{Eq:Def-Recoupling-Constants}
 \end{eqnarray}
where we defined the 3-, 4-, and 5-point CGCs $A_{...}^{(...)}$ and the recoupling constants $\tilde{C}_{...}^{...}$. 
%
The 3-point coefficients $A$ differ from the $G$ coefficients, because they couple an axial-vector current instead of a Goldstone boson to the two baryons. Nevertheless, they only differ in their particle-specific normalization. All needed coefficients $A_{\bar{B}B}^{(i)}$ are given in Tables \ref{Tab:Clebsches3pointA88} - \ref{Tab:Clebsches3pointA1010}.
We note that there are some correlations among such coefficients. For instance we 
find
\begin{eqnarray}
&& C_{LQ,i}^{(\bar{B}B),5} = C_{LQ,i}^{(\bar{B}B),4} \Big|_{ g^T_{D,F}  \to 8\sqrt{3}\,g^F_{D,F} \; {\rm and}\; g^T_1 \to 0 }\, \quad {\rm for} \quad {L \in [8]} \,,
\nonumber\\
&& C_{QR,i}^{(\bar{B}B),5} = C_{QR,i}^{(\bar{B}B),4} \Big|_{ g^T_{D,F}  \to 8\sqrt{3}\,g^F_{D,F} \; {\rm and}\; g^T_1 \to 0 }\, \quad {\rm for} \quad {R \in [8]} \,,
\nonumber\\
&& C_{LQ,i}^{(\bar{B}B),3} = C_{LQ,i}^{(\bar{B}B),2} \Big|_{ f^A_i  \to 0\; {\rm but}\; f^A_2 \to 16 f^F_F }\, \quad \qquad {\rm for} \quad {L \in [10]} \,,
\nonumber\\
&& C_{QR,i}^{(\bar{B}B),3} = C_{QR,i}^{(\bar{B}B),2} \Big|_{  f^A_i  \to 0\; {\rm but}\; f^A_2 \to 16 f^F_F  }\, \quad \qquad {\rm for} \quad {R \in [10]} \,,
\nonumber\\
&&C_{LQR,i}^{(\bar{B}B),2} = C_{LQR,i}^{(\bar{B}B),1}
\Big|_{C_A \to -2\,f^F_E } \,.
\end{eqnarray}

The 5-point CGCs $A_{Q,i}^{(\bar{B}B)}$ can directly be linked to the 3-point coefficients $A_{\bar{B}B}^{(i)}$ (see Table \ref{Tab:Clebsches5point}).
4-point vertices contributing to the bubble diagrams require some more attention. First of all, several parts of the Lagrangians $\mathcal{L}_\text{kin}$ \eqref{Eq:Def-Lkin} and $\mathcal{L}^{(2)}$ \eqref{Eq:Def-L2split} contribute. This is labeled by the parameter $n$, the exact mapping can be found in Table \ref{Tab:FeynmanDiagrams2}. The 4-point CGCs $A_{QL,(\bar{B})}^{(i B),n}$ depend, in addition to the connected particles $L$, $Q$, $B$, and the current $i$, on the isospin of the outgoing baryon $\bar{B}$. All CGCs $A_{QL,(\bar{B})}^{(i B),n}$ can be found in the Appendix in Tables \ref{Tab:Clebsches4point8n1} - \ref{Tab:Clebsches4point10n2} and Eqs. \eqref{Eq:Clebsch-Vector-Couplings} and \eqref{Eq:Clebsch-n-running-for-transitions}. 
The CGCs $A_{QR,(B)}^{(i\bar{B}),n}$ are linked to $A_{QL,(\bar{B})}^{(i B),n}$ via the relations given in Eq. \eqref{Eq:Clebsch-4-point-vertices-RvsL}.

It turns out that the CGCs $G$ and $A$ are not enough to describe the full flavor structure of all bubble and triangle diagrams. We need additional `recoupling constants' $\tilde{C}$, which compensate for the use of isospin multiplets in the strict isospin limit. Of course the goal is to be consistent with results in particle basis. We have explicitly checked that this is achieved by the inclusion of the recoupling constants $\tilde{C}$. 
The recoupling constants of the bubble diagrams  $\tilde{C}_{i}^{\bar{B}B}$ depend on external particles only and are given in Table \eqref{Tab:Recouplingconstants-Bubble}. 
The situation turns slightly more sophisticated for triangle diagrams. The recoupling constants $\tilde{C}_{\bar{B}B,LR}^{iQ}$ depend on all involved particles and are shown in Eqs. \eqref{Eq:Recouplingsconstants-Triangle-Eta}, \eqref{Eq:Recouplingsconstants-Triangle-Kbar} and Tables \ref{Tab:Recouplingconstants-Triangle-Pi} and \ref{Tab:Recouplingconstants-Triangle-K}.

\subsection{Unrenormalized Results}
We reduce the expressions for the amplitudes $J_{...}^A$ from Eq. \eqref{Eq:GA-all-contributions-of-diagrams} by using a Passarino-Veltman reduction scheme \cite{Lutz:2020dfi}. Rewriting our amplitudes in terms of Passarino-Veltman basis integrals  and reduced integrals \eqref{Eq:Def-Integrals-after-PVR} after cancelling the occurring kinematic singularities \cite{Lutz:2020dfi}, we arrive at the intermediate result in Eq. \eqref{Eq:PVR-unrenormalized-results-after-PVR} for amplitudes $J_{...}^A$. We apply the notation of Ref. \cite{Lutz:2020dfi} and find
\begin{eqnarray}
&& J^A = 1 \,,
\nonumber\\
&& J_Q^A = K_Q^{(Q)}\,I_Q \,,
\nonumber\\
&& J_{LQ}^{A,n} = K_{LQ}^{(Q),n}I_Q+K_{LQ}^{(L),n}I_L+K_{LQ}^{(LQ),n}I_{LQ}(M_{\bar{p}}) \,,
\nonumber\\
&& J_{QR}^{A,n} = K_{QR}^{(Q),n}I_Q+K_{QR}^{(R),n}I_R+K_{QR}^{(QR),n}I_{QR}(M_{p}) \,,
\nonumber\\
&& J_{LQR}^{A,n} = K_{LQR}^{(Q),n}\,I_Q+K_{LQR}^{(L),n}\,I_L+K_{LQR}^{(R),n}\,I_R
\nonumber\\
&& \hspace{1.15cm}  +K_{LQR}^{(LQ),n}\,I_{LQ}(M_{\bar{p}})
+K_{LQR}^{(QR),n}\,I_{QR}(M_{p})+K_{LQR}^{(LR),n}\,I_{LR}(t)
+K_{LQR}^{'(LR),n}\,\frac{\Delta I_{LR}(t)}{t}
\nonumber\\
&& \hspace{1.15cm} +K_{LQR}^{''(LR),n}\,\frac{\Delta\Delta I_{LR}(t)}{t^2}
+K_{LQR}^{(LQR),n}\,I_{LQR}(t)
+K_{LQR}^{'(LQR),n}\frac{\Delta I_{LQR}(t)}{t-(M_{\bar{p}}-M_p)^2} \,, 
\label{Eq:PVR-unrenormalized-results-after-PVR}
\end{eqnarray}
where the amplitude $J_{abc}^A$\, , which originates from a diagram with internal particles $a$, $b$, $c$, is expressed by a combination of pre-factor and integral $K_{abc}^{(ABC)} I_{ABC}$. Note that the Passarino-Veltman reduction scheme leads to the appearance of integrals with internal particles $ABC$, which differ from the internal particles of the original diagram $abc$. 
The pre-factors $K_{abc}^{(ABC)}$ are too long to be shown here in full detail. The pre-factors $K'$ and $K''$ are linked to the reduced integrals $\Delta I_{...}$ and $\Delta\Delta I_{...}$,
\begin{eqnarray}
&& I_R=\int \frac{d^dl}{(2\pi)^d}\,\frac{i\,\mu^{4-d}}{l^2-M_R^2} \,,
\nonumber\\
&&I_Q=\int \frac{d^dl}{(2\pi)^d}\,\frac{i\,\mu^{4-d}}{l^2-m_Q^2} \,,
\nonumber\\
&&I_{L}=\int \frac{d^dl}{(2\pi)^d}\,\frac{i\mu^{4-d}}{l^2-M_{L}^2} \,,
\hspace{0.4cm}
\nonumber\\
&&I_{LQ}\big(\bar{p}^2\big)=\int \frac{d^dl}{(2\pi)^d}\,
\frac{-i\,\mu^{4-d}}{((l-\bar{p})^2-M_{L}^2)(l^2-m_Q^2)} \,,
\nonumber\\
&&I_{LR}(t)=\int \frac{d^dl}{(2\pi)^d}\,
\frac{-i\,\mu^{4-d}}{((l-(\bar{p}-p))^2-M_{L}^2)(l^2-M_R^2)}\,,
\nonumber\\
&&I_{QR}\big(p^2\big)=\int \frac{d^dl}{(2\pi)^d}\,
\frac{-i\mu^{4-d}}{(l^2-m_Q^2)((l-p)^2-M_R^2)} \,,
\nonumber\\
&&I_{LQR}\big(\bar{p}^2,p^2,t\big)  = \int \frac{d^dl}{(2\pi)^d}\,
\frac{i\,\mu^{4-d}}{((l-\bar{p})^2-M_{L}^2)(l^2-m_Q^2)((l-p)^2-M_R^2)} \,,
\nonumber \\
&& \Delta I_{LR}(t) = I_{LR}(t) - I_{LR}(t=0)
 \nonumber \\
&& \Delta\Delta I_{LR}(t) = I_{LR}(t) - I_{LR}(t=0) - t \;\frac{\partial}{\partial t} I_{LR}(t)|_{t=0}\,,
 \nonumber \\
 && \Delta I_{LQR}(t) = I_{LQR}(t)-I_{LQR}\big(t=(M_{\bar{p}}-M_p)^2\big) \,, 
 \label{Eq:Def-Integrals-after-PVR}
\end{eqnarray}
with 
  \begin{eqnarray}
&&I_{LR}(t=0)=\frac{I_R-I_L}{M_L^2-M_R^2}\,, 
\nonumber\\ 
&&\frac{\partial}{\partial t} I_{LR}(t)\Big|_{t=0}=
\frac{ \big( (d-4) \, M_L^2-d \,   M_R^2\big)}
{d \, (M_L^2-M_R^2)^3} \, I_L+\frac{ \big(d \, M_L^2-(d-4) \, M_R^2\big)}
{d \, (M_L^2-M_R^2)^3} \, I_R\,,
\nonumber \\
 &&  I_{LQR}\big(t=(M_{\bar{p}}-M_p)^2\big)= 
 \nonumber \\
 && \hspace{1cm} 
 \frac{M_{\bar{p}} \Big(I_{LQ}(M_{\bar{p}}) - I_{LR}\big((M_{\bar{p}}-M_p)^2\big)\Big) - M_p \Big(I_{QR}(M_p) - I_{LR}\big((M_{\bar{p}}-M_p)^2\big)\Big)}{M_p \, (M_{\bar{p}}^2 + m_Q^2 - M_L^2) - M_{\bar{p}} \, (M_p^2 + m_Q^2 - M_R^2)} \, .
 \label{Eq:PVR-kin-sing-cancelled}
\end{eqnarray}

\section{Renormalization and Power Counting}
\label{Section:Renormalization}

Applying the general strategy of Refs. \cite{Lutz:2018cqo, Lutz:2020dfi} we will use the following basic guidelines for renormalization and power counting:
\begin{itemize}
 \item the use of on-shell masses in the loop contributions, 
 \item introduction of a consistent power-counting scheme with special attention to the decuplet-octet mass difference, 
 \item no expansion of the integrals $I$, but the use of appropriate subtractions in order to preserve power counting, 
 \item a consistent expansion of the pre-factors $K$ in appropriate variables. 
\end{itemize}

In Eq. \eqref{Eq:PVR-unrenormalized-results-after-PVR} the integrals $I_{...}$ and their pre-factors $K_{...}^{(...)}$ depend on both external masses, $M_{\bar{p}}$ and $M_p$, and internal ones $M_L$, $M_R$, and $m_Q$. There are different approaches on how to treat these internal masses. Ledwig et al. \cite{Ledwig:2014rfa} use physical meson masses and present two options for the baryon masses. Firstly, the octet and decuplet masses are set to their values in the chiral limit ($M$ and $M+\Delta$), whereas their second option is the average over physical octet and the decuplet masses. Whether any of these simplifications is appropriate to describe QCD lattice data at unphysically large meson masses to high precision appears questionable. 

Here we will use on-shell masses in the loop contributions, following the path demonstrated for baryon masses \cite{Lutz:2018cqo}. 
It is a self-consistent approach, which is based on the eight coupled equations for the baryon octet and decuplet masses:
\begin{eqnarray}
&& \hspace{-0.4cm} M_N = M + \tilde{\Sigma}_N \big(M_N, M_\Lambda, M_\Sigma, M_\Xi, M_{\Delta_\mu}, M_{\Sigma_\mu}, M_{\Xi_\mu}, M_{\Omega_\mu}, m_\pi, m_K, m_{\bar{K}}, m_\eta\big)\,,
 \nonumber\\
&& \hspace{-0.4cm} M_\Sigma = M + \tilde{\Sigma}_\Sigma \big(M_N, M_\Lambda, M_\Sigma, M_\Xi, M_{\Delta_\mu}, M_{\Sigma_\mu}, M_{\Xi_\mu}, M_{\Omega_\mu}, m_\pi, m_K, m_{\bar{K}}, m_\eta\big)\,,
\nonumber\\
&& \hspace{-0.4cm} M_\Lambda = ...
\nonumber\\
&& \hspace{-0.4cm} \vdots
\label{Eq:On-shell-masses-visualization}
\nonumber\\
&& \hspace{-0.4cm} M_{\Omega_\mu}= M + \Delta + \tilde{\Sigma}_{\Omega_\mu} \big(M_N, M_\Lambda, M_\Sigma, M_\Xi, M_{\Delta_\mu}, M_{\Sigma_\mu}, M_{\Xi_\mu}, M_{\Omega_\mu}, m_\pi, m_K, m_{\bar{K}}, m_\eta\big) \,,
 \end{eqnarray}
 where $\tilde{\Sigma}_A$ is the self-energy to the mass of an octet or decuplet particle $A$, which depends on internal loop masses.  
 In conventional $\chi$PT, e.g. Ref. \cite{Bernard:2009mw, Ren:2013dzt}, the masses on the right-hand side of Eq. \eqref{Eq:On-shell-masses-visualization} are expanded in order to decouple the equations. 
 In contrast to this, Lutz et al. \cite{Lutz:2018cqo} keep the full structure  on the right-hand side of Eq. \eqref{Eq:On-shell-masses-visualization} and solve the set of coupled equations. 
This self-consistent approach leads to a very good description of octet and decuplet masses up to large meson masses of 600 MeV \cite{Guo:2019nyp}.
We expect that the use of these accurate on-shell masses allows an improved description of the loop contributions to the axial-vector form factors as well. 

The use of on-shell masses requires a clear strategy regarding power counting. In principle, the triangle integral $I_{LQR}(M_{\bar{p}}, M_p)$ can involve four different baryon masses, $M_{\bar{p}}$, $M_p$, $M_L$, and $M_R$. Although $M_{\bar{p}}$ and $M_p$ are restricted to octet masses, the fact that $M_L$ and $M_R$ can be octet and decuplet masses complicates the question how to deal with baryon mass differences in our framework.  
In contrast, previous works expanded the masses in the loop contributions and were therefore mainly confronted with the treatment of the non-zero mass difference between the decuplet and the octet baryons in the chiral limit $\Delta$.
The small-scale expansion \cite{Hemmert:1997ye} uses an expansion in $\Delta$. Using the explicit example of the bubble integral $I_{QR}(M_p)$, the poor convergence of this expansion has been demonstrated in Ref. \cite{Lutz:2018cqo}. Therefore, we do not expand in $\Delta$. 
Our main strategy is to subtract the terms that violate power counting in any of the two regions $ m_Q\, < \,\Delta$ and $m_Q \, \simeq \, \Delta$. 
%
To cope with the four different baryon masses, we rewrite them in terms of mass differences $\delta_B$, $\delta_L$, $\delta_R$: 
\begin{eqnarray}
&&\delta_B=M_{\bar{p}}-M_p\,, \hspace{2.65cm}
\delta_L=M_L-M_{\bar{p}}\,(1+\gamma_L)\,,
\nonumber\\
&&M_B=\frac{M_{\bar{p}}+M_p}{2}\,, \hspace{2.39cm}
 \delta_R=M_R-M_p\,(1+\gamma_R)\,,
 \nonumber\\
 \label{Eq:PowerCounting-new-variables}
&&\gamma_L =\begin{cases}
  0 \ &\mbox{if } M_L\in[8]\,, \\
   \Delta/M\ &\mbox{if } M_L\in[10]\,,
\end{cases}\hspace{0.7cm}
\gamma_R =\begin{cases}
  0 \ &\mbox{if } M_R\in[8]\,, \\
   \Delta/M\ &\mbox{if } M_R\in[10]\,.
\end{cases}
  \end{eqnarray}
The subtractions $\gamma_L$ and $\gamma_R$ are chosen in such a way that $\delta_L$ and $\delta_R$ vanish in the chiral limit for all cases $L/R \in \{[8],[10]\}$. As we restrict external particles to the baryon octet, $\delta_B$ vanishes in the chiral limit without any subtractions needed. 

In the following we turn to the explicit renormalization rules for the amplitudes in Eq. \eqref{Eq:PVR-unrenormalized-results-after-PVR}:
\begin{eqnarray}
J_{...}^A \rightarrow \bar{J}_{...}^A \hspace{1.5cm} \text{with} \hspace{1.5cm}
 I_{...} \rightarrow \bar{I}_{...} \hspace{0.5cm} \text{and} \hspace{0.5cm} K_{...}^{(...)} \rightarrow \bar{K}_{...}^{(...)}\,.
\end{eqnarray}
Eq. \eqref{Eq:PVR-unrenormalized-results-after-PVR} can be rewritten in terms of renormalized integrals $\bar{I}_{...}$ and pre-factors $\bar{K}_{...}^{(...)}$:
\begin{eqnarray}
&& \bar{J}^A = 1 \,,
\nonumber\\
&& \bar{J}_Q^A = \bar{K}_Q^{(Q)}\,\bar{I}_Q \,,
\nonumber\\
&& \bar{J}_{LQ}^{A,n} = \bar{K}_{LQ}^{(Q),n}\bar{I}_Q+\bar{K}_{LQ}^{(LQ),n}\bar{I}_{LQ}(M_{\bar{p}}) \,,
\nonumber\\
&& \bar{J}_{QR}^{A,n} = \bar{K}_{QR}^{(Q),n}\bar{I}_Q+\bar{K}_{QR}^{(QR),n}\bar{I}_{QR}(M_{p}) \,,
\nonumber\\
&& \bar{J}_{LQR}^{A,n} = \bar{K}_{LQR}^{(Q),n}\,\bar{I}_Q
+\bar{K}_{LQR}^{(LQ),n}\,\bar{I}_{LQ}(M_{\bar{p}})
+\bar{K}_{LQR}^{(QR),n}\,\bar{I}_{QR}(M_{p})
\nonumber\\
&& \hspace{1.15cm} 
+\bar{K}_{LQR}^{(LQR),n}\,\bar{I}_{LQR}(t)
+\bar{K}_{LQR}^{'(LQR),n}\frac{\Delta \bar{I}_{LQR}(t)}{t-(M_{\bar{p}}-M_p)^2} \,.
\label{Eq:Results-form-factor-after-renormlization1}
\end{eqnarray}
Note that the integrals $I_L$, $I_R$, $I_{LR}$, $\Delta I_{LR}$, and $\Delta \Delta I_{LR}$ do not appear in Eq. \eqref{Eq:Results-form-factor-after-renormlization1} anymore (see Eq. \eqref{Eq:Renormalization-vanishing-integrals} and the discussion above). 

We apply as power-counting rules:
\begin{eqnarray}
&& m_Q^2 \sim Q^2\,,\hspace{1cm} t\sim Q^2\,,\hspace{1.18cm}  M_B\sim Q^0\,, 
 \nonumber\\
&&\delta_B\sim Q^2\,, \hspace{1.15cm}\delta_L\sim Q^2\,,\hspace{1cm}
  \delta_R\sim Q^2\,.
  \label{Eq:Power-Counting-Rules}
\end{eqnarray} 
The pre-factors $\bar{K}_{...}^{(...)}$ are obtained from $K_{...}^{(...)}$ by subtracting power-counting violating terms $\delta K_{...}^{(...)}$:
\begin{eqnarray}
 \bar{K}_{...}^{(...)}=K_{...}^{(...)}-\delta K_{...}^{(...)}\,,
\end{eqnarray}
All factors $K_{...}^{(...)}$, $\bar{K}_{...}^{(...)}$ and $\delta K_{...}^{(...)}$ are rewritten in terms of the new variables, defined in Eq. \eqref{Eq:PowerCounting-new-variables}. The unrenormalized, full factors $K_{...}^{(...)}$ are expanded following the power-counting rules in Eq. \eqref{Eq:Power-Counting-Rules}. The renormalized coefficients $\bar{K}_{...}^{(...)}$ are explicitly given in the Appendix \eqref{Eq:K-factors-Q} - \eqref{Eq:K-factors-LQR} and involve contributions proportional to factors $\alpha_{ij}$, which are functions of the ratio $\Delta/M$ only and normalized to $\alpha_{ij}\rightarrow 1$ for $\Delta\rightarrow 0$. An expansion in $\Delta/M$, like performed in Ref. \cite{Hemmert:1997ye}, would lead to $\alpha_{ij}\rightarrow 1$ at leading order. We insist on keeping the full factors $\alpha_{ij}$ and demonstrate in Ref. \cite{Lutz:2020dfi} that $\alpha_{ij}=1$ is a very poor approximation for some $ij$.  Eq. \eqref{Eq:alphas-explicit} in the Appendix \ref{Chapter:Appendix-K-factors} lists all $\alpha_{ij}$. 
Power-counting violating terms $\delta K_{...}^{(...)}$ are subtracted and denoted by an index 0: $\alpha_{0j}\rightarrow 0$ and $\alpha_{j0}\rightarrow 0$.
All power-counting violating, and therefore subtracted terms, $\delta K_{...}^{(...)}$ are given in Eqs. \eqref{Eq:K-factors-Q-subtracted}, \eqref{Eq:K-factors-LQ-QR-subtracted}, and \eqref{Eq:K-factors-LQR-subtracted}.

The renormalized integrals $\bar{I}_{...}$, which we want to keep unchanged, if power counting allows, are not expressed in the new variables. We will develop their explicit structure  in the following. 

Integrals that do not involve any meson mass consist of large scales only, because we do not consider $\Delta$ as a small scale. Therefore, it was argued in Ref. \cite{Semke:2005sn} that these contributions should be fully subtracted, so they do not appear in Eq. \eqref{Eq:Results-form-factor-after-renormlization1} anymore:
\begin{eqnarray}
  \bar{I}_L = \bar{I}_R = \bar{I}_{LR} = \Delta \bar{I}_{LR} = \Delta \Delta \bar{I}_{LR} = 0\,.
  \label{Eq:Renormalization-vanishing-integrals}
\end{eqnarray}
The reduced integrals $\Delta I_{LR}$ and $\Delta \Delta I_{LR}$ were defined in Eqs. \eqref{Eq:Def-Integrals-after-PVR}.

The renormalized meson tadpole $\bar{I}_Q$ is of order $Q^2$,
\begin{eqnarray}
 &&\bar{I}_Q= \frac{m_Q^2}{16 \, \pi^2} \, \log\Bigg[\frac{m_Q^2}{\mu^2}\Bigg]\,, 
 \label{Eq:Renormalization-Tadpole-renormalized}
\end{eqnarray}
It is the only remaining source of the renormalization scale $\mu$. 
We expand its pre-factor $K_{...}^{(Q)}$, but consider only order-$Q^0$ terms, since higher-order terms are expected to be renormalized by counterterms of order $Q^4$ and higher. 
As diagrams with internal decuplet are not expected to contribute to this order, their contributions $\delta K_{...}^{(Q)}$, proportional to $\alpha_{0j}$ with $j\in\{1,2,3,4\}$, are subtracted via
\begin{eqnarray}
&& \delta K_{L\in[10] Q R\in[8]}^{(Q)}=
\delta  K_{L\in[8] Q R\in[10]}^{(Q)}=\frac{1}{3} \, \alpha_{01}\,,
 \nonumber\\
&&\delta  K_{L\in[10] Q R\in[10]}^{(Q)}=-\frac{4}{9} \, \alpha_{02}\,,
 \nonumber\\
&&  \delta K_{L\in[10] Q }^{(Q),n=1}=
 \delta K_{Q R\in[10]}^{(Q),n=1}=\frac{10}{9} \, \frac{\Delta}{M} \, \alpha_{03} \, M_B\,,
 \nonumber\\
&& \delta K_{L\in[10] Q }^{(Q),n=2}=
  \delta K_{ Q R\in[10]}^{(Q),n=2}=-\frac{2}{9} \, \frac{\Delta}{M} \, \alpha_{04} \, M_B\,.
\label{Eq:K-factors-Q-subtracted}
\end{eqnarray}

The discussion becomes more complex for the bubble integrals evaluated at external baryon masses $\bar{I}_{LQ}(M_{\bar{p}})$ and $\bar{I}_{QR}(M_p)$. If $L, R \in[8]$ no subtractions are necessary, but in case $L, R \in[10]$ power-counting violating terms arise. Therefore we subtract all $Q^0$-contributions $\delta K_{...}^{(LQ)}$ and $\delta K_{...}^{(QR)}$ in Eq. \eqref{Eq:K-factors-LQ-QR-subtracted} by the rule $\alpha_{j0}\rightarrow 0$ for $j=1-5$. For instance
\begin{eqnarray}
 && \delta K_{L\in[8] Q R\in[10]}^{(LQ)}= \delta K_{L\in[10] Q R\in[8]}^{(QR)}=-\frac{5}{12} \, \frac{\Delta}{M} \, \alpha_{10} \, M_B^2 ,
\nonumber\\
&& \delta K_{L\in[10] Q R\in[8]}^{(LQ)}= \delta K_{L\in[8] Q R\in[10]}^{(QR)}=
-\frac{5}{12} \, \frac{\Delta}{M} \, \alpha_{20} \, M_B^2 \,,
\nonumber\\
&& \delta K_{L\in[10] Q R\in[10]}^{(LQ)}= \delta K_{L\in[10] Q R\in[10]}^{(QR)} = \frac{7}{9} \, \frac{\Delta}{M} \, \alpha_{30} \, M_B^2 \,,
\nonumber\\
&&\delta K_{L\in[10] Q }^{(LQ),n=1}=\delta K_{Q R\in[10]}^{(QR),n=1}=-\frac{20}{9} \, \frac{\Delta^2}{M^2} \, \alpha_{40} \, M_B^3 \,,
\nonumber\\
&&\delta K_{L\in[10] Q }^{(LQ),n=2}=\delta K_{ Q R\in[10]}^{(QR),n=2}=\frac{4}{9} \, \frac{\Delta^2}{M^2} \, \alpha_{50} \, M_B^3 \,.
\label{Eq:K-factors-LQ-QR-subtracted}
\end{eqnarray}
 These subtractions are not enough to eliminate all power-counting violating contributions. Additional subtractions are needed for the integrals $\bar{I}_{LQ}$ and $\bar{I}_{QR}$. We therefore introduce the subtractions  $\gamma_{\bar{B}}^L$ and $\gamma_B^R$, which assure the correct power counting 
 $\bar{I}_{L\in[8]Q}/\bar{I}_{QR\in[8]}\sim Q^1$ and $\bar{I}_{L\in[10]Q}/\bar{I}_{QR\in[10]}\sim Q^2$ in the chiral domain \cite{Lutz:2018cqo}:
\begin{eqnarray}
&&\hspace{-0.5cm}\bar{I}_{LQ}(M_{\bar{p}})=I_{LQ}(M_{\bar{p}})+\frac{I_{L}}{M_L^2}-\frac{1-\gamma_{\bar{B}}^L}{16 \, \pi^2}\,, 
\nonumber\\
 &&\hspace{-0.5cm}\bar{I}_{QR}(M_p)=I_{QR}(M_p)+\frac{I_{R}}{M_R^2}-\frac{1-\gamma_B^R}{16 \, \pi^2}\,,
\label{Eq:Renormalization-BubbleLQ&QR-unrenormalized}
 \end{eqnarray}
 where $\gamma_{\bar{B}}^L$ and $\gamma_B^R$ depend on baryon masses in the chiral limit only:
 \begin{eqnarray}
  &&  \gamma_B^R =  \begin{cases}
  0 \ &\mbox{if } \, R\in[8]\,, \\
 -\dfrac{2\,M\,\Delta+\Delta^2}{M^2}\, \log\bigg[\dfrac{2\,M\,\Delta+\Delta^2}{(M+\Delta)^2}\bigg] &\mbox{if } \, R\in[10]\,.
\end{cases} 
   \label{Eq:Renormalization-gammaLbarB&BR}
 \end{eqnarray}
The explicit form of $\bar{I}_{QR}(p^2=M_p^2)$ has been determined in Ref. \cite{Lutz:2018cqo}:
\begin {eqnarray}
&& \hspace{-1cm}\bar{I}_{Q R}\big(M_p\big)=
 \frac{1}{16\pi^2} \,\Bigg[\gamma_B^R-\bigg(\frac{1}{2}+\frac{m_Q ^2-M_R^2}{2\,M_p^2}\bigg)
 \log\,\bigg[\frac{m_Q^2}{M_R^2}\bigg]
 \nonumber\\ 
&& \hspace{-1cm}\quad +\,\frac{p_{Q R}}{M_p}\bigg(\log\,\bigg[1-\frac{M_p^2-2\,p_{Q R}\,M_p}{m_Q ^2+M_R^2}\bigg]
- \log\,\bigg[1-\frac{M_p^2+2\,p_{Q R}\,M_p}{m_Q ^2+M_R^2}\bigg]\bigg)\Bigg]\,, 
\nonumber \\
&& \hspace{-1cm} p_{Q R}^2  =  \frac{M_p^2}{4}-\frac{M_R^2+m_{Q}^2}{2}+\frac{(M_R^2-m_Q^2)^2}{4\,M_p^2} \, .
\label{Eq:Renormalization-BubbleLQ&QR-renormalized}
\end{eqnarray}

It is left to discuss the contributions of the triangle integrals $I_{LQR}(t)$ and $\Delta I_{LQR}(t)$. Analogous to the bubbles, power-counting violating terms only appear in the presence of internal decuplet particles. This is true for the cases $L\in[10], R\in[8]$ as well as for $L\in[8], R\in[10]$, and $L, R\in[10]$. Therefore, the order-$Q^0$ contributions $\delta K_{...}^{(LQR)}$ in Eq. \eqref{Eq:K-factors-LQR-subtracted}, associated with $\alpha_{j0}\rightarrow 0$ with $j\in\{6,7,8\}$, are subtracted. For instance

\begin{eqnarray}
&& \delta K_{L\in[10] Q R\in[8]}^{(LQR),1}=\delta K_{L\in[8] Q R\in[10]}^{(LQR),1}=\frac{5}{6} \,
 \frac{\Delta^2}{M^2} \, \alpha_{60} \, M_B^4\,,
\nonumber\\
&&\delta K_{L\in[10] Q R\in[8]}^{'(LQR),1}= \delta K_{L\in[8] Q R\in[10]}^{'(LQR),1}=\frac{2}{3} \, \frac{\Delta^2}{M^2} \, \alpha_{70} \, M_B^6\,,
\nonumber\\
&&\delta K_{L\in[10] Q R\in[10]}^{(LQR),1}=-\frac{4}{3} \, \frac{\Delta^2}{M^2} \, \alpha_{80} \, M_B^4\,.
\label{Eq:K-factors-LQR-subtracted}
\end{eqnarray}
As the triangle integrals $I_{LQR}(t)$ and $\Delta I_{LQR}(t)$, defined in Eq. \eqref{Eq:Def-Integrals-after-PVR}, are finite for $d=4$, renormalization is not needed. Only power-counting violating contributions of the integrals need to be eliminated by introducing the subtractions $\gamma_{LQR}$ and $\gamma'_{LQR}$ \cite{Lutz:2020dfi}:
\begin{eqnarray}
&&  \bar I_{LQR}(t)= I_{LQR}(t)- \gamma_{LQR}\,, 
\label{Eq:Renormalization-Triangle-LQR}
\\
&& \nonumber\gamma_{L\in[10]QR\in[8]}=\gamma_{L\in[8]QR\in[10]}=
\nonumber\\
&&
\hspace{2.19cm} = -\frac{1}{16\pi^2\,M^2}\log\bigg[\frac{2\,M\,\Delta+\Delta^2}{(M+\Delta)^2}\bigg]
+\frac{1}{16\pi^2\,(2\,M\,\Delta+\Delta^2)}\log\bigg[\frac{(M+\Delta)^2}{M^2}\bigg]\,,
\nonumber \\ 
\nonumber
&&\gamma_{L\in[8]QR\in[8]}=0\,,
\quad \gamma_{L\in[10]QR\in[10]} = -\frac{1}{16\pi^2\,M^2}\log\bigg[\frac{2\,M\,\Delta+\Delta^2}{(M+\Delta)^2}\bigg]\,,
\\ \nonumber\\ 
 && \Delta\bar{I}_{LQR}(t)=\bar{I}_{LQR} (t)-\bar{I}_{L QR} \big((M_{\bar{p}}-M_p)^2\big)-\big(t-(M_{\bar{p}}-M_p)^2\big)\,\gamma'_{LQR}\,, 
  \label{Eq:Renormalization-Triangle-LQR-reduced} \\
 &&\gamma'_{L\in[10]QR\in[8]}
 =\gamma'_{L\in[8]QR\in[10]}
=-\frac{1}{96\pi^2\, M^4\, \Delta^3\, (2\,M + \Delta)^3}
\;\Bigg(\Delta^3 \,(2\,M + \Delta)^3 \log\bigg[\frac{\Delta}{M}\bigg]
\nonumber\\
&& \hspace{1cm} - 2\, (M + \Delta)^2 \log\bigg[\frac{M + \Delta}{M}\bigg] 
    (4M^4 - 2 M^3 \Delta + 3M^2 \Delta^2  + 4 M \Delta^3 + \Delta^4)   
   \nonumber\\
   &&\hspace{1cm} +\Delta \,(2\,M + \Delta)\bigg(M^2\,(4\,M^2 + 2 \, M\,\Delta + \Delta^2) 
   + \Delta^2 \,(2\, M + \Delta)^2 \log\bigg[\frac{2\,M + \Delta}{M}\bigg]\bigg)\Bigg)\,,
   \nonumber\\ \nonumber
 &&
 \gamma'_{L\in[8]QR\in[8]}=\gamma'_{L\in[10]QR\in[10]}=0\,. 
 \end{eqnarray}
The subtractions $\gamma_{LQR}$ and $\gamma'_{LQR}$ are analogous to the subtraction term $\gamma^B_R$ introduced in the  bubble function. With Eqs. \eqref{Eq:Renormalization-Triangle-LQR} and \eqref{Eq:Renormalization-Triangle-LQR-reduced} we obtain 
$ \bar I_{L\in[8]QR\in[8]}(t)\sim Q^0$, \\
$ \bar I_{L\in[8]QR\in[10]}(t)\sim Q^1$, and $ \bar I_{L\in[10]QR\in[10]}(t)\sim Q^2$ and 
$ \Delta \bar I_{L\in[8]QR\in[8]}(t)/\big(t- (M_{\bar{p}}-M_p)^2\big)\sim Q^0$, 
$ \Delta \bar I_{L\in[8]QR\in[10]}(t)/\big(t- (M_{\bar{p}}-M_p)^2\big)\sim Q^1$, and $ \Delta \bar I_{L\in[10]QR\in[10]}(t)/\big(t- (M_{\bar{p}}-M_p)^2\big)\sim Q^2$
in the chiral domain. 
The subtractions are crucial, since otherwise the explicit evaluation of a class of two-loop diagrams would be needed \cite{Lutz:2020dfi, Long:2009wq}.
%
An explicit expression for the triangle integral at $t= (M_{\bar{p}}-M_p)^2$ was given in Eq. \eqref{Eq:PVR-kin-sing-cancelled}. For general values of the momentum transfer $t$ we use the Feynman parameterization: 
\begin{eqnarray}
&&\bar{I}_{LQR}(t)=-
\int_0^1\int_0^{1-u}  dv \, du  \, \frac{1}{(4\pi)^2 \, \Omega^2}-\gamma_{LQR}\,,
\nonumber\\
&&\text{with} \hspace{0.3cm}\Omega^2=-m_Q^2+v \, \big((1-v) \, M_p^2-M_R^2+m_Q^2\big)
\nonumber\\
&& \hspace{1.2cm} +u \, \big((1-u) \, M_{\bar{p}}^2-M_L^2+m_Q^2\big)+u \, v \, \big(t-M_{\bar{p}}^2-M_p^2\big)\,, 
\label{Eq:Renormalization_Triangle-LQR-Feynman}
\end{eqnarray} 
from which also the reduced triangle integral $\Delta \bar{I}_{LQR}(t)$ follows with Eq. \eqref{Eq:Renormalization-Triangle-LQR-reduced}.

Our renormalized and power-counting respecting result for the axial-vector form factors $G_{A,i}^{\bar{B}B}(t)$ is summarized in Eq. \eqref{Eq:Results-form-factor-after-renormlization2}. It uses the renormalized amplitudes from Eq. \eqref{Eq:Results-form-factor-after-renormlization1} plugged in Eq. \eqref{Eq:GA-all-contributions-of-diagrams} and rearranged by renormalized integrals instead of diagrams: 
\begin{eqnarray}
&&\hspace{-0.5cm}(2\,f)^2\,G_{A,i}^{\bar{B}B}(t)=(2\,f)^2\,\Big(\sqrt{Z_{\bar{B}}Z_B}\, C_{i}^{(\bar{B}B)}+  t\,T_{i}^{(\bar{B}B)}+  2\,B_0\,X_{i}^{(\bar{B}B)}\Big)
\nonumber\\
&&+\sum_{Q\in [8]}\bar{I}_Q\Bigg(C_{Q,i}^{(\bar{B}B)}\,\bar{K}_Q^{(Q)} 
+ \sum_{L\in [8],[10]}
G^{(\bar{B})}_{QL}\,C_{LQ,i}^{(\bar{B}B), n}\bar{K}_{LQ}^{(Q),n}
\nonumber\\
&& \hspace{0.3cm} + \sum_{R\in [8],[10]}
C^{(\bar{B}B),n}_{QR,i}\,G^{(B)}_{QR}\bar{K}_{QR}^{(Q),n}
+\sum_{L,R\in [8],[10]}\hspace{-0.3cm}G^{(\bar{B})}_{QL}\,
C_{LQR,i}^{(\bar{B}B),n}\,G^{(B)}_{QR}\bar{K}_{LQR}^{(Q),n}\Bigg)
\nonumber\\
&&+\sum_{Q\in [8]}\sum_{L\in[8],[10]}\bar{I}_{LQ}(M_{\bar{p}})
\Bigg(
G^{(\bar{B})}_{QL}\,C_{LQ,i}^{(\bar{B}B), n}\bar{K}_{LQ}^{(LQ),n} +\sum_{R\in [8],[10]}G^{(\bar{B})}_{QL}\,
C_{LQR,i}^{(\bar{B}B),n}\,G^{(B)}_{QR}\bar{K}_{LQR}^{(LQ),n}\Bigg)
\nonumber\\
&&+\sum_{Q\in [8]}\sum_{R\in [8],[10]}\bar{I}_{QR}(M_p)\Bigg(
C^{(\bar{B}B),n}_{QR,i}\,G^{(B)}_{QR}\bar{K}_{QR}^{(QR),n}
 +\sum_{L\in [8],[10]}G^{(\bar{B})}_{QL}\,
C_{LQR,i}^{(\bar{B}B),n}\,G^{(B)}_{QR}\bar{K}_{LQR}^{(QR),n}\Bigg)
\nonumber\\
&&+\sum_{Q\in [8]}\sum_{L,R\in [8],[10]}\bar{I}_{LQR}(t) \Bigg(G^{(\bar{B})}_{QL}\,C_{LQR,i}^{(\bar{B}B),n}\,G^{(B)}_{QR}\bar{K}_{LQR}^{(LQR),n}\Bigg)
\nonumber\\
&&+\sum_{Q\in [8]}\sum_{L,R\in [8],[10]}\frac{\Delta\bar{I}_{LQR}(t)}{t-(M_{\bar{p}}-M_p)^2}\Bigg(G^{(\bar{B})}_{QL}\,C_{LQR,i}^{(\bar{B}B),n}\,G^{(B)}_{QR}\bar{K}_{LQR}^{'(LQR),n}\Bigg)\,,
\label{Eq:Results-form-factor-after-renormlization2}
\end{eqnarray}
where a summation of repeated indices $n$ is understood.

\section{Numerical Analysis of Convergence Properties}
\subsection{Convergence Properties at the Physical Point}

The 12 independent axial-vector form factors of the baryon octet in Eq. \eqref{Eq:Results-form-factor-after-renormlization2} are ready to be compared to experimental and QCD lattice data. In Table \ref{Tab:Cabibbo-Model} we have shown the six existing experimental data points. Unfortunately, the available QCD lattice data for the hyperons $\Sigma$, $\Xi$, and $\Lambda$ is very limited. In Refs. \cite{Bali:2019svt, Lin:2007ap, Gockeler:2011ze}, results for the form factors $G_{A,a_\mu^\pi}^{\Sigma \Sigma}(t=0)$ and $G_{A,a_\mu^\pi}^{\Xi \Xi}(t=0)$ have been presented. However, these data points are not enough to determine all LECs of our chiral Lagrangian. 
\begin{table}[ht]
\centering
\renewcommand{\arraystretch}{1.0}
\begin{tabular}{| c  c | c  c |}\hline
masses & PDG \cite{Tanabashi:2018oca} & \multicolumn{2}{|c|}{parameters from \cite{Lutz:2018cqo}, \cite{Guo:2019nyp}}
\\  \hline 
$m_\pi$          [MeV] & 137 &  $F$            & 0.48  \\
$m_K$            [MeV] & 494 &  $D$            & 0.75  \\
$m_\eta$         [MeV] & 548 &  $C$            & 1.50  \\
\cline{1-2}
$M_N$            [MeV] & 939 &  $H$            & 2.06  \\   
\cline{3-4}
$M_\Lambda$      [MeV] &1116 &  $f$ [MeV]      & 92.4  \\
$M_\Sigma$       [MeV] &1193 &  $M$ [MeV]      & 866.2 \\ 
$M_\Xi$          [MeV] &1318 &  $\Delta$ [MeV] & 382.7 \\ 
\cline{1-2}
$M_{\Delta_\mu}$ [MeV] &1232 &  $\mu$ [MeV]    & 770   \\ 
\cline{3-4}
$M_{\Sigma_\mu}$ [MeV] &1385 &  $Z_N$          & 1.118 \\ 
$M_{\Xi_\mu}$    [MeV] &1533 &  $Z_\Lambda$    & 2.064 \\ 
$M_{\Omega_\mu}$ [MeV] &1672 &  $Z_\Sigma$     & 2.507 \\
                       &     &  $Z_\Xi$        & 3.423
\\ \hline
\end{tabular}
\caption{Physical masses \cite{Tanabashi:2018oca} and input parameters for the numerical estimates in flavor-SU(3). The LECs refer to `set 1' of Ref. \cite{Guo:2019nyp}, whereas the factors $Z_B$, with $B\in[8]$, are taken from Ref. \cite{Lutz:2018cqo}. }
\label{Tab:Numerical-Estimate-SU3-Input-Physical}
\end{table}
On the other hand, further investigations of axial-vector form factors were announced in Ref. \cite{Bali:2019svt}. Until this data becomes available, we use the LECs determined from fits of the baryon octet and decuplet masses \cite{Guo:2019nyp} and test some properties of our results for the axial-vector form factors from Eq.  \eqref{Eq:Results-form-factor-after-renormlization2} at a more elementary level.  

\begin{figure}[ht]
\centering
\includegraphics[width=160mm]{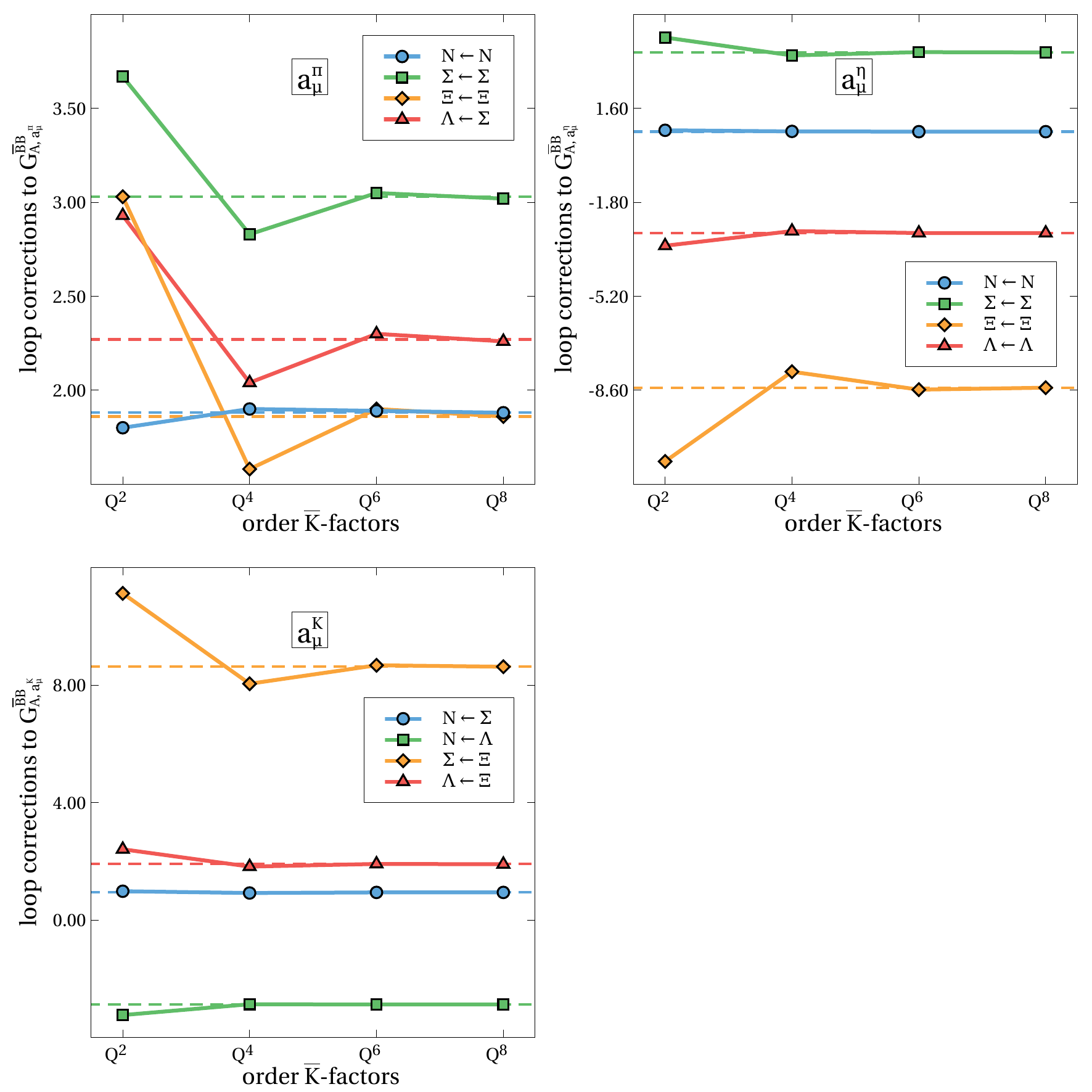}
\caption{Convergence properties of the kinematic factors $\bar{K}$. We show their contributions to $G_{A,i}^{\bar{B}B}\big(t = (M_{\bar{p}} - M_p)^2\big)$ for the chiral orders $Q^2$, $Q^4$, $Q^6$, and $Q^8$ as points and their unexpanded value as a dashed line. }
\label{Fig:PlotSU3ConvergenceK1100}
\end{figure}

First we scrutinize the convergence properties of the kinematic factors $\bar{K}$ of the one-loop contributions in Eq. \eqref{Eq:Results-form-factor-after-renormlization2}, which are expanded according to the power-counting scheme presented in Eq. \eqref{Eq:Power-Counting-Rules}. We use isospin-averaged physical on-shell masses in the loop contributions, as listed in the left column in Table \ref{Tab:Numerical-Estimate-SU3-Input-Physical}. 

In Figure \ref{Fig:PlotSU3ConvergenceK1100} we show the one-loop contributions to the axial-vector form factors $G_{A,i}^{\bar{B}B}\big(t = (M_{\bar{p}} - M_p)^2\big)$, but restrict the used LECs to the lowest order from $\mathcal{L}^{(1)}$ in Eq.  \eqref{Eq:Def-L1}. We do not take into account the LECs from $\mathcal{L}^{(2)}$ and $\mathcal{L}^{(3)}$ in Eqs. \eqref{Eq:Def-L2split} and \eqref{Eq:Def-L3split}, because not all of them are known. Furthermore, we add the order-$Q^2$ impact of the wave-function renormalization to the tree-level contributions $\big(\sqrt{Z_{\bar{B}}Z_B}-1\big) C_{i}^{(\bar{B}B)}$. The used parameters are shown in the right column of Table \ref{Tab:Numerical-Estimate-SU3-Input-Physical}.

We divide the 12 independent processes ($\bar{B} \leftarrow B$) in Figure \ref{Fig:PlotSU3ConvergenceK1100} with respect to their axial-vector current. In the two top panels the strangeness-conserving axial-vector currents $a_\mu^\pi$ and $a_\mu^\eta$ are shown and the strangeness-changing current $a_\mu^K$ is presented in the bottom of Figure \ref{Fig:PlotSU3ConvergenceK1100}. Dashed lines represent the full, non-expanded expression for the kinematic factors $\bar{K}$, whereas the points give the values for the expanded factors up to chiral order $Q^2$, $Q^4$, $Q^6$, and $Q^8$. These orders refer to the kinematic factors $\bar{K}$ only, the order of the referring integrals  $\bar{I}$ is not taken into account here. This restricts the discussion to the convergence properties of the kinematic factors $\bar{K}$. A discussion of the convergence properties of the loop contributions in general is performed below. The higher-order terms $Q^6$ and $Q^8$ reproduce the full contributions $\bar{K}$ for all processes to high accuracy, the order-$Q^4$ terms show reasonable results. However, only taking into account the order-$Q^2$ terms does not appear desirable, except for the processes with at least one external nucleon. In general, we find large loop contributions compared to the tree-level values of Table \ref{Tab:Cabibbo-Model}.

\begin{figure}[ht]
\centering
\includegraphics[width=160mm]{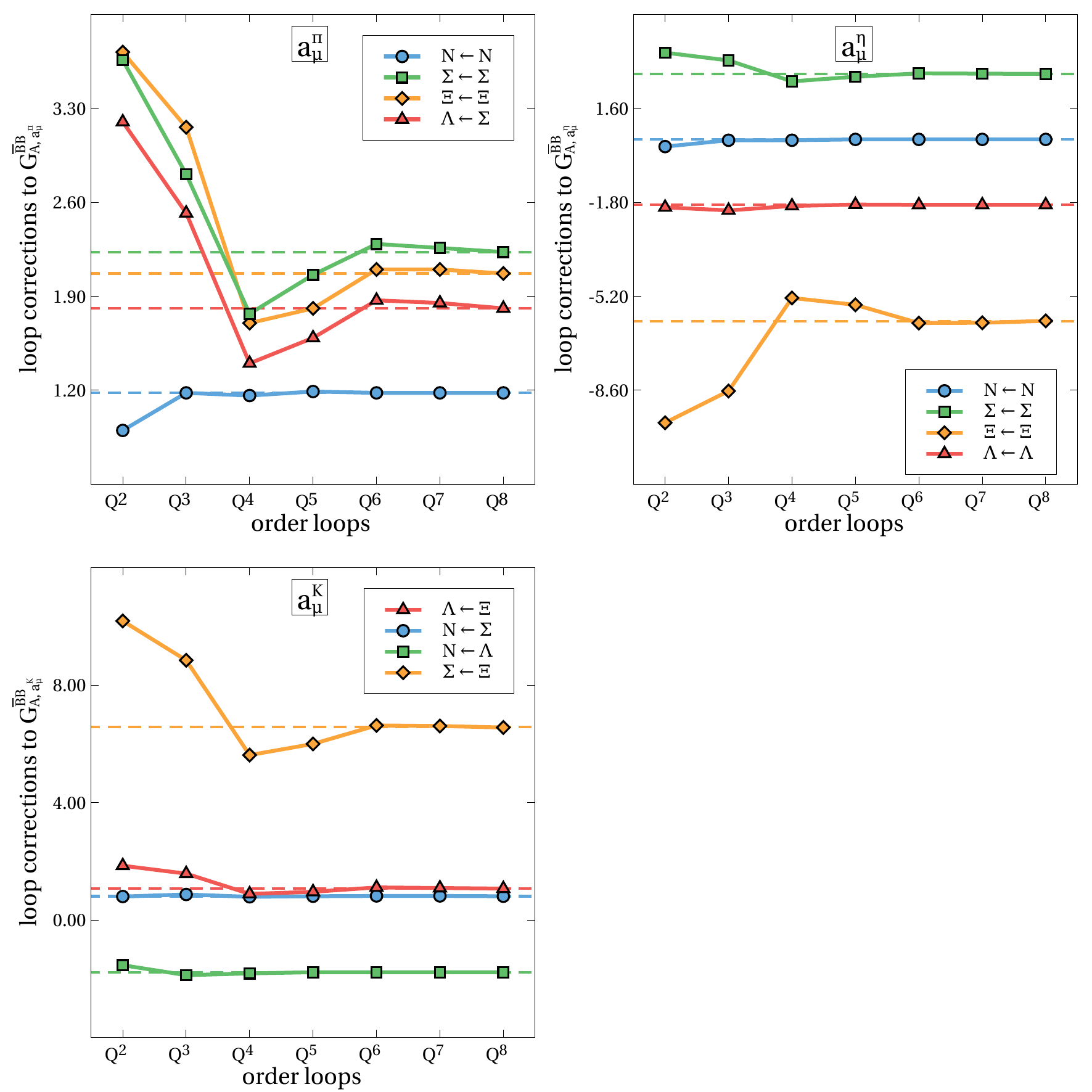}
\caption{Loop corrections to $G_{A,i}^{\bar{B}B}\big(t = (M_{\bar{p}} - M_p)^2\big)$ for different axial-vector currents $i$. The LECs $g^{(S)}$, $g^{(V)}$, $g^{(T)}$, $g^{(F)}$, $f^{(F)}$, and $f^{(A)}$ are not included. The points refer to the expansion to the corresponding chiral order, whereas the dashed lines show the full contributions. }
\label{Fig:PlotSU3ConvergenceL1100}
\end{figure}

In order to scrutinize the full convergence properties of our chiral expansion, we also take into account the chiral power of the integrals $\bar{I}_Q \sim Q^2$, $\bar{I}_{LQ}/\bar{I}_{QR} \sim Q^1$, and $\bar{I}_{LQR} \sim Q^0$. We use the momentum transfer $t=\big(M_{\bar{p}}-M_p\big)^2$, implying that $\Delta \bar{I}_{LQR}\Big(\big(M_{\bar{p}}-M_p\big)^2\Big) =0$. The tadpole integral $\bar{I}_Q$ still involves the renormalization scale $\mu$ \eqref{Eq:Renormalization-Tadpole-renormalized}. Therefore all contributions $\bar{K}^{(Q)}\bar{I}_Q \sim Q^4$ or higher are neglected, since they renormalize higher-order counterterms. 
The combined effect of $\bar{K}$ and $\bar{I}$ on the loop contributions to the axial-vector form factors are shown in Figure \ref{Fig:PlotSU3ConvergenceL1100}.

The convergence properties in Figure \ref{Fig:PlotSU3ConvergenceL1100} are worse than the ones in Figure \ref{Fig:PlotSU3ConvergenceK1100}. Nevertheless, our power-counting scheme is well converging. The chiral order $Q^5$ describes the full contributions to acceptable accuracy, whereas for some processes even lower orders are sufficient. We continue the discussion for the orders $Q^2$ - $Q^5$ and include additional terms from the Lagrangian $\mathcal{L}^{(2)}$ in Eq. \eqref{Eq:Def-L2split}. 

In the following we also consider the LECs of the higher order Lagrangians from Eqs. \eqref{Eq:Def-L2} and \eqref{Eq:Def-L3}. Note that the LECs from the Lagrangian $\mathcal{L}^{(2)}$ directly contribute to the loop contributions, whereas the LECs from $\mathcal{L}^{(3)}$ do not interfere with the loops. At first we include the LECs $g^{(S)}$, $g^{(V)}$, $g^{(T)}$, and $f^{(A)}$ from the Lagrangian $\mathcal{L}^{(2)}$, which directly contribute to the loop contributions of the order $Q^3$. The LECs $g^{(F)}$ and $f^{(F)}$ from Eq. \eqref{Eq:Def-L2} are not considered in this work, whereas the LECs $g^{(\chi)}$ and $g^{(R)}$ from $\mathcal{L}^{(3)}$ will be added later. 

The constants with scalar and vector structure $g^{(S)}$ and $g^{(V)}$ have been determined in Ref. \cite{Guo:2019nyp}, whereas the LECs $g^{(T)}$ and $f^{(A)}$ are unknown. Since the success of Cabibbo's model \cite{Cabibbo:2003cu} is compatible with small loop contributions, we determine the LECs $g^{(T)}$ and $f^{(A)}$ using a fit, which minimizes the loop contributions of all processes. This is performed for all orders $Q^3$, $Q^4$, and $Q^5$ separately, resulting in three sets of determined LECs, which are shown in Table \ref{Tab:Numerical-Estimate-SU3-g-LECs}. The size of the minimized one-loop contributions is characterized by the chi-square per number of processes $\chi^2/N_p$, with $N_p=12$, and the largest remaining one-loop contribution (`max').
\begin{table}[hb]
\centering
\renewcommand{\arraystretch}{1.0}
\begin{tabular}{| c  c | c  c  c  c |}\hline
LEC & from \cite{Guo:2019nyp}  & LEC  & $Q^3$ & $Q^4$ & $Q^5$ 
\\  \hline 
$g_0^{(S)}$  [GeV$^{-1}$] &  $-8.8678$  & $g_1^{(T)}$  [GeV$^{-1}$] &    $0.87$  &    $1.67$  &    $4.32$ \\ 
$g_1^{(S)}$  [GeV$^{-1}$] &   $0.8058$  & $g_D^{(T)}$  [GeV$^{-1}$] &   $-6.35$  &   $-4.13$  &   $-6.10$ \\ 
$g_D^{(S)}$  [GeV$^{-1}$] &  $-1.4485$  & $g_F^{(T)}$  [GeV$^{-1}$] &   $-8.22$  &   $-9.47$  &   $-7.76$ \\ 
$g_F^{(S)}$  [GeV$^{-1}$] &  $-5.1101$  & $f_1^{(A)}$  [GeV$^{-1}$] &    $1.82$  &    $1.67$  &    $6.35$ \\ 
$g_0^{(V)}$  [GeV$^{-2}$] &  $-0.3710$  & $f_2^{(A)}$  [GeV$^{-1}$] &    $0.71$  &    $0.53$  &    $0.39$ \\ 
$g_1^{(V)}$  [GeV$^{-2}$] &  $-7.2709$  & $f_3^{(A)}$  [GeV$^{-1}$] &   $-9.88$  &   $-4.65$  &   $-8.92$ \\ 
$g_D^{(V)}$  [GeV$^{-2}$] &   $10.002$  & $f_4^{(A)}$  [GeV$^{-1}$] &    $1.14$  &    $0.36$  &   $-0.34$ \\ 
\cline{3-6}
$g_F^{(V)}$  [GeV$^{-2}$] &  $-2.8688$  & $\chi^2/N_p$              &    $0.78$  &    $0.24$  &    $0.30$  \\
                          &           & max                       &    $1.94$  &    $1.05$  &    $1.20$                                                            
\\  \hline
\end{tabular}
\caption{Left column: input LECs for the numerical estimates in flavor-SU(3) from Ref. \cite{Guo:2019nyp}. Right column: LECs as determined by our fit from the minimization of the loop contributions of order $Q^3$, $Q^4$, and $Q^5$. The magnitude of the remaining one-loop contributions is described by the chi-square per number of processes $\chi^2/N_p$ and the maximal value `max'.  }
\label{Tab:Numerical-Estimate-SU3-g-LECs}
\end{table}
The LECs $g^{(T)}$ and $f^{(A)}$, determined by our fit, are reasonably small. We do not give errors, but expect some range of $\pm 3$ for the LECs and observe quite large deviations of the LECs for different orders. This is not surprising, since our estimate is not intended to be seen as a rigorous fit, for which, apart from accurate QCD lattice data, also the LECs $g^{(F)}$, $f^{(F)}$, $g^{(R)}$, and $g^{(\chi)}$ need to be taken into account. 

\begin{figure}[ht]
\centering
\includegraphics[width=160mm]{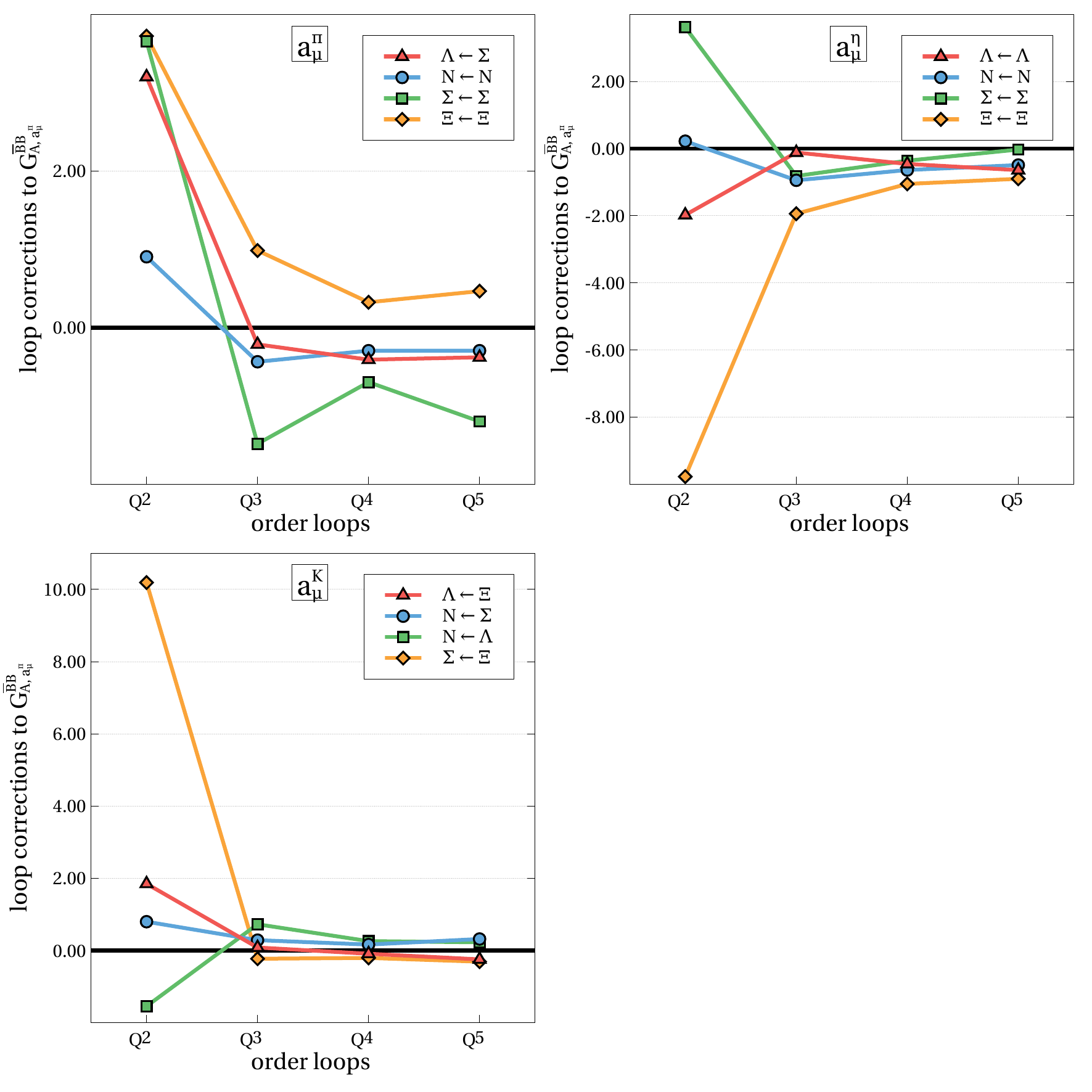}
\caption{Loop corrections to $G_{A,i}^{\bar{B}B}\big(t = (M_{\bar{p}} - M_p)^2\big)$ up to order $Q^5$ like in Figure \ref{Fig:PlotSU3ConvergenceL1100}, but with effect of the LECs $g^{(S)}$, $g^{(V)}$, $g^{(T)}$, and $f^{(A)}$ from Table \ref{Tab:Numerical-Estimate-SU3-g-LECs}, which contribute to order $Q^3$ and higher. }
\label{Fig:PlotSU3ConvergenceL1113}
\end{figure}

In Figure \ref{Fig:PlotSU3ConvergenceL1113} we reproduce the loop contributions of orders $Q^2$ - $Q^5$ of Figure \ref{Fig:PlotSU3ConvergenceL1100}, but this time with the contributions of $g^{(S)}$, $g^{(V)}$, $g^{(T)}$, and $f^{(A)}$ from Table \ref{Tab:Numerical-Estimate-SU3-g-LECs}. 
\noindent Since these LECs become relevant only at order $Q^3$, the order-$Q^2$ contributions remain unchanged compared to Figure \ref{Fig:PlotSU3ConvergenceL1100}. For the contributions of order $Q^3$, $Q^4$, and $Q^5$ we find that the loop contributions are drastically reduced in all processes. We observe that contributions are most efficiently minimized at order $Q^4$, but also the contributions at order $Q^5$ and, with a few exceptions, at order $Q^3$ are small. 
Therefore, we conclude that it is possible to reduce the loop contributions to the axial-vector form factors for all processes by appropriate choices for the LECs $g^{(T)}$ and $f^{(A)}$ at all orders.  

As a last step, we turn on the LECs of the counterterms of the Lagrangian $\mathcal{L}^{(3)}$ in Eq. \eqref{Eq:Def-L3split}. Unlike the LECs $g^{(S)}$, $g^{(V)}$, $g^{(T)}$, and $f^{(A)}$, the LECs $g^{(\chi)}$ and $g^{(R)}$ do not enter as loop contributions but as counterterms. 
 We find that it is possible to further reduce the one-loop contributions with these counterterms and show the resulting LECs $g_{1-7}^{(\chi)}$ and $g_{D/F}^{(R)}$ in Table \ref{Tab:Numerical-Estimate-SU3-Fit-gchi-gR}. The LECs are reasonably small, fulfilling $|g_{1-7}^{(\chi)}|<3$ and $|g_{D/F}^{(R)}|<7$.
The parameters $\chi^2/N_p$ and the maximal remaining contribution `max' are significantly smaller than the corresponding ones without the LECs $g_{1-7}^{(\chi)}$ and $g_{D/F}^{(R)}$ in Table \ref{Tab:Numerical-Estimate-SU3-g-LECs}. 
\begin{table}[ht]
\centering
\renewcommand{\arraystretch}{1.0}
\begin{tabular}{| c c c c | c c c c |}\hline
order & $Q^3$ & $Q^4$ & $Q^5$ & order & $Q^3$ & $Q^4$ & $Q^5$ \\ \hline
$g_1^{(\chi)}$  & $-0.46$ & $-0.18$ & $-0.58$ &
$g_5^{(\chi)}$  & $-2.44$ & $-1.14$ & $-0.25$ \\
$g_2^{(\chi)}$  & $-1.87$ & $-0.76$ & $-0.88$ &
$g_6^{(\chi)}$  &  $2.43$ & $-0.13$ & $-0.29$ \\
$g_3^{(\chi)}$  & $-2.64$ & $-1.27$ & $-2.01$ &
$g_7^{(\chi)}$  &  $2.27$ &  $1.11$ &  $1.86$ \\
$g_4^{(\chi)}$  & $-0.43$ & $-0.47$ & $-0.69$ & 
$g_D^{(R)}$     &  $6.71$ &  $4.38$ &  $6.75$ \\
\cline{1-4}
$\chi^2/N_p$    &  $0.06$ &  $0.02$ &  $0.01$ &
$g_F^{(R)}$     &  $6.84$ &  $3.83$ &  $4.94$ \\
max             &  $0.61$ &  $0.35$ &  $0.28$ &&&&
\\ \hline
\end{tabular}
\caption{LECs as determined by our fit. We also give the chi-square per number of processes $\chi^2/N_p$ and the largest remaining one-loop contribution `max'. }
\label{Tab:Numerical-Estimate-SU3-Fit-gchi-gR}
\end{table}

\subsection{Convergence Properties in the Flavor-SU(3) Limit}
In the previous Section, we showed that the inclusion of the LECs $g^{(S)}$, $g^{(V)}$, $g^{(T)}$, and $f^{(A)}$, and additionally the LECs $g^{(R)}$ and $g^{(\chi)}$, leads to strong cancellations amongst the various loop contributions, when the physical masses of Table \ref{Tab:Numerical-Estimate-SU3-Input-Physical} are used. Here we want to scrutinize whether similar effects can be observed for other on-shell baryon masses. We choose a lattice QCD ensemble in the unphysical flavor-SU(3) limit, which is characterized by $m_u = m_d = m_s$ \cite{Bietenholz:2011qq}. 
 \begin{table}[ht]
\centering
\renewcommand{\arraystretch}{1.0}
\begin{tabular}{|c |c |c |}\hline
mass & lattice ensemble \cite{Bietenholz:2011qq}
\\  \hline 
$m_\pi$, $m_K$, $m_\eta$    [MeV]                                          & 391 \\
$M_N$, $M_\Lambda$, $M_\Sigma$, $M_\Xi$ [MeV]                              &1124 \\ 
$M_{\Delta_\mu}$, $M_{\Sigma_\mu}$, $M_{\Xi_\mu}$, $M_{\Omega_\mu}$ [MeV]  &1438 \\
$Z_N$, $Z_\Lambda$, $Z_\Sigma$, $Z_\Xi$                                    &2.28
\\ \hline
\end{tabular}
\caption{QCD lattice masses in the flavor-SU(3) limit \cite{Bietenholz:2011qq}. }
\label{Tab:Numerical-Estimate-SU3-InputFlavor}
\end{table}
This leads to equal masses of the Goldstone bosons, within the baryon octet and within the baryon decuplet. The exact masses and the wave-function renormalization factor $Z$ are given in Table \ref{Tab:Numerical-Estimate-SU3-InputFlavor}.

\begin{figure}[ht]
\centering
\includegraphics[width=160mm]{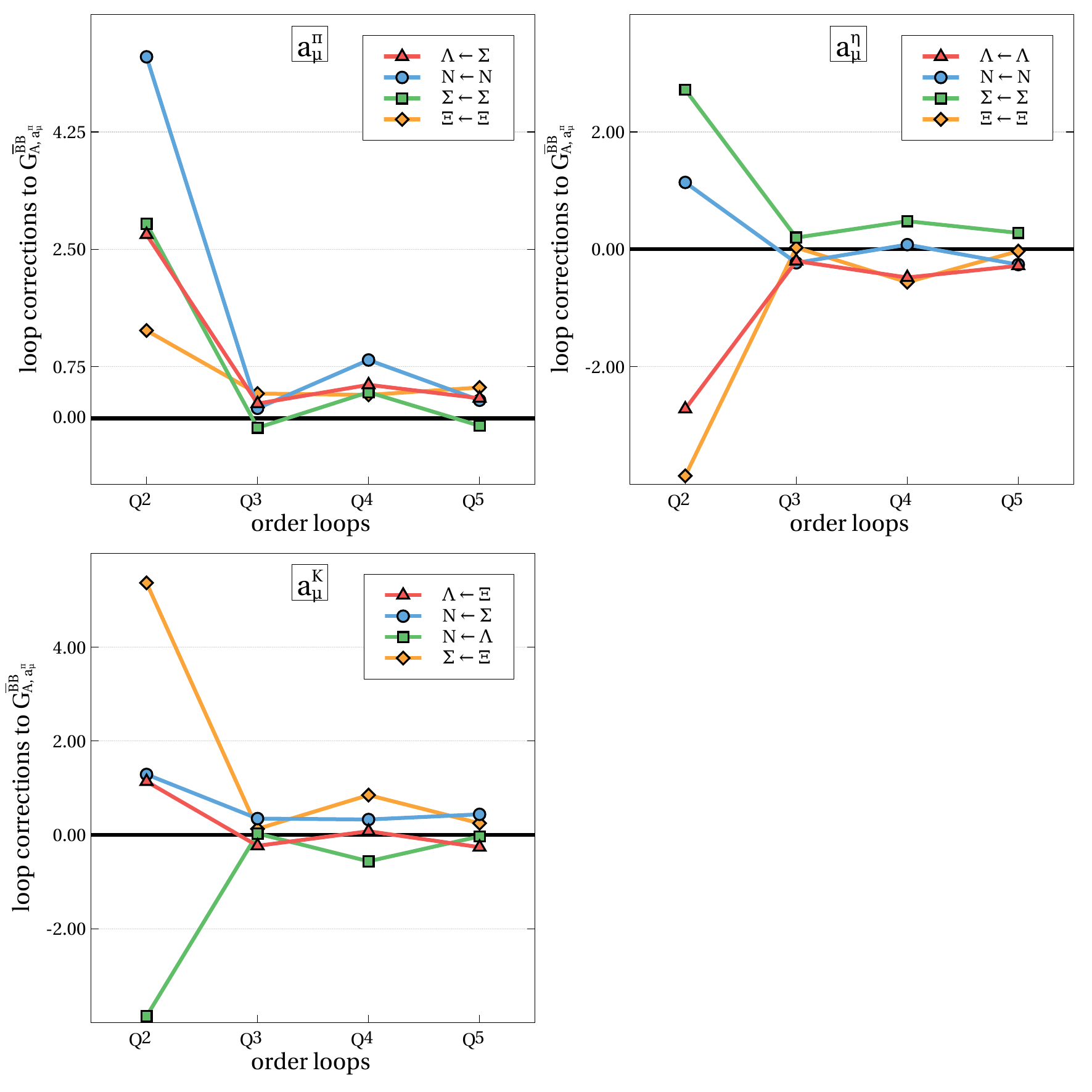}
\caption{Loop corrections to $G_{A,i}^{\bar{B}B}\big(t = (M_{\bar{p}} - M_p)^2\big)$ in the flavor-SU(3) limit, with the masses from Table \ref{Tab:Numerical-Estimate-SU3-InputFlavor}. The LECs $g^{(S)}$, $g^{(V)}$, $g^{(T)}$, and $f^{(A)}$ are taken from Table \ref{Tab:Numerical-Estimate-SU3-g-LECs}. They are not contributing to the order $Q^2$. The LECs $g^{(\chi)}$ and $g^{(R)}$ are not included here.  }
\label{Fig:PlotSU3ConvergenceL1213}
\end{figure}

At first, we neglect the effect of the LECs $g^{(R)}$ and $g^{(\chi)}$. In Figure \ref{Fig:PlotSU3ConvergenceL1213} the one-loop contributions are shown. At order $Q^2$, the LECs $g^{(S)}$, $g^{(V)}$, $g^{(T)}$, and $f^{(A)}$ are not effective, so the loop contributions are rather large. For the higher-order terms $Q^3$, $Q^4$, and $Q^5$ the, to the relevant order corresponding, LECs $g^{(T)}$ and $f^{(A)}$ from Table \ref{Tab:Numerical-Estimate-SU3-g-LECs} are included. This procedure evidently reduces the loop contributions for all processes. The resulting terms of order $Q^3$ and $Q^5$ are very small, the order-$Q^4$ contributions only slightly larger. We want to stress here that these contributions were not directly minimized by a fit, but only by the use of the LECs of Table \ref{Tab:Numerical-Estimate-SU3-g-LECs}. 

In the next step, we also include the LECs $g^{(\chi)}$ and $g^{(R)}$ from the Lagrangian $\mathcal{L}^{(3)}$. We use their values from Table \ref{Tab:Numerical-Estimate-SU3-Fit-gchi-gR} and perform a comparison of important parameters of the one-loop contributions between the cases with and without the counterterms $g^{(R)}$ and $g^{(\chi)}$. We give the chi-square per number of processes $\chi^2/N_p$ and the largest remaining one-loop contribution in Table \ref{Tab:Numerical-Estimate-SU3-InputFlavorLimit}. For the orders $Q^3$ and $Q^5$ the inclusion of $g^{(\chi)}$ and $g^{(R)}$ further reduces the one-loop contribution, whereas this is not the case for the order $Q^4$. A reason for that could be the appearance of additional counterterms turning relevant at order $Q^4$. 

The numerical estimates, given in this Section, show good convergence effects, both for the kinematic factors $\bar{K}$ and the one-loop contributions to the axial-vector form factors in general. With estimates for the LECs $g^{(T)}$, $f^{(A)}$, $g^{(\chi)}$, and $g^{(R)}$ we find large cancellations of the loop effects for the chiral order $Q^3$ and higher. We conclude that the comparison with QCD lattice data should be performed at the order-$Q^3$ level.

\begin{table}[ht]
\centering
\renewcommand{\arraystretch}{1.0}
\begin{tabular}{| c  c | c | c | c | c c |c c c |c c | c |}\hline
&& $Q^3$ &$Q^4$ &$Q^5$  \\ \hline
with $g_{1-7}^{(\chi)}=g_{D/F}^{(R)}=0$ & $\chi^2/N_p$ & 0.04 & 0.26 & 0.08 \\ 
& max & 0.35 & 0.85 & 0.44 \\ \hline
with $g_{1-7}^{(\chi)}$ and $g_{D/F}^{(R)}$ from Table \ref{Tab:Numerical-Estimate-SU3-Fit-gchi-gR} & $\chi^2/N_p$ & 0.02 & 0.31 & 0.04 \\ 
& max & 0.23 & 0.92 & 0.31\\ \hline
\end{tabular}
\caption{Parameters which quantify the success of the reduction of one-loop contributions in the flavor-SU(3) limit, $\chi^2/N_p$ and the maximal remaining contribution `max'. }
\label{Tab:Numerical-Estimate-SU3-InputFlavorLimit}
\end{table}

\section{Summary and Outlook}
In this paper, we investigated the axial-vector currents of the baryon octet in flavor-SU(3) chiral perturbation theory. We established analytical results for the axial-vector form factors at N$^2$LO that include one-loop contributions. The decuplet baryons were considered as an explicit degree of freedom. Since we explored the use of on-shell masses in the one-loop contributions, particular attention was paid to a power-counting scheme that leads to good convergence properties. It was demonstrated that the use of on-shell masses fulfils this condition. Furthermore, the use of on-shell masses with their finite-volume dependence implies an implicit finite-volume effect of the axial-vector form factors. While our analytical results are prepared to be confronted with QCD lattice data, a rich basis of QCD lattice data is so far only available in flavor-SU(2). However, we expect that additional flavor-SU(3) QCD lattice data will be published within the next few years.

Unfortunately, for most of the flavor-SU(3) QCD lattice ensembles, which were used to fit the baryon octet and decuplet masses to high accuracy \cite{Guo:2019nyp}, axial-vector form factors are not available yet. Once the data basis is significantly enlarged further investigations are planned. Our goal is a global fit of octet and decuplet masses and the axial-vector form factors, as presented in Ref. \cite{Lutz:2020dfi} in flavor-SU(2). The large amount of low-energy constants can be reduced by appropriate large-$N_c$ relations, to be established in future work. 
%

\section{Acknowledgments}
We thank Stefan Leupold for helpful discussions. 

\clearpage

\appendix

\section{Amplitudes}
\label{Chapter:Appendix-Amplitudes}

The following formulas for $J_{...}^P$ and $J_{...}^I$ project onto the form factors $G_{P,i}^{\bar{B}B}(t)$ and $G_{I,i}^{\bar{B}B}(t)$  like $J_{...}^A$ does onto $G_{A,i}^{\bar{B}B}(t)$ \eqref{Eq:Projection-JA}. The form factors $G_{P,i}^{\bar{B}B}(t)$ and $G_{I,i}^{\bar{B}B}(t)$ are defined in Eq. \eqref{Eq:Def-Axial-Form-Factor-SU3}.
\begin{eqnarray}
&&J_{...}^P=
\frac{M_{\bar{p}}\, M_p \, (M_{\bar{p}}+M_p)^2}
{\sqrt{2} \, (d-2) \, (M_{\bar{p}} \, M_p-\bar{p}\cdot p) \, (M_{\bar{p}} \, M_p+\bar{p}\cdot p)} \times
\nonumber\\ 
&& \hspace{1cm} \tr\bigg[\gamma_{\mu} \, \gamma_5 \, 
\frac{\bar{\slashed{p}}+M_{\bar{p}}}{2 \, M_{\bar{p}}} \, J^{\mu}_{...} \, \frac{\slashed{p}+M_p}{2 \, M_p}\bigg]
\nonumber\\
&&\hspace{0.2cm}+\frac{M_{\bar{p}} \, M_p \, (M_{\bar{p}}+M_p)^2\big((d-1) \, (M_{\bar{p}}^2+M_p^2)+2 \, M_{\bar{p}} \, M_p+2\, (d-2) \, \bar{p}\cdot p\big)}
{2\, \sqrt{2} \, (d-2) \, (M_{\bar{p}}\, M_p-\bar{p}\cdot p)^2 \, (M_{\bar{p}} \, M_p+\bar{p}\cdot p)}\times
\nonumber\\
&& \hspace{1cm}\tr\bigg[\frac{\bar{p}_{\mu}-p_{\mu}}{M_{\bar{p}}+M_p} \, \gamma_5\;
\frac{\bar{\slashed{p}}+M_{\bar{p}}}{2\, M_{\bar{p}}}\,  J^{\mu}_{...} \, \frac{\slashed{p}+M_p}{2 \, M_p}\bigg]
\nonumber\\
&&\hspace{0.2cm}+\frac{(d-1)\, M_{\bar{p}} \, M_p \, (M_{\bar{p}}+M_p)^4}
{2 \, \sqrt{2} \, (d-2) \, (M_{\bar{p}} \, M_p-\bar{p}\cdot p)^2 \, (M_{\bar{p}} \, M_p+\bar{p}\cdot p)} \times
\nonumber\\
&&\hspace{1cm}\tr\bigg[\frac{(\slashed{\bar{p}}+\slashed{p})\,(\bar{p}_\mu + p_{\mu})}{(M_{\bar{p}}+M_p)^2}\,
\gamma_5 \, \frac{\bar{\slashed{p}}+M_{\bar{p}}}{2 \, M_{\bar{p}}} \, J^{\mu}_{...}\frac{\slashed{p}+M_p}{2 \, M_p}\bigg]\,,
\label{JPallcontributionsnew}
\\
\nonumber\\
&&J_{...}^I=
-\frac{M_{\bar{p}} \, M_p \, (M_{\bar{p}}+M_p)^2}
{\sqrt{2} \, (d-2) \, (M_{\bar{p}} \, M_p-\bar{p}\cdot p) \, (M_{\bar{p}} \, M_p+\bar{p}\cdot p)} \times
\nonumber\\
&& \hspace{1cm} \tr\bigg[\gamma_{\mu}\, \gamma_5 \, 
\frac{\bar{\slashed{p}}+M_{\bar{p}}}{2 \, M_{\bar{p}}} \, J^{\mu}_{...} \, \frac{\slashed{p}+M_p}{2 \, M_p}\bigg]
\nonumber\\
&&\hspace{0.2cm} -\frac{(d-1)\, M_{\bar{p}} \, M_p \, (M_{\bar{p}}+M_p)^4}
{2 \, \sqrt{2} \, (d-2) \, (M_{\bar{p}} \, M_p-\bar{p}\cdot p)^2 \, (M_{\bar{p}} \, M_p+\bar{p}\cdot p)} \times
\nonumber\\
&& \hspace{1cm}
\tr\bigg[\frac{\bar{p}_{\mu}-p_{\mu}}{M_{\bar{p}}+M_p}\, \gamma_5\;
\frac{\bar{\slashed{p}}+M_{\bar{p}}}{2 \, M_{\bar{p}}}  \, J^{\mu}_{...} \, \frac{\slashed{p}+M_p}{2 \, M_p}\bigg]
\nonumber\\
&& \hspace{0.2cm} -\frac{M_{\bar{p}} \, M_p \, (M_{\bar{p}}+M_p)^4((d-1) \, (M_{\bar{p}}^2+M_p^2)-2 \, M_{\bar{p}} \, M_p-2 \, (d-2) \, \bar{p}\cdot p)}
{2 \, \sqrt{2} \, (d-2) \, (M_{\bar{p}}-M_p)^2 \, (M_{\bar{p}}M_p-\bar{p}\cdot p)^2 \, (M_{\bar{p}} \, M_p+\bar{p}\cdot p)} \;\times 
\nonumber \\
&&\; \hspace{1cm}\tr\bigg[\frac{(\slashed{\bar{p}}+\slashed{p})(\bar{p}_\mu + p_{\mu})}{(M_{\bar{p}}+M_p)^2}\, \gamma_5\;
\frac{\bar{\slashed{p}}+M_{\bar{p}}}{2 \, M_{\bar{p}}} \, J^{\mu}_{...} \, \frac{\slashed{p}+M_p}{2 \, M_p}\bigg]\,.
\label{JIallcontributionsnew}
\end{eqnarray}

In the following, we give the amplitudes of all Feynman diagrams $J_{...}^\mu$, which enter Eq. \eqref{Eq:GA-all-contributions-of-diagrams}  in form of $J_{...}^A$ after the projection with formula \eqref{Eq:Projection-JA}.
%
We use the following propagators in $d$ dimensions \cite{Lutz:2001yb, Semke:2005sn}, which are derived from $\mathcal{L}_\text{kin}$ \eqref{Eq:Def-Lkin} \cite{Haberzett:1998rw}: 
\begin{itemize}
\item Spin-0 Goldstone boson:
\begin{eqnarray}
 D_Q(k)=1/(k^2-m_Q^2+i\epsilon)\,,
\label{Eq:Spin0Propagator}
 \end{eqnarray}
\item Spin-1/2 (octet) baryon:
\begin{eqnarray}
 S_R(k)=1/(\slashed{k}-M_{R}+i\epsilon)\;,
 \label{Eq:Spin1/2Propagator}
\end{eqnarray}
\item Spin-3/2 (decuplet) baryon: 
\begin{eqnarray}
\hspace{-0.85cm} S_R^{\mu\nu}(k)=\frac{-1}{\slashed{k}-M_{R}+i\epsilon}
\bigg(g^{\mu\nu}-\frac{\gamma^{\mu}
\gamma^{\nu}}{d-1}+\frac{(k^{\mu}\gamma^{\nu}-k^{\nu}\gamma^{\mu})}{(d-1)M_R}-
\frac{(d-2)k^{\mu}k^{\nu}}{(d-1)M_R^2}\bigg).
\label{Eq:Spin3/2Propagator}
\end{eqnarray}
\end{itemize}
The Feynman diagrams associated with the following amplitudes are given in Tables \ref{Tab:FeynmanDiagrams1}-\ref{Tab:FeynmanDiagrams3}.
 \allowdisplaybreaks[1] 
\begin{eqnarray}
&& J^{\mu}= \frac{1}{\sqrt{2}} \, \gamma^{\mu} \, \gamma_5 \,,
\label{Eq:Amplitude-Tree}
\\
&& J^{\mu}= \frac{1}{\sqrt{2}\,t}\big(\gamma^{\nu}\,\gamma_5\, 
  (-\bar{p}_\nu\bar{p}^\mu +\bar{p}^\mu p_\nu+\bar{p}_\nu p^\mu-p_\nu p^\mu )+q^2\gamma^{\mu}\gamma_5\big) \,,
  \label{Eq:Amplitude-Counter-Term}
\\
&& J_{Q}^{\mu}=  \frac{i}{\sqrt{2}}\int \frac{d^d l}{(2\pi)^{d}} \, \gamma^{\mu} \, \gamma_5 \, D_Q(l)\,,
  \label{Eq:Amplitude-Tadpole}
\\
&& J^{\mu, 1}_{LQ}= -\frac{i}{\sqrt{2}}\int \frac{d^d l}{(2\pi)^{d}}\,\slashed{l}\,\gamma_5 \, S_L(\bar{p}-l) \, D_Q(l) \, \gamma^{\mu} 
\,,
  \label{Eq:Amplitude-BubbleL8-1}
\\
&& J^{\mu,2}_{LQ}= -\frac{i}{\sqrt{2}}\int \frac{d^d l}{(2\pi)^{d}} \, \slashed{l} \, \gamma_5 \, S_L(\bar{p}-l) \, D_Q(l) \, l^{\mu} \,,
  \label{Eq:Amplitude-BubbleL8-2}
\\  
&& J^{\mu,3}_{LQ}= -\frac{i}{\sqrt{2}}\int \frac{d^d l}{(2\pi)^{d}} \, \slashed{l}\, \gamma_5 \, S_L(\bar{p}-l) \, D_Q(l) \times
\nonumber \\
&& \hspace{2cm} \big((\gamma^{\mu} \, p\cdot l+\slashed{l}\, p^{\mu}) + (\gamma^{\mu} \, (\bar{p}-l)\cdot l+\slashed{l}\,(\bar{p}-l)^{\mu})\big)\,,
  \label{Eq:Amplitude-BubbleL8-3}
\\
&& J^{\mu,4}_{LQ}= -\frac{i}{\sqrt{2}}\int \frac{d^d l}{(2\pi)^{d}} \, \slashed{l} \, \gamma_5 \, S_L(\bar{p}-l) \, D_Q(l)\, i \, \sigma^{\mu\nu} \, l_{\nu}\,,
  \label{Eq:Amplitude-BubbleL8-4}
\\
&& J^{\mu,5}_{LQ}= \frac{i}{\sqrt{2}}\int \frac{d^d l}{(2\pi)^{d}} \, \slashed{l} \, \gamma_5 \, S_L(\bar{p}-l) \, D_Q(l)\, i \, \sigma^{\mu\nu} \, (\bar p -p)_{\nu}\,,
  \label{Eq:Amplitude-BubbleL8-5}  
  \\
&& J^{\mu,1}_{QR}=  \frac{i}{\sqrt{2}}\int \frac{d^d l}{(2\pi)^{d}} \, \gamma^{\mu} \, D_Q(l) \, S_R(p-l) \, \slashed{l} \, \gamma_5  \,,
  \label{Eq:Amplitude-BubbleR8-1}
  \\
&& J^{\mu,2}_{QR}= - \frac{i}{\sqrt{2}}\int \frac{d^d l}{(2\pi)^{d}} \, l^{\mu} \,
 D_Q(l) \, S_R(p-l) \, \slashed{l} \, \gamma_5\,,   
 \label{Eq:Amplitude-BubbleR8-2}
 \\
 && J^{\mu,3}_{QR}=  -\frac{i}{\sqrt{2}}\int \frac{d^d l}{(2\pi)^{d}} \, \Big((l\cdot (p-l) \, \gamma^{\mu}+\slashed{l} \, (p-l)^{\mu}) 
 \nonumber \\ 
&& \hspace{2cm} + (\gamma^{\mu} \, \bar{p}\cdot l+\slashed{l} \, \bar{p}^{\mu})\Big) 
 \, D_Q(l) \, S_R(p-l) \, \slashed{l} \, \gamma_5 \,,
 \label{Eq:Amplitude-BubbleR8-3}
\\
&& J^{\mu,4}_{QR}=  -\frac{i}{\sqrt{2}}\int \frac{d^d l}{(2\pi)^{d}} \,i \,\sigma^{\mu\nu} \, l_{\nu} \, D_Q(l) \, S_R(p-l) \, \slashed{l} \, \gamma_5 \,, 
 \label{Eq:Amplitude-BubbleR8-4}
 \\
&& J^{\mu,5}_{QR}=  \frac{i}{\sqrt{2}}\int \frac{d^d l}{(2\pi)^{d}} \,i \,\sigma^{\mu\nu} \, (\bar p -p)_{\nu} \, D_Q(l) \, S_R(p-l) \, \slashed{l} \, \gamma_5 \,, 
 \label{Eq:Amplitude-BubbleR8-5}
 \\
&& J_{LQ}^{\mu,1}=\frac{i}{\sqrt{2}} \hspace{-0.05cm} \int \hspace{-0.1cm} \frac{d^dl}{(2\pi)^d} l_{\nu} \big(S_L^{\nu\mu}(\bar{p}-l)\slashed{l} + S_L^{\nu\alpha}(\bar{p}-l)l_{\alpha}\gamma^{\mu}\big) \gamma_5 D_Q(l) ,
  \label{Eq:Amplitude-BubbleL10-1}
 \\  
&& J_{LQ}^{\mu,2}=\frac{i}{\sqrt{2}} \hspace{-0.05cm} \int \hspace{-0.1cm} \frac{d^dl}{(2\pi)^d} l_{\nu} \big(S_L^{\nu\mu}(\bar{p}-l)\slashed{l} - S_L^{\nu\alpha}(\bar{p}-l)l_{\alpha}\gamma^{\mu}\big) \gamma_5 D_Q(l) ,
  \label{Eq:Amplitude-BubbleL10-2}
\\  
&& J_{LQ}^{\mu,3}=\frac{i}{\sqrt{2}} \hspace{-0.05cm} \int \hspace{-0.1cm} \frac{d^dl}{(2\pi)^d} l_{\nu} \big(S_L^{\nu\alpha}(\bar{p}-l)\gamma^\mu 
- S_L^{\nu\mu}(\bar{p}-l) \gamma^{\alpha} \big) (\bar p-p)_\alpha\,\gamma_5 D_Q(l) ,
  \label{Eq:Amplitude-BubbleL10-3}  
  \\
&& J_{LQ}^{\mu,4}= J_{LQ}^{\mu,5}= 0\,,
  \label{Eq:Amplitude-BubbleL10-3-4}
  \\
&&  J_{QR}^{\mu,1}= \hspace{-0.07cm} \frac{i}{\sqrt{2}} \hspace{-0.08cm} \int \hspace{-0.15cm} \frac{d^dl}{(2\pi)^d}\big(\slashed{l}\gamma_5  S_R^{\mu\nu}(p\hspace{-0.02cm} - \hspace{-0.02cm} l) \hspace{-0.02cm} + \hspace{-0.02cm} \gamma^{\mu}\gamma_5 l_{\alpha} S_R^{\alpha\nu}(p \hspace{-0.02cm} - \hspace{-0.02cm} l)\big) l_{\nu}D_Q(l),
  \label{Eq:Amplitude-BubbleR10-1}
   \\
&&  J_{QR}^{\mu,2}= \hspace{-0.07cm} \frac{i}{\sqrt{2}} \hspace{-0.08cm} \int \hspace{-0.15cm} \frac{d^dl}{(2\pi)^d}\big(\slashed{l}\gamma_5  S_R^{\mu\nu}(p\hspace{-0.02cm} - \hspace{-0.02cm} l) \hspace{-0.02cm} - \hspace{-0.02cm} \gamma^{\mu}\gamma_5 l_{\alpha} S_R^{\alpha\nu}(p \hspace{-0.02cm} - \hspace{-0.02cm} l)\big) l_{\nu}D_Q(l),
  \label{Eq:Amplitude-BubbleR10-2} \\
&&  J_{QR}^{\mu,3}= \hspace{-0.07cm} \frac{i}{\sqrt{2}} \hspace{-0.08cm} \int \hspace{-0.15cm} \frac{d^dl}{(2\pi)^d}\,\gamma_5\,(\bar p -p)_\alpha\,\big(\gamma^\alpha\, S_R^{\mu\nu}(p-l)\hspace{-0.02cm} -  \gamma^\mu S_R^{\alpha\nu}(p \hspace{-0.02cm} - \hspace{-0.02cm} l)\big)\, l_{\nu}D_Q(l),
  \label{Eq:Amplitude-BubbleR10-3} 
  \\
&& J_{QR}^{\mu,4}= J_{QR}^{\mu,5}= 0\,,
  \label{Eq:Amplitude-BubbleR10-3-4}  
\\
&& J^{\mu, 1}_{LQR}= \frac{i}{\sqrt{2}}\int \frac{d^d l}{(2\pi)^{d}} \, \slashed{l}  \gamma_5  \, S_L(\bar{p}-l) \, \gamma^{\mu}  \gamma_5 \, D_Q(l) \, S_R(p-l) \, \slashed{l} \gamma_5 \, ,
  \label{Eq:Amplitude-TriangleL8R8} 
\\
&& J^{\mu, 1}_{LQR}=\frac{i}{\sqrt{2}} \int \frac{d^d l}{(2\pi)^{d}} \, l_{\sigma}\, 
S_L^{\sigma\mu}(\bar{p}-l) \, D_Q(l) \, S_R(p-l) \, \slashed{l} \, \gamma_5\,,
  \label{Eq:Amplitude-TriangleL10R8} 
\\  
&& J^{\mu, 1}_{LQR}= \frac{i}{\sqrt{2}}\int \frac{d^d l}{(2\pi)^{d}} \, \slashed{l} \, \gamma_5 \, S_L(\bar{p}-l) \,  D_Q(l) \, S_R^{\mu\tau}(p-l) \, l_{\tau} \,,  
\label{Eq:Amplitude-TriangleL8R10} 
\\
&& J^{\mu,1}_{LQR}=\frac{i}{\sqrt{2}}\int \frac{d^d l}{(2\pi)^{d}}\,l_{\sigma}\,
S_L^{\sigma\alpha}(\bar{p}-l) \, D_Q(l) \, \gamma^{\mu} \, \gamma_5 \, S^R_{\alpha\beta}(p-l) \,l^{\beta}\,,
\label{Eq:Amplitude-TriangleL10R10} 
\\
&& J^{\mu, 2}_{LQR}=\frac{i}{\sqrt{2}} \int \frac{d^d l}{(2\pi)^{d}} \, l_{\sigma}\, 
\big( S_L^{\sigma\alpha}(\bar{p}-l)\,\gamma^\mu
\nonumber\\
&& \qquad \qquad -\,S_L^{\sigma\mu }(\bar{p}-l)\,\gamma^\alpha\big) (\bar p -p)_\alpha\, D_Q(l) \, S_R(p-l) \, \slashed{l} \, \gamma_5\,,
  \label{Eq:Amplitude-TriangleL10R8-2} 
\\  
&& J^{\mu, 2}_{LQR}= \frac{i}{\sqrt{2}}\int \frac{d^d l}{(2\pi)^{d}} \, \slashed{l} \, \gamma_5 \, S_L(\bar{p}-l) \,  D_Q(l) \, \big( \gamma^\alpha\,S_R^{\mu\tau}(p-l) 
\nonumber\\
&& \qquad \qquad -\,\gamma^\mu\,S_R^{\alpha\tau}(p-l)  \big)\,(\bar p -p)_\alpha\, l_{\tau} \,.  
\label{Eq:Amplitude-TriangleL8R10-2} 
\end{eqnarray}

\section{CGCs without Axial-Vector Current}
\label{Chapter:Clebsch-Coefficients}
In this part we give the 3-point Clebsch-Gordan coefficients $G_{QR}^{(B)}$ with the Goldstone boson $Q$ and the baryons $B,R\in[8], [10]$.
\begin{table}[ht]
\centering
\renewcommand{\arraystretch}{1.0}
\begin{tabular}{|l l l|}\hline
\rule{0pt}{3ex}
$G_{\pi N}^{(N)}=\sqrt{3}(D+F)$&
 $G_{\pi \Sigma}^{(\Lambda)}=2D$&
 $G_{\pi \Lambda}^{(\Sigma)}=\frac{2}{\sqrt{3}}D$
  \\ 
  \, $G_{\eta N}^{(N)}=-\frac{1}{\sqrt{3}}(D-3F)$&
  $G_{\bar{K} N}^{(\Lambda)}=-\sqrt{\frac{2}{3}}(D+3F)$&
 $G_{\pi \Sigma}^{(\Sigma)}=-2\sqrt{2}F$
 \\ 
 \,  $G_{K \Lambda}^{(N)}=-\frac{1}{\sqrt{3}}(D+3F)$&
  $G_{\eta \Lambda}^{(\Lambda)}=-\frac{2}{\sqrt{3}}D$&
 $G_{\bar{K} N}^{(\Sigma)}=\sqrt{2}(D-F)$
 \\ 
 \,   $G_{K \Sigma}^{(N)}=\sqrt{3}(D-F)$&
$G_{K \Xi}^{(\Lambda)}=\sqrt{\frac{2}{3}}(D-3F)$&
 $G_{\eta \Sigma}^{(\Sigma)}=\frac{2}{\sqrt{3}}D$
 \\  \cline{1-2}
 \rule{0pt}{3ex}
 $G_{\pi \Xi}^{(\Xi)}=-\sqrt{3}(D-F)$&
  $G_{\bar{K}\Lambda}^{(\Xi)}=-\frac{1}{\sqrt{3}}(D-3F)$&
$G_{K \Xi}^{(\Sigma)}=\sqrt{2}(D+F)$
\\
\, $G_{\bar{K}\Sigma}^{(\Xi)}=-\sqrt{3}(D+F)$&
 $G_{\eta\Xi}^{(\Xi)}=-\frac{1}{\sqrt{3}}(D+3F)$&
\\ \hline
\end{tabular}
\caption{CGCs $G_{QR}^{(B)}$ with  $B,R\in[8]$.}
\label{Tab:Clebsches3pointG88}
\end{table} 

\begin{table}[ht]
\centering
\renewcommand{\arraystretch}{1.0}
\begin{tabular}{|l l l l|}\hline
\rule{0pt}{3ex}
$G_{\pi N}^{(\Delta_{\mu})}=\sqrt{2}C$&
 $G_{\pi \Lambda}^{(\Sigma_{\mu})}=-C$&
$G_{\pi \Xi}^{(\Xi_{\mu})}=-C$&
 $G_{\bar{K} \Xi}^{(\Omega_{\mu})}=-2C$\\ 
\,  $G_{K\Sigma}^{(\Delta_{\mu})}=-\sqrt{2}C$&
   $G_{\pi\Sigma}^{(\Sigma_{\mu})}=-\sqrt{\frac{2}{3}}C$&
 $G_{\bar{K}\Lambda}^{(\Xi_{\mu})}=C$&   \\ 
&
  $G_{\bar{K} N}^{(\Sigma_{\mu})}=\sqrt{\frac{2}{3}}C$&
 $G_{\bar{K} \Sigma}^{(\Xi_{\mu})}=C$&
 \\ 
   &
  $G_{\eta \Sigma}^{(\Sigma_{\mu})}=C$&
 $G_{\eta \Xi}^{(\Xi_{\mu})}=-C$&
\\  
  &
   $G_{K\Xi}^{(\Sigma_{\mu})}=-\sqrt{\frac{2}{3}}C$&
 &
\\  \hline
\end{tabular}
\caption{CGCs $G_{QR}^{(B)}$ with $B\in[10]$, denoted by $\mu$,  and $R\in [8]$.}
\label{Tab:Clebsches3pointG108}
\end{table} 

\begin{table}[ht]
\centering
\renewcommand{\arraystretch}{1.0}
\begin{tabular}{|l l l l|}\hline
\rule{0pt}{3ex}
   $G_{\pi \Delta_{\mu}}^{(N)}=2C$&
  $G_{\pi\Sigma_{\mu}}^{(\Lambda)}=-\sqrt{3}C$&
 $G_{\pi \Sigma_{\mu}}^{(\Sigma)}=-\sqrt{\frac{2}{3}}C$&
 $G_{\pi\Xi_{\mu}}^{(\Xi)}=-C$\\ 
\, $G_{K \Sigma_{\mu}}^{(N)}=C$&
  $G_{K\Xi_{\mu}}^{(\Lambda)}=-\sqrt{2}C$&
 $G_{\bar{K} \Delta_{\mu}}^{(\Sigma)}=-\sqrt{\frac{8}{3}}C$&
 $G_{\eta\Xi_{\mu}}^{(\Xi)}=-C$\\ 
  &
 &
 $G_{\eta \Sigma_{\mu}}^{(\Sigma)}=C$&
 $G_{\bar{K}\Sigma_{\mu}}^{(\Xi)}=C$\\ 
   &
 &
$G_{K \Xi_{\mu}}^{(\Sigma)}=-\sqrt{\frac{2}{3}}C$&
 $G_{K\Omega_{\mu}}^{(\Xi)}=-\sqrt{2}C$\\ 
 \hline
\end{tabular}
\caption{CGCs $G_{QR}^{(B)}$ with $B\in[8]$ and $R\in [10]$, denoted by $\mu$.}
\label{Tab:Clebsches3pointG810}
\end{table} 

\begin{table}[ht]
\centering
\renewcommand{\arraystretch}{1.0}
\begin{tabular}{|l l l l|}\hline
\rule{0pt}{3ex}
$G_{\pi \Delta_{\mu}}^{(\Delta_{\mu})}=-\sqrt{\frac{5}{3}}H$&
$G_{\pi \Sigma_{\mu}}^{(\Sigma_{\mu})}=\frac{2\sqrt{2}}{3}H$&
 $G_{\pi \Xi_{\mu}}^{(\Xi_{\mu})}=-\frac{1}{\sqrt{3}}H$&
 $G_{\bar{K} \Xi_{\mu}}^{(\Omega_{\mu})}=-\frac{2}{\sqrt{3}}H$\\ 
\,  $G_{\eta\Delta_{\mu}}^{(\Delta_{\mu})}=-\frac{1}{\sqrt{3}}H$&
  $G_{\bar{K}\Delta_{\mu}}^{(\Sigma_{\mu})}=-\frac{2\sqrt{2}}{3}H$&
 $G_{\bar{K}\Sigma_{\mu}}^{(\Xi_{\mu})}=-\frac{2}{\sqrt{3}}H$& 
  $G_{\eta\Omega_{\mu}}^{(\Omega_{\mu})}=\frac{2}{\sqrt{3}}H$\\ 
\, $G_{K \Sigma_{\mu}}^{(\Delta_{\mu})}=-\sqrt{\frac{2}{3}}H$&
  $G_{\eta\Sigma_{\mu}}^{(\Sigma_{\mu})}=0$&
$G_{K\Omega_{\mu}}^{(\Xi_{\mu})}=-\sqrt{\frac{2}{3}}H$&
 \\ 
   &
 $G_{K\Xi_{\mu}}^{(\Sigma_{\mu})}=\frac{2\sqrt{2}}{3}H$&
 $G_{\eta \Xi_{\mu}}^{(\Xi_{\mu})}=\frac{1}{\sqrt{3}}H$ &
\\  
 \hline
\end{tabular}
\caption{CGCs $G_{QR}^{(B)}$ with $B, R\in [10]$, denoted by $\mu$.}
\label{Tab:Clebsches3pointG1010}
\end{table} 


\clearpage

\section{CGCs of Counterterms}

Here we give the CGCs of the counterterms $X^{(\bar{B}B)}_{i}$ and $T^{(\bar{B}B)}_{i}$.

\begin{table}[ht]
\centering
\renewcommand{\arraystretch}{1.0}
\begin{tabular}{|l|}\hline
\rule{0pt}{3ex}
$X^{(NN)}_{a_{\mu}^{\eta}}=
\frac{1}{\sqrt{6}} \, \Big(-g_1^{(\chi)}-g_2^{(\chi)}+3 \, g_3^{(\chi)}+3 \, g_4^{(\chi)}+2 \, g_5^{(\chi)}+6\, g_7^{(\chi)} \Big) \,m $
\\ \hspace{1.4cm}
$+\frac{1}{\sqrt{6}}\, \Big(-g_1^{(\chi)}+g_2^{(\chi)}+3\,g_3^{(\chi)}-3\,g_4^{(\chi)}-2\,g_5^{(\chi)}+3\,g_7^{(\chi)}\Big) \, m_s$
\\
\, $X^{(\Sigma\Sigma)}_{a_{\mu}^{\eta}}=
\sqrt{\frac{2}{3}}\,\Big(2\, g_1^{(\chi)}+g_5^{(\chi)}\Big) \, m
-\sqrt{\frac{2}{3}}\, g_5^{(\chi)}\,m_s$
\\
\, $X^{(\Xi\Xi)}_{a_{\mu}^{\eta}}=
-\frac{1}{\sqrt{6}}\, \Big(g_1^{(\chi)}-g_2^{(\chi)}+3 \, g_3^{(\chi)}-3 \, g_4^{(\chi)}-2 \, g_5^{(\chi)}+6 \, g_7^{(\chi)}\Big) \, m $
\\ \hspace{1.4cm} 
$-\frac{1}{\sqrt{6}} \, \Big(g_1^{(\chi)}+g_2^{(\chi)}+3 \, g_3^{(\chi)}+3 \, g_4^{(\chi)}+2 \, g_5^{(\chi)}+3 \, g_7^{(\chi)}\Big) \, m_s$
\\
\, $X^{(\Lambda\Lambda)}_{a_{\mu}^{\eta}}=
\frac{\sqrt{2}}{3\sqrt{3}}\, \Big(2 \, g_1^{(\chi)}+3 \, g_5^{(\chi)}+3 \, g_6^{(\chi)}\Big) \, m
-\frac{\sqrt{2}}{3\sqrt{3}} \, \Big(8 \, g_1^{(\chi)}+3 \, g_5^{(\chi)}+3 \, g_6^{(\chi)}\Big)\, m_s$ 
\\ \hline \rule{0pt}{3ex}
\hspace{-0.18cm}
\, $X^{(NN)}_{a_{\mu}^{\pi}}=
\Big(g_1^{(\chi)}+g_2^{(\chi)}+g_3^{(\chi)}+g_4^{(\chi)}+2 \, g_7^{(\chi)}\Big) \, m$
\\ \hspace{1.4cm}
$+\Big(g_1^{(\chi)}-g_2^{(\chi)}+g_3^{(\chi)}-g_4^{(\chi)}+g_7^{(\chi)}\Big) \,m_s$
\\
\, $X^{(\Sigma\Sigma)}_{a_{\mu}^{\pi}}=
2\, \sqrt{2}\, \Big(g_3^{(\chi)}+g_7^{(\chi)}\Big) \, m
+\sqrt{2} \, g_7^{(\chi)}\, m_s$
\\
\, $X^{(\Xi\Xi)}_{a_{\mu}^{\pi}}=
\Big(g_1^{(\chi)}-g_2^{(\chi)}-g_3^{(\chi)}+g_4^{(\chi)}-2 \, g_7^{(\chi)}\Big)\, m$
\\ \hspace{1.4cm}
$\Big(g_1^{(\chi)}+g_2^{(\chi)}-g_3^{(\chi)}-g_4^{(\chi)}-g_7^{(\chi)}\Big)\, m_s$
\\
\, $X^{(\Sigma\Lambda)}_{a_{\mu}^{\pi}}=X^{(\Lambda\Sigma)}_{a_{\mu}^{\pi}}=
\frac{1}{\sqrt{6}}\, \Big(4 \, g_1^{(\chi)}+g_6^{(\chi)}\Big)\, m
-\frac{1}{\sqrt{6}}\,g_6^{(\chi)}\, m_s$
\\ \hline \rule{0pt}{3ex}
\hspace{-0.18cm}
\, $X^{(N\Sigma)}_{a_{\mu}^K}=X^{(\Sigma N)}_{a_{\mu}^{\bar{K}}}=
\frac{1}{2}\, \Big(3 \, g_1^{(\chi)}+g_2^{(\chi)}-3 \, g_3^{(\chi)}-g_4^{(\chi)}-4 \, g_7^{(\chi)}\Big)\, m$
\\ \hspace{1.9cm}
$+\frac{1}{2}\, \Big(g_1^{(\chi)}-g_2^{(\chi)}-g_3^{(\chi)}+g_4^{(\chi)}-2 \, g_7^{(\chi)}\Big)m_s$
\\
\, $X^{(N\Lambda)}_{a_{\mu}^K}=X^{(\Lambda N)}_{a_{\mu}^{\bar{K}}}=
-\frac{1}{2\sqrt{6}} \Big(-g_1^{(\chi)}+g_2^{(\chi)}+5  g_3^{(\chi)}+3  g_4^{(\chi)}-2  g_6^{(\chi)}+12  g_7^{(\chi)}\Big) \, m$ 
\\ \hspace{1.9cm}
$-\frac{1}{2\sqrt{6}}\,\Big(5 \, g_1^{(\chi)}-g_2^{(\chi)}+7 \, g_3^{(\chi)}-3 \, g_4^{(\chi)}+2 \, g_6^{(\chi)}+6 \, g_7^{(\chi)}\Big)\, m_s$
\\ 
\, $X^{(\Sigma\Xi)}_{a_{\mu}^K}=X^{(\Xi\Sigma)}_{a_{\mu}^{\bar{K}}}=
\frac{1}{2}\, \Big(3 \, g_1^{(\chi)}-g_2^{(\chi)}+3 \, g_3^{(\chi)}-g_4^{(\chi)}+4 \, g_7^{(\chi)}\Big) \, m$
\\ \hspace{2.4cm}
$+\frac{1}{2} \, \Big(g_1^{(\chi)}+g_2^{(\chi)}+g_3^{(\chi)}+g_4^{(\chi)}+2 \, g_7^{(\chi)}\Big) \, m_s$
\\
\, $X^{(\Lambda\Xi)}_{a_{\mu}^K}=X^{(\Xi\Lambda)}_{a_{\mu}^{\bar{K}}}=
\frac{1}{2\sqrt{6}}\, \Big(g_1^{(\chi)}+g_2^{(\chi)}+5  g_3^{(\chi)}-3  g_4^{(\chi)}+2  g_6^{(\chi)}+12  g_7^{(\chi)}\Big)\, m$
\\ \hspace{2.4cm}
$+\frac{1}{2\sqrt{6}} \,\Big(-5  g_1^{(\chi)}-g_2^{(\chi)}+7  g_3^{(\chi)}+3 g_4^{(\chi)}-2  g_6^{(\chi)}+6  g_7^{(\chi)} \Big) \, m_s$
\\ \hline
\end{tabular}
\caption{CGCs $X^{(\bar{B}B)}_{i}$
with the axial-vector current $i$ and the baryons $\bar{B},B\in[8]$. }
\label{Tab:Clebsches-Counterterms-X}
\end{table}

\begin{table}[ht]
\centering
\renewcommand{\arraystretch}{1.0}
\begin{tabular}{|ll|}\hline
\rule{0pt}{3ex}
$T^{(NN)}_{a_{\mu}^{\eta}}=
-\sqrt{\frac{2}{3}}\,\Big(g_D^{(R)}-3 \, g_F^{(R)}\Big)$& 
$T^{(\Sigma\Sigma)}_{a_{\mu}^{\eta}}=
2\, \sqrt{\frac{2}{3}}\, g_D^{(R)}$
\\ 
\, $T^{(\Xi\Xi)}_{a_{\mu}^{\eta}}=
-\sqrt{\frac{2}{3}}\, \Big(g_D^{(R)}+3 \, g_F^{(R)}\Big)$&
$T^{(\Lambda\Lambda)}_{a_{\mu}^{\eta}}=
-2\, \sqrt{\frac{2}{3}} \, g_D^{(R)}$
\\ \hline
\rule{0pt}{3ex}
$T^{(NN)}_{a_{\mu}^{\pi}}=
2\, \Big(g_D^{(R)}+g_F^{(R)}\Big)$&
$T^{(\Sigma\Sigma)}_{a_{\mu}^{\pi}}=
2 \, \sqrt{2} \, g_F^{(R)}$
\\
\, $T^{(\Xi\Xi)}_{a_{\mu}^{\pi}}=
2\, \Big(g_D^{(R)}-g_F^{(R)}\Big)$&
$T^{(\Sigma\Lambda)}_{a_{\mu}^{\pi}}=T^{(\Lambda\Sigma)}_{a_{\mu}^{\pi}}=
2 \, \sqrt{\frac{2}{3}} \, g_D^{(R)}$
\\ \hline
\rule{0pt}{3ex}
$T^{(N\Sigma)}_{a_{\mu}^K}=T^{(\Sigma N)}_{a_{\mu}^{\bar{K}}}=
2\Big(g_D^{(R)}-g_F^{(R)}\Big)$&
$T^{(N\Lambda)}_{a_{\mu}^K}=T^{(\Lambda N)}_{a_{\mu}^{\bar{K}}}=
-\sqrt{\frac{2}{3}}\Big(g_D^{(R)}+3g_F^{(R)}\Big)$ 
\\ 
\, $T^{(\Sigma\Xi)}_{a_{\mu}^K}= T^{(\Xi\Sigma)}_{a_{\mu}^{\bar{K}}}=
2\Big(g_D^{(R)}+g_F^{(R)}\Big)$ &
$T^{(\Lambda\Xi)}_{a_{\mu}^K}=T^{(\Xi\Lambda)}_{a_{\mu}^{\bar{K}}}=
-\sqrt{\frac{2}{3}}\Big(g_D^{(R)}-3g_F^{(R)}\Big)$ 
\\ \hline
\end{tabular}
\caption{CGCs $T^{(\bar{B}B)}_{i}$
with the axial-vector current $i$ and the baryons $\bar{B},B\in[8]$.}
\label{Tab:Clebsches-Counterterms-T}
\end{table}   
      
\newpage
      
\section{CGCs with Axial-Vector Current}
In this Section we show all CGCs $A$ with axial-vector current, the 3-point coefficients in Tables \ref{Tab:Clebsches3pointA88} - \ref{Tab:Clebsches3pointA1010}, the 4-point coefficients in Tables \ref{Tab:Clebsches4point8n1} - \ref{Tab:Clebsches4point10n2} and the 5-point CGCs in Table \ref{Tab:Clebsches5point}.

\begin{table}[ht]
\centering
\renewcommand{\arraystretch}{1.0}
\begin{tabular}{|l l l|} \hline 
\rule{0pt}{4ex}
$A_{N N}^{(a_{\mu}^{\eta})}=-\frac{1}{\sqrt{6}}(D-3F)$&
$A_{N N}^{(a_{\mu}^{\pi})}=D+F$&
$A_{N \Sigma}^{(a_{\mu}^{K})}=A_{\Sigma N}^{(a_{\mu}^{\bar{K}})}=D-F$
\\
 $A_{\Sigma\Sigma}^{(a_{\mu}^{\eta})}=\sqrt{\frac{2}{3}}D$&
 $A_{\Sigma \Sigma}^{(a_{\mu}^{\pi})}=\sqrt{2}F$&
 $A_{N \Lambda}^{(a_{\mu}^{K})}=A_{\Lambda N}^{(a_{\mu}^{\bar{K}})}=-\frac{1}{\sqrt{6}}(D+3F)$
 \\
 $A_{\Xi\Xi}^{(a_{\mu}^{\eta})}=-\frac{1}{\sqrt{6}}(D+3F)$& 
 $A_{\Xi \Xi}^{(a_{\mu}^{\pi})}=D-F$&
 $A_{\Sigma \Xi}^{(a_{\mu}^{K})}=A_{\Xi\Sigma}^{(a_{\mu}^{\bar{K}})}=D+F$
\\ 
 $A_{\Lambda \Lambda}^{(a_{\mu}^{\eta})}=-\sqrt{\frac{2}{3}}D$&
 $A_{\Sigma \Lambda}^{(a_{\mu}^{\pi})}=\sqrt{\frac{2}{3}}D$&
 $A_{\Lambda\Xi}^{(a_{\mu}^{K})}=A_{\Xi\Lambda}^{(a_{\mu}^{\bar{K}})}=-\frac{1}{\sqrt{6}}(D-3F)$
 \\  
& $A_{\Lambda \Sigma}^{(a_{\mu}^{\pi})}=\sqrt{\frac{2}{3}}D$&
 \\  \hline
\end{tabular}
\caption{CGCs $A_{\bar{B}B}^{(i)}$ with axial-vector current $i$ and $\bar{B}, B\in[8]$. }
\label{Tab:Clebsches3pointA88}
\end{table}

\begin{table}[ht]
\centering
\renewcommand{\arraystretch}{1.0}
\begin{tabular}{|l l l l|}\hline
\rule{0pt}{4ex}
$A_{\Sigma_{\mu} \Sigma}^{(a_{\mu}^{\eta})}=\frac{1}{\sqrt{2}}C$&
$A_{\Delta_{\mu} N}^{(a_{\mu}^{\pi})}=C$&
$A_{\Delta_{\mu}\Sigma}^{(a_{\mu}^{K})}=-\frac{1}{\sqrt{3}}C$&
$A_{\Sigma_{\mu} N}^{(a_{\mu}^{\bar{K}})}=\frac{1}{\sqrt{3}}C$
\\
\,  $A_{\Xi_{\mu} \Xi}^{(a_{\mu}^{\eta})}=-\frac{1}{\sqrt{2}}C$&
 $A_{\Sigma_{\mu}\Sigma}^{(a_{\mu}^{\pi})}=\frac{1}{\sqrt{6}}C$&
 $A_{\Sigma_{\mu}\Xi}^{(a_{\mu}^{K})}=-\frac{1}{\sqrt{3}}C$&
 $A_{\Xi_{\mu} \Sigma}^{(a_{\mu}^{\bar{K}})}=-\frac{1}{\sqrt{3}}C$
 \\
& 
$A_{\Xi_{\mu} \Xi}^{(a_{\mu}^{\pi})}=\frac{1}{\sqrt{3}}C$&
&
$A_{\Xi_{\mu}\Lambda}^{(a_{\mu}^{\bar{K}})}=\frac{1}{\sqrt{2}}C$
\\
 &
 $A_{\Sigma_{\mu} \Lambda}^{(a_{\mu}^{\pi})}=-\frac{1}{\sqrt{2}}C$&
 &
 $A_{\Omega_{\mu} \Xi}^{(a_{\mu}^{\bar{K}})}=-C$
 \\ \hline
\end{tabular}
\caption{CGCs $A_{\bar{B}B}^{(i)}$ with axial-vector current $i$, $\bar{B}\in[10]$, denoted by $\mu$, and $B\in[8]$.}
\label{Tab:Clebsches3pointA108}
\end{table}

\begin{table}[ht]
\centering
\renewcommand{\arraystretch}{1.0}
\begin{tabular}{|l l l l|} \hline
\rule{0pt}{4ex}
$A_{\Sigma \Sigma_{\mu}}^{(a_{\mu}^{\eta})}=\frac{1}{\sqrt{2}}C$&
$A_{N \Delta_{\mu}}^{(a_{\mu}^{\pi})}=C$&
$A_{N \Sigma_{\mu}}^{(a_{\mu}^{K})}=\frac{1}{\sqrt{3}}C$&
$A_{\Sigma \Delta_{\mu}}^{(a_{\mu}^{\bar{K}})}=-\frac{1}{\sqrt{3}}C$
\\
\, $A_{\Xi\Xi_{\mu}}^{(a_{\mu}^{\eta})}=-\frac{1}{\sqrt{2}}C$&
 $A_{\Sigma \Sigma_{\mu}}^{(a_{\mu}^{\pi})}=\frac{1}{\sqrt{6}}C$&
 $A_{\Sigma \Xi_{\mu}}^{(a_{\mu}^{K})}=-\frac{1}{\sqrt{3}}C$&
 $A_{\Xi\Sigma_{\mu}}^{(a_{\mu}^{\bar{K}})}=-\frac{1}{\sqrt{3}}C$
 \\
&
\, $A_{\Xi\Xi_{\mu}}^{(a_{\mu}^{\pi})}=\frac{1}{\sqrt{3}}C$&
$A_{\Lambda\Xi_{\mu}}^{(a_{\mu}^{K})}=\frac{1}{\sqrt{2}}C$&
\\
 &
 $A_{\Lambda\Sigma_{\mu}}^{(a_{\mu}^{\pi})}=-\frac{1}{\sqrt{2}}C$&
 $A_{\Xi\Omega_{\mu}}^{(a_{\mu}^{K})}=-C$&
\\  \hline
\end{tabular}
\caption{CGCs $A_{\bar{B}B}^{(i)}$ with axial-vector current $i$, $\bar{B}\in[8]$ and $B\in[10]$, denoted by $\mu$.}
\label{Tab:Clebsches3pointA810}
\end{table}

\begin{table}[ht]
\centering
\renewcommand{\arraystretch}{1.0}
\begin{tabular}{|l l l l|}\hline
\rule{0pt}{4ex}
$A_{\Delta_{\mu}\Delta_{\mu}}^{(a_{\mu}^{\eta})}=-\frac{1}{\sqrt{6}}H$&
$A_{\Delta_{\mu} \Delta_{\mu}}^{(a_{\mu}^{\pi})}=-\frac{1}{\sqrt{3}}H$&
$A_{\Delta_{\mu} \Sigma_{\mu}}^{(a_{\mu}^{K})}=-\frac{1}{3}H$&
$A_{\Sigma_{\mu}\Delta_{\mu}}^{(a_{\mu}^{\bar{K}})}=-\frac{1}{3}H$
\\
\, $A_{\Sigma_{\mu}\Sigma_{\mu}}^{(a_{\mu}^{\eta})}=0$&
 $A_{\Sigma_{\mu} \Sigma_{\mu}}^{(a_{\mu}^{\pi})}=-\frac{\sqrt{2}}{3}H$&
 $A_{\Sigma_{\mu}\Xi_{\mu}}^{(a_{\mu}^{K})}=\frac{2}{3}H$&
 $A_{\Xi_{\mu}\Sigma_{\mu}}^{(a_{\mu}^{\bar{K}})}=\frac{2}{3}H$
 \\
\, $A_{\Xi_{\mu} \Xi_{\mu}}^{(a_{\mu}^{\eta})}=\frac{1}{\sqrt{6}}H$& 
$A_{\Xi_{\mu} \Xi_{\mu}}^{(a_{\mu}^{\pi})}=\frac{1}{3}H$&
$A_{\Xi_{\mu}\Omega_{\mu}}^{(a_{\mu}^{K})}=-\frac{1}{\sqrt{3}}H$&
$A_{\Omega_{\mu} \Xi_{\mu}}^{(a_{\mu}^{\bar{K}})}=-\frac{1}{\sqrt{3}}H$
\\ 
 \, $A_{\Omega_{\mu}\Omega_{\mu}}^{(a_{\mu}^{\eta})}=\sqrt{\frac{2}{3}}H$&
 &
 &
 \\ \hline
\end{tabular}
\caption{CGCs $A_{\bar{B}B}^{(i)}$ with axial-vector current $i$ and $\bar{B},B\in[10]$, denoted by $\mu$.}
\label{Tab:Clebsches3pointA1010}
\end{table}

\begin{table}[ht]
\centering
\renewcommand{\arraystretch}{1.0}
\begin{tabular}{|l  l  l |}\hline
\rule{0pt}{4ex}
 $A_{\pi, a_{\mu}^{\eta}}^{(\bar{B} B)}=0$&
$A_{\pi, a_{\mu}^{\pi}}^{(\bar{B} B)}=-4A_{\bar{B}B}^{(a_{\mu}^{\pi})}$&
$A_{\pi, a_{\mu}^{K}}^{(\bar{B} B)}=-\frac{3}{2}A_{\bar{B}B}^{(a_{\mu}^{K})}$
\\ 
\, $A_{K, a_{\mu}^{\eta}}^{(\bar{B} B)}=-3A_{\bar{B}B}^{(a_{\mu}^{\eta})}$&
$A_{K, a_{\mu}^{\pi}}^{(\bar{B} B)}=-A_{\bar{B}B}^{(a_{\mu}^{\pi})}$&
$A_{K, a_{\mu}^{K}}^{(\bar{B} B)}=A_{\bar{K}, a_{\mu}^{\bar{K}}}^{(\bar{B} B)}=0$
\\ 
\, $A_{\bar{K}, a_{\mu}^{\eta}}^{(\bar{B} B)}=-3A_{\bar{B}B}^{(a_{\mu}^{\eta})}$&
$A_{\bar{K}, a_{\mu}^{\pi}}^{(\bar{B} B)}=-A_{\bar{B}B}^{(a_{\mu}^{\pi})}$&
$A_{\bar{K}, a_{\mu}^{K}}^{(\bar{B} B)}=-3A_{\bar{B}B}^{(a_{\mu}^{K})}$
\\ 
\, $A_{\eta, a_{\mu}^{\eta}}^{(\bar{B} B)}=0$&
$A_{\eta, a_{\mu}^{\pi}}^{(\bar{B} B)}=0$&
$A_{\eta, a_{\mu}^{K}}^{(\bar{B} B)}=-\frac{3}{2}A_{\bar{B}B}^{(a_{\mu}^{K})}$
\\ \hline
\rule{0pt}{4ex}
 $A_{\pi, a_{\mu}^{\bar{K}}}^{(\bar{B} B)}=-\frac{3}{2}A_{\bar{B}B}^{(a_{\mu}^{\bar{K}})}$&
$A_{K, a_{\mu}^{\bar{K}}}^{(\bar{B} B)}=-3A_{\bar{B}B}^{(a_{\mu}^{\bar{K}})}$&
$A_{\eta, a_{\mu}^{\bar{K}}}^{(\bar{B} B)}=-\frac{3}{2}A_{\bar{B}B}^{(a_{\mu}^{\bar{K}})}$
\\ \hline
\end{tabular}
\caption{5-point vertex CGCs $A_{Q,i}^{(\bar{B} B)}$ 
with the axial-vector current $i$, the Goldstone boson $Q$, and the baryons $\bar{B},B\in[8],[10]$.}
\label{Tab:Clebsches5point}
\end{table}

\begin{table}[ht]
\centering
\renewcommand{\arraystretch}{1.0}
\begin{tabular}{|l l l|}\hline

\parbox[c]{3cm}{
$A_{\pi N, (N)}^{(a_{\mu}^{\pi} N),1}=-2\sqrt{2}$ \\
$A_{\pi N, (N)}^{(a_{\mu}^{K} \Lambda),1}=\frac{3}{\sqrt{2}}$ \\
$A_{\pi N, (N)}^{(a_{\mu}^{K}\Sigma),1}=-\frac{1}{\sqrt{2}}$ \\
$A_{\eta N, (N)}^{(a_{\mu}^{K}\Lambda),1}=\frac{3}{\sqrt{2}}$ \\
$A_{\eta N, (N)}^{(a_{\mu}^{K}\Sigma),1}=\frac{3}{\sqrt{2}}$ \\
$A_{K\Sigma, (N)}^{(a_{\mu}^{K}\Sigma),1}=-2\sqrt{2}$ }

&

\parbox[c]{3cm}{
$ A_{\pi \Sigma, (\Lambda)}^{(a_{\mu}^{\pi} \Sigma),1}=-4\sqrt{2}$ \\
$ A_{\pi \Sigma, (\Lambda)}^{(a_{\mu}^{\bar{K}}N),1}=-\sqrt{3}$ \\
$ A_{\pi \Sigma, (\Lambda)}^{(a_{\mu}^{K} \Xi),1}=\sqrt{3}$ \\
$ A_{\bar{K} N, (\Lambda)}^{(a_{\mu}^{\bar{K}} N),1}=-3\sqrt{2}$ \\
$ A_{\bar{K} N, (\Lambda)}^{(a_{\mu}^{\eta}\Lambda),1}=-3$ \\
$ A_{\eta\Lambda, (\Lambda)}^{(a_{\mu}^{K}\Xi),1}=3$ \\
$ A_{K\Xi, (\Lambda)}^{(a_{\mu}^{K}\Xi),1}=-3\sqrt{2}$ }

& 
\rule{0pt}{22ex}

\parbox[c]{3cm}{
$ A_{\pi\Lambda, (\Sigma)}^{(a_{\mu}^{\bar{K}} N),1}=-\sqrt{3}$ \\
$ A_{\pi\Lambda, (\Sigma)}^{(a_{\mu}^{K}\Xi),1}=-\sqrt{3}$ \\
$ A_{\pi\Sigma, (\Sigma)}^{(a_{\mu}^{\pi}\Sigma),1}=-2\sqrt{2}$ \\
$ A_{\pi\Sigma, (\Sigma)}^{(a_{\mu}^{\bar{K}}N),1}=-\sqrt{2}$ \\
$ A_{\pi\Sigma, (\Sigma)}^{(a_{\mu}^{K}\Xi),1}=\sqrt{2}$ \\
$ A_{\bar{K}N, (\Sigma)}^{(a_{\mu}^{\bar{K}}N),1}=-\sqrt{2}$ \\
$ A_{\bar{K}N, (\Sigma)}^{(a_{\mu}^{\eta}\Sigma),1}=-\sqrt{3}$ \\
$ A_{\eta\Sigma, (\Sigma)}^{(a_{\mu}^{K}\Xi),1}=-\sqrt{3}$ \\
$ A_{K\Xi, (\Sigma)}^{(a_{\mu}^{K}\Xi),1}=-\sqrt{2}$ }

\\ \hline

\rule{0pt}{6ex}

\parbox[c]{3cm}{
$A_{\pi\Xi, (\Xi)}^{(a_{\mu}^{\pi}\Xi),1}=-2\sqrt{2}$ \\
$A_{\pi\Xi, (\Xi)}^{(a_{\mu}^{\bar{K}}\Lambda),1}=-\frac{3}{\sqrt{2}}$}
&
\parbox[c]{3cm}{$A_{\pi\Xi, (\Xi)}^{(a_{\mu}^{\bar{K}}\Sigma),1}=-\frac{1}{\sqrt{2}}$ \\
$A_{\bar{K}\Lambda, (\Xi)}^{(a_{\mu}^{\eta}\Xi),1}=\frac{3}{\sqrt{2}}$ }
&
\parbox[c]{3cm}{$A_{\bar{K}\Sigma, (\Xi)}^{(a_{\mu}^{\bar{K}}\Sigma),1}=-2\sqrt{2}$ \\
$A_{\bar{K}\Sigma, (\Xi)}^{(a_{\mu}^{\eta}\Xi),1}=-\frac{3}{\sqrt{2}}$ }
\\
\hline
\end{tabular}
\caption{CGCs $A_{QL,(\bar{B})}^{(iB),1}$ with the axial-vector-current index $i$,
meson $Q$, and the baryons $\bar{B},L,B\in[8]$.
All CGCs satisfy: $A_{QL,(\bar{B})}^{(iB),1}=A_{iB,(\bar{B})}^{(QL),1}$\,.
Not all contributions are shown here.}
\label{Tab:Clebsches4point8n1}
\end{table}

\begin{table}[ht]
\centering\small
\renewcommand{\arraystretch}{1.0}
\begin{tabular}{| l  l |}\hline
\rule{0pt}{27ex}

\parbox[l]{7cm}{
$ A_{\pi N, (N)}^{(a_{\mu}^{\pi} N),2}=-\frac{1}{\sqrt{2}}(2g_0^{(S)}+g_D^{(S)}+g_F^{(S)})$ \\
$ A_{\pi N, (N)}^{(a_{\mu}^{\eta} N),2}=-\frac{1}{\sqrt{2}}(g_D^{(S)}+g_F^{(S)})$ \\
$ A_{\pi N, (N)}^{(a_{\mu}^{K} \Lambda),2}=\frac{1}{2\sqrt{2}}(g_D^{(S)}+3g_F^{(S)})$ \\
$ A_{\pi N, (N)}^{(a_{\mu}^{K}\Sigma),2}=-\frac{1}{2\sqrt{2}}(2g_1^{(S)}-g_D^{(S)}+g_F^{(S)})$ \\
$ A_{\eta N, (N)}^{(a_{\mu}^{\eta} N),2}=\frac{-1}{3\sqrt{2}}(6g_0^{(S)}+5g_D^{(S)}-3g_F^{(S)})$  \\
$ A_{\eta N, (N)}^{(a_{\mu}^{K}\Lambda),2}=\frac{-1}{6\sqrt{2}}(6g_1^{(S)}+g_D^{(S)}+3g_F^{(S)})$ \\
$ A_{\eta N, (N)}^{(a_{\mu}^{K}\Sigma),2}=\frac{1}{2\sqrt{2}}(g_D^{(S)}-g_F^{(S)})$ \\
$ A_{K\Lambda, (N)}^{(a_{\mu}^{K}\Lambda),2}=-\sqrt{2}g_0^{(S)}-\frac{5}{3\sqrt{2}}g_D^{(S)}$ \\
$ A_{K\Lambda, (N)}^{(a_{\mu}^{K}\Sigma),2}=-\frac{1}{\sqrt{2}}g_D^{(S)}$ \\
$ A_{K\Sigma, (N)}^{(a_{\mu}^{K}\Sigma),2}=-\frac{1}{\sqrt{2}}(2g_0^{(S)}+g_D^{(S)}-2g_F^{(S)}) $ }

&

\parbox[l]{7cm}{
$ A_{\pi \Sigma, (\Lambda)}^{(a_{\mu}^{\pi} \Sigma),2}=-\sqrt{2}(g_0^{(S)}+2g_1^{(S)}+g_D^{(S)})$ \\
$ A_{\pi \Sigma, (\Lambda)}^{(a_{\mu}^{\bar{K}}N),2}=\frac{\sqrt{3}}{2}(-2g_1^{(S)}-g_D^{(S)}+g_F^{(S)})$ \\
$ A_{\pi \Sigma, (\Lambda)}^{(a_{\mu}^{\eta}\Lambda),2}=-\frac{1}{\sqrt{6}}(3g_1^{(S)}+2g_D^{(S)})$  \\
$ A_{\pi \Sigma, (\Lambda)}^{(a_{\mu}^{K} \Xi),2}=\frac{\sqrt{3}}{2}(2g_1^{(S)}+g_D^{(S)}+g_F^{(S)})$ \\
$ A_{\bar{K} N, (\Lambda)}^{(a_{\mu}^{\bar{K}} N),2}=\frac{-1}{\sqrt{2}}(2g_0^{(S)}+2g_1^{(S)}+3g_D^{(S)}+g_F^{(S)})$  \\
$ A_{\bar{K} N, (\Lambda)}^{(a_{\mu}^{\eta}\Lambda),2}=-\frac{1}{6}(6g_1^{(S)}+g_D^{(S)}+3g_F^{(S)})$   \\
$ A_{\bar{K} N, (\Lambda)}^{(a_{\mu}^{K}\Xi),2}=\frac{3}{\sqrt{2}}g_1^{(S)}$ \\
$ A_{\eta\Lambda, (\Lambda)}^{(a_{\mu}^{\eta}\Lambda),2}=-\sqrt{2}(g_0^{(S)}+g_1^{(S)}+g_D^{(S)})$ \\
$ A_{\eta\Lambda, (\Lambda)}^{(a_{\mu}^{K}\Xi),2}=\frac{1}{6}(6g_1^{(S)}+g_D^{(S)}-3g_F^{(S)})$ \\
$ A_{K\Xi, (\Lambda)}^{(a_{\mu}^{K}\Xi),2}=\frac{-1}{\sqrt{2}}(2g_0^{(S)}+2g_1^{(S)}+3g_D^{(S)}-g_F^{(S)})$  }

\\ \hline \rule{0pt}{29ex}

\hspace{-0.2cm}
\parbox[c]{7cm}{
$ A_{\pi\Lambda, (\Sigma)}^{(a_{\mu}^{\pi} \Lambda),2}=-\frac{\sqrt{2}}{3}(3g_0^{(S)}+g_D^{(S)})$ \\
$ A_{\pi\Lambda, (\Sigma)}^{(a_{\mu}^{\bar{K}} N),2}=\frac{1}{2\sqrt{3}}(g_D^{(S)}+3g_F^{(S)})$ \\
$ A_{\pi\Lambda, (\Sigma)}^{(a_{\mu}^{\eta} \Sigma),2}=-\frac{1}{\sqrt{2}}g_1^{(S)}-\frac{\sqrt{2}}{3}g_D^{(S)}$ \\
$ A_{\pi\Lambda, (\Sigma)}^{(a_{\mu}^{K}\Xi),2}=\frac{1}{2\sqrt{3}}(g_D^{(S)}-3g_F^{(S)})$ \\
$ A_{\pi\Sigma, (\Sigma)}^{(a_{\mu}^{\pi}\Sigma),2}=\frac{-1}{\sqrt{2}}(2g_0^{(S)}-g_1^{(S)}+2g_D^{(S)})$  \\
$ A_{\pi\Sigma, (\Sigma)}^{(a_{\mu}^{\bar{K}}N),2}=\frac{1}{\sqrt{2}}(-g_D^{(S)}+g_F^{(S)})$ \\
$ A_{\pi\Sigma, (\Sigma)}^{(a_{\mu}^{\eta}\Sigma),2}=\frac{2}{\sqrt{3}}g_F^{(S)}$ \\
$ A_{\pi\Sigma, (\Sigma)}^{(a_{\mu}^{K}\Xi),2}=\frac{1}{\sqrt{2}}(g_D^{(S)}+g_F^{(S)})$ \\
$ A_{\bar{K}N, (\Sigma)}^{(a_{\mu}^{\bar{K}}N),2}=\frac{-1}{\sqrt{2}}(2g_0^{(S)}+g_D^{(S)}-g_F^{(S)})$   \\
$ A_{\bar{K}N, (\Sigma)}^{(a_{\mu}^{\eta}\Sigma),2}=\frac{1}{2\sqrt{3}}(g_D^{(S)}-g_F^{(S)})$ \\
$ A_{\bar{K}N, (\Sigma)}^{(a_{\mu}^{K}\Xi),2}=-\frac{1}{\sqrt{2}}g_1^{(S)}$ \\
$ A_{\eta\Sigma, (\Sigma)}^{(a_{\mu}^{\eta}\Sigma),2}=-\frac{\sqrt{2}}{3}(3g_0^{(S)}+g_D^{(S)})$ 
 $ A_{\eta\Sigma, (\Sigma)}^{(a_{\mu}^{K}\Xi),2}=\frac{1}{2\sqrt{3}}(g_D^{(S)}+g_F^{(S)})$ \\
 $ A_{K\Xi, (\Sigma)}^{(a_{\mu}^{K}\Xi),2}=\frac{-1}{\sqrt{2}}(2g_0^{(S)}+g_D^{(S)}+g_F^{(S)}) $  
}

&

\hspace{-0.3cm}
\parbox[c]{7cm}{
\parbox[c]{7cm}{
$ A_{\pi\Xi, (\Xi)}^{(a_{\mu}^{\pi}\Xi),2}=-\frac{1}{\sqrt{2}}(2g_0^{(S)}+g_D^{(S)}-g_F^{(S)})$ \\
$ A_{\pi\Xi, (\Xi)}^{(a_{\mu}^{\bar{K}}\Lambda),2}=\frac{1}{2\sqrt{2}}(-g_D^{(S)}+3g_F^{(S)})$ \\
$ A_{\pi\Xi, (\Xi)}^{(a_{\mu}^{\bar{K}}\Sigma),2}=\frac{1}{2\sqrt{2}}(-2g_1^{(S)}+g_D^{(S)}+g_F^{(S)})$ \\
$ A_{\pi\Xi, (\Xi)}^{(a_{\mu}^{\eta}\Xi),2}=\frac{1}{\sqrt{2}}(g_D^{(S)}-g_F^{(S)})$ \\
$ A_{\bar{K}\Lambda, (\Xi)}^{(a_{\mu}^{\bar{K}}\Lambda),2}=-\sqrt{2}g_0^{(S)}-\frac{5}{3\sqrt{2}}g_D^{(S)}$ \\
$ A_{\bar{K}\Lambda, (\Xi)}^{(a_{\mu}^{\bar{K}}\Sigma),2}=\frac{1}{\sqrt{2}}g_D^{(S)}$ \\
$ A_{\bar{K}\Lambda, (\Xi)}^{(a_{\mu}^{\eta}\Xi),2}=\frac{1}{6\sqrt{2}}(-6g_1^{(S)}-g_D^{(S)}+3g_F^{(S)})$ \\
$ A_{\bar{K}\Sigma, (\Xi)}^{(a_{\mu}^{\bar{K}}\Sigma),2}=-\frac{1}{\sqrt{2}}(2g_0^{(S)}+g_D^{(S)}+2g_F^{(S)})$ \\
$ A_{\bar{K}\Sigma, (\Xi)}^{(a_{\mu}^{\eta}\Xi),2}=-\frac{1}{2\sqrt{2}}(g_D^{(S)}+g_F^{(S)})$ \\
$ A_{\eta\Xi, (\Xi)}^{(a_{\mu}^{\eta}\Xi),2}=-\frac{1}{\sqrt{2}}(2g_0^{(S)}+\frac{5}{3}g_D^{(S)}+g_F^{(S)} )$}
\\
%
}
\\ \hline
\end{tabular}
\caption{CGCs $A_{QL,(\bar{B})}^{(iB),2}$  with the axial-vector-current index $i$, meson $Q$, and the baryons $\bar{B},L,B\in[8]$. 
All CGCs satisfy: $A_{QL,(\bar{B})}^{(iB),2}=A_{iB,(\bar{B})}^{(QL),2}$.
Not all contributions are shown here. }
\label{Tab:Clebsches4point8n2}
\end{table}

\begin{table}[ht]
\centering
\renewcommand{\arraystretch}{1.0}
\begin{tabular}{| l  l |}\hline
\rule{0pt}{24ex}
\hspace{-0.2cm}
\parbox[c]{7cm}{
$ A_{\pi N, (N)}^{(a_{\mu}^{\pi} N),4}=\sqrt{2}(g_D^{(T)}+g_F^{(T)})$ \\
$ A_{\pi N, (N)}^{(a_{\mu}^{\eta} N),4}=0$ \\
$ A_{\pi N, (N)}^{(a_{\mu}^{K} \Lambda),4}=\frac{1}{2\sqrt{2}}(-g_D^{(T)}-3g_F^{(T)})$ \\
$ A_{\pi N, (N)}^{(a_{\mu}^{K}\Sigma),4}=\frac{1}{2\sqrt{2}}(-2g_1^{(T)}-g_D^{(T)}+g_F^{(T)})$ \\
$ A_{\eta N, (N)}^{(a_{\mu}^{\eta} N),4}=0$ \\
$ A_{\eta N, (N)}^{(a_{\mu}^{K}\Lambda),4}=-\frac{1}{2\sqrt{2}}(2g_1^{(T)}
+g_D^{(T)}+3g_F^{(T)})$ \\
$ A_{\eta N, (N)}^{(a_{\mu}^{K}\Sigma),4}=\frac{3}{2\sqrt{2}}(g_D^{(T)}-g_F^{(T)})$ \\
$ A_{K\Lambda, (N)}^{(a_{\mu}^{K}\Lambda),4}=\frac{1}{\sqrt{2}}g_D^{(T)}$ \\
$ A_{K\Lambda, (N)}^{(a_{\mu}^{K}\Sigma),4}=-\frac{1}{\sqrt{2}}g_D^{(T)}$ \\
$ A_{K\Sigma, (N)}^{(a_{\mu}^{K}\Sigma),4}=\frac{1}{\sqrt{2}}(-g_D^{(T)}+2g_F^{(T)}) $}

&

\hspace{-0.2cm}
\parbox[c]{7cm}{
$ A_{\pi \Sigma, (\Lambda)}^{(a_{\mu}^{\pi} \Sigma),4}=\sqrt{2}(g_1^{(T)}+2g_F^{(T)})$ \\
$ A_{\pi \Sigma, (\Lambda)}^{(a_{\mu}^{\bar{K}}N),4}=\frac{\sqrt{3}}{2}(2g_1^{(T)}-g_D^{(T)}+g_F^{(T)})$ \\
$ A_{\pi \Sigma, (\Lambda)}^{(a_{\mu}^{\eta}\Lambda),4}=\sqrt{\frac{3}{2}}g_1^{(T)}$ \\
$ A_{\pi \Sigma, (\Lambda)}^{(a_{\mu}^{K} \Xi),4}=-\frac{\sqrt{3}}{2}(2g_1^{(T)}+g_D^{(T)}+g_F^{(T)})$ \\
$ A_{\bar{K} N, (\Lambda)}^{(a_{\mu}^{\bar{K}} N),4}=\frac{1}{\sqrt{2}}(2g_1^{(T)}+g_D^{(T)}+3g_F^{(T)})$ \\
$ A_{\bar{K} N, (\Lambda)}^{(a_{\mu}^{\eta}\Lambda),4}=\frac{1}{2}(2g_1^{(T)}+g_D^{(T)}+3g_F^{(T)})$ \\
$ A_{\bar{K} N, (\Lambda)}^{(a_{\mu}^{K}\Xi),4}=-\frac{1}{\sqrt{2}}g_1^{(T)}$ \\
$ A_{\eta\Lambda, (\Lambda)}^{(a_{\mu}^{\eta}\Lambda),4}=0$ \\
$ A_{\eta\Lambda, (\Lambda)}^{(a_{\mu}^{K}\Xi),4}=\frac{1}{2}(-2g_1^{(T)}+g_D^{(T)}-3g_F^{(T)})$ \\
$ A_{K\Xi, (\Lambda)}^{(a_{\mu}^{K}\Xi),4}=\frac{1}{\sqrt{2}}(2g_1^{(T)}-g_D^{(T)}+3g_F^{(T)})$ }

\\ \hline \rule{0pt}{29ex}

\hspace{-0.2cm}
\parbox[c]{7cm}{
$ A_{\pi\Lambda, (\Sigma)}^{(a_{\mu}^{\pi} \Sigma),4}=-\frac{2}{\sqrt{3}}g_D^{(T)}$ \\
$ A_{\pi\Lambda, (\Sigma)}^{(a_{\mu}^{\bar{K}} N),4}=\frac{1}{2\sqrt{3}}(g_D^{(T)}+3g_F^{(T)})$ \\
$ A_{\pi\Lambda, (\Sigma)}^{(a_{\mu}^{\eta} \Sigma),4}=-\frac{1}{\sqrt{2}}g_1^{(T)}$ \\
$ A_{\pi\Lambda, (\Sigma)}^{(a_{\mu}^{K}\Xi),4}=-\frac{1}{2\sqrt{3}}(g_D^{(T)}-3g_F^{(T)})$ \\
$ A_{\pi\Sigma, (\Sigma)}^{(a_{\mu}^{\pi}\Sigma),4}=\frac{1}{\sqrt{2}}(g_1^{(T)}+2g_F^{(T)})$ \\
$ A_{\pi\Sigma, (\Sigma)}^{(a_{\mu}^{\bar{K}}N),4}=\frac{1}{\sqrt{2}}(-g_D^{(T)}+g_F^{(T)})$ \\
$ A_{\pi\Sigma, (\Sigma)}^{(a_{\mu}^{K}\Xi),4}=-\frac{1}{\sqrt{2}}(g_D^{(T)}+g_F^{(T)})$ \\
$ A_{\bar{K}N, (\Sigma)}^{(a_{\mu}^{\bar{K}}N),4}=\frac{1}{\sqrt{2}}(-g_D^{(T)}+g_F^{(T)})$ \\
$ A_{\bar{K}N, (\Sigma)}^{(a_{\mu}^{\eta}\Sigma),4}=\frac{\sqrt{3}}{2}(-g_D^{(T)}+g_F^{(T)})$ \\
$ A_{\bar{K}N, (\Sigma)}^{(a_{\mu}^{K}\Xi),4}=-\frac{1}{\sqrt{2}}g_1^{(T)}$ \\
$ A_{\eta\Sigma, (\Sigma)}^{(a_{\mu}^{K}\Xi),4}=\frac{\sqrt{3}}{2}(g_D^{(T)}+g_F^{(T)})$ \\
$ A_{K\Xi, (\Sigma)}^{(a_{\mu}^{K}\Xi),4}=\frac{1}{\sqrt{2}}(g_D^{(T)}+g_F^{(T)}) $ }

&

\hspace{-0.2cm}
\parbox[c]{7cm}{
$ A_{\pi\Xi, (\Xi)}^{(a_{\mu}^{\pi}\Xi),4}=\sqrt{2}(-g_D^{(T)}+g_F^{(T)})$ \\
$ A_{\pi\Xi, (\Xi)}^{(a_{\mu}^{\bar{K}}\Lambda),4}=\frac{1}{2\sqrt{2}}(-g_D^{(T)}+3g_F^{(T)})$ \\
$ A_{\pi\Xi, (\Xi)}^{(a_{\mu}^{\bar{K}}\Sigma),4}=\frac{-1}{2\sqrt{2}}(2g_1^{(T)}-g_D^{(T)}-g_F^{(T)})$ \\
$ A_{\pi\Xi, (\Xi)}^{(a_{\mu}^{\eta}\Xi),4}=0$ \\
$ A_{\bar{K}\Lambda, (\Xi)}^{(a_{\mu}^{\bar{K}}\Lambda),4}=-\frac{1}{\sqrt{2}}g_D^{(T)}$ \\
$ A_{\bar{K}\Lambda, (\Xi)}^{(a_{\mu}^{\bar{K}}\Sigma),4}=-\frac{1}{\sqrt{2}}g_D^{(T)}$ \\
$ A_{\bar{K}\Lambda, (\Xi)}^{(a_{\mu}^{\eta}\Xi),4}=\frac{-1}{2\sqrt{2}}(2g_1^{(T)}-g_D^{(T)}+3g_F^{(T)})$ \\
$ A_{\bar{K}\Sigma, (\Xi)}^{(a_{\mu}^{\bar{K}}\Sigma),4}=\frac{1}{\sqrt{2}}(g_D^{(T)}+2g_F^{(T)})$ \\
$ A_{\bar{K}\Sigma, (\Xi)}^{(a_{\mu}^{\eta}\Xi),4}=\frac{3}{2\sqrt{2}}(g_D^{(T)}+g_F^{(T)})$ \\
$ A_{\eta\Xi, (\Xi)}^{(a_{\mu}^{\eta}\Xi),4}=0 $}

\\ \hline
\end{tabular}
\caption{CGCs $A_{QL,(\bar{B})}^{(iB),4}$ with $L\in[8]$, 
all CGCs satisfy:  $A_{QL,(\bar{B})}^{(iB),4}=A_{iB,(\bar{B})}^{(QL),4}$. Not all contributions are shown here. }
\label{Tab:Clebsches4point8n4}
\end{table}

\begin{table}[ht]
\centering \small
\renewcommand{\arraystretch}{1.0}
\begin{tabular}{| l  l |}\hline
\rule{0pt}{20ex}

\parbox[c]{7cm}{
$ A_{\pi \Delta_\mu, (N)}^{(a_{\mu}^{\pi} N),1}=\frac{5}{2\sqrt{6}}\, f_3^{(A)}$ \\
$ A_{\pi \Delta_\mu, (N)}^{(a_{\mu}^{\eta} N),1}=\frac{1}{2\sqrt{6}}\, \big(-2 \, f_1^{(A)}+f_3^{(A)}\big)$ \\
$ A_{\pi \Delta_\mu, (N)}^{(a_{\mu}^{K} \Lambda),1}=-\frac{\sqrt{3}}{2\sqrt{2}}\, f_3^{(A)}$ \\
$ A_{\pi \Delta_\mu, (N)}^{(a_{\mu}^{K} \Sigma),1}=\frac{1}{2\sqrt{6}} \, \big(2 \, f_1^{(A)}+f_3^{(A)}\big)$ \\
$ A_{K \Sigma_\mu, (N)}^{(a_{\mu}^{\pi} N),1}=\frac{1}{4\sqrt{6}}\, \big(f_1^{(A)}+5 \, f_3^{(A)}\big)$ \\
$ A_{K \Sigma_\mu, (N)}^{(a_{\mu}^{\eta} N),1}=\frac{1}{4\sqrt{6}} \, \big(f_1^{(A)}+f_3^{(A)}\big)$ \\
$ A_{K \Sigma_\mu, (N)}^{(a_{\mu}^{K} \Lambda),1}=\frac{1}{4\sqrt{6}} \, \big(3 \, f_1^{(A)}-3 \, f_3^{(A)}\big)$ \\
$ A_{K \Sigma_\mu, (N)}^{(a_{\mu}^{K} \Sigma),1}=\frac{1}{4\sqrt{6}} \, \big(3 \, f_1^{(A)}+f_3^{(A)}\big) $}

&

\parbox[c]{7cm}{
$ A_{\pi \Sigma_\mu, (\Lambda)}^{(a_{\mu}^{\pi} \Sigma),1}=-\frac{1}{2\sqrt{6}}\, \big(f_1^{(A)}+3 \, f_3^{(A)}\big)$ \\
$ A_{\pi \Sigma_\mu, (\Lambda)}^{(a_{\mu}^{\bar{K}} N),1}=-\frac{1}{4} \, \big(f_1^{(A)}-3 \, f_3^{(A)}\big)$ \\
$ A_{\pi \Sigma_\mu, (\Lambda)}^{(a_{\mu}^{\eta} \Lambda),1}=\frac{1}{2\sqrt{2}} \, \big(f_1^{(A)}+f_3^{(A)}\big)$ \\
$ A_{\pi \Sigma_\mu, (\Lambda)}^{(a_{\mu}^{K} \Xi),1}=\frac{1}{4}\, \big(-f_1^{(A)}+f_3^{(A)}\big)$ \\
$ A_{K \Xi_\mu, (\Lambda)}^{(a_{\mu}^{\pi} \Sigma),1}=-\frac{1}{4} \, \big(f_1^{(A)}+2 \, f_3^{(A)}\big)$ \\
$ A_{K \Xi_\mu, (\Lambda)}^{(a_{\mu}^{\bar{K}} N),1}=\frac{\sqrt{3}}{2\sqrt{2}} \, f_3^{(A)}$ \\
$ A_{K \Xi_\mu, (\Lambda)}^{(a_{\mu}^{\eta} \Lambda),1}=-\frac{1}{4\sqrt{3}}\, \big(f_1^{(A)}-2 \, f_3^{(A)}\big)$ \\
$ A_{K \Xi_\mu, (\Lambda)}^{(a_{\mu}^{K} \Xi),1}=\frac{1}{2\sqrt{6}}\, \big(-2 \, f_1^{(A)}+f_3^{(A)}\big)$ }

\\ \hline

\rule{0pt}{41ex}

\parbox[c]{7cm}{
$ A_{\pi \Sigma_\mu, (\Sigma)}^{(a_{\mu}^{\pi} \Sigma),1}=\frac{1}{2\sqrt{6}}\, \big(-f_1^{(A)}+2 \, f_3^{(A)}\big)$ \\
$ A_{\pi \Sigma_\mu, (\Sigma)}^{(a_{\mu}^{\bar{K}} N),1}=-\frac{1}{2\sqrt{6}}\, \big(f_1^{(A)}+f_3^{(A)}\big)$ \\
$ A_{\pi \Sigma_\mu, (\Sigma)}^{(a_{\mu}^{\eta} \Sigma),1}=\frac{1}{6} \,\big( f_1^{(A)} - 2\, f_3^{(A)}\big)$ \\
$ A_{\pi \Sigma_\mu, (\Sigma)}^{(a_{\mu}^{K} \Xi),1}=-\frac{1}{2\sqrt{6}} \, \big(f_1^{(A)}+f_3^{(A)}\big)$ \\
$ A_{\bar{K} \Delta_\mu, (\Sigma)}^{(a_{\mu}^{\pi} \Lambda),1}=-\frac{1}{2}\, f_3^{(A)}$ \\
$ A_{\bar{K} \Delta_\mu, (\Sigma)}^{(a_{\mu}^{\pi} \Sigma),1}=-\frac{1}{2\sqrt{6}}\, \big(f_1^{(A)}-2 \, f_3^{(A)}\big)$ \\
$ A_{\bar{K} \Delta_\mu, (\Sigma)}^{(a_{\mu}^{\bar{K}} N),1}=-\frac{1}{\sqrt{6}} \, f_1^{(A)}$ \\
$ A_{\bar{K} \Delta_\mu, (\Sigma)}^{(a_{\mu}^{\eta} \Sigma),1}=-\frac{1}{6} \, \big(f_1^{(A)}+f_3^{(A)}\big)$ \\
$ A_{\bar{K} \Delta_\mu, (\Sigma)}^{(a_{\mu}^{K} \Xi),1}=-\frac{\sqrt{2}}{\sqrt{3}} \, f_3^{(A)}$ \\
$ A_{\eta \Sigma_\mu, (\Sigma)}^{(a_{\mu}^{\pi} \Lambda),1}=\frac{1}{2\sqrt{6}}\, \big(f_1^{(A)}+f_3^{(A)}\big)$ \\
$ A_{\eta \Sigma_\mu, (\Sigma)}^{(a_{\mu}^{\pi} \Sigma),1}=\frac{1}{6} \, \big(f_1^{(A)}-2 \, f_3^{(A)}\big)$ \\
$ A_{\eta \Sigma_\mu, (\Sigma)}^{(a_{\mu}^{\bar{K}} N),1}=\frac{1}{12}\, \big(f_1^{(A)}+f_3^{(A)}\big)$ \\
$ A_{\eta \Sigma_\mu, (\Sigma)}^{(a_{\mu}^{\eta} \Sigma),1}=\frac{1}{2\sqrt{6}} \, \big(f_1^{(A)}+f_3^{(A)}\big)$ \\
$ A_{\eta \Sigma_\mu, (\Sigma)}^{(a_{\mu}^{K} \Xi),1}=\frac{1}{12} \, \big(-f_1^{(A)}+5 \, f_3^{(A)}\big)$ \\
$ A_{K \Xi_\mu, (\Sigma)}^{(a_{\mu}^{\pi} \Lambda),1}=-\frac{1}{4} \, f_1^{(A)}$ \\
$ A_{K \Xi_\mu, (\Sigma)}^{(a_{\mu}^{\pi} \Sigma),1}=\frac{1}{2\sqrt{6}}\, \big(-f_1^{(A)}+2 \, f_3^{(A)}\big)$ \\
$ A_{K \Xi_\mu, (\Sigma)}^{(a_{\mu}^{\bar{K}} N),1}=-\frac{1}{2\sqrt{6}} \, f_3^{(A)}$ 
}

&

\parbox[c]{7cm}{
$ A_{\pi \Xi_\mu, (\Xi)}^{(a_{\mu}^{\pi} \Xi),1}=\frac{1}{2\sqrt{6}}\, f_1^{(A)}$ \\
$ A_{\pi \Xi_\mu, (\Xi)}^{(a_{\mu}^{\bar{K}} \Lambda),1}=\frac{\sqrt{3}}{4\sqrt{2}}\, f_1^{(A)}$ \\
$ A_{\pi \Xi_\mu, (\Xi)}^{(a_{\mu}^{\bar{K}} \Sigma),1}=-\frac{1}{4\sqrt{6}} \, \big(f_1^{(A)}-6 \, f_3^{(A)}\big)$ \\
$ A_{\pi \Xi_\mu, (\Xi)}^{(a_{\mu}^{\eta} \Xi),1}=\frac{1}{2\sqrt{6}}\, \big(f_1^{(A)}+f_3^{(A)}\big)$ \\
$ A_{\bar{K} \Sigma_\mu, (\Xi)}^{(a_{\mu}^{\pi} \Xi),1}=-\frac{1}{4\sqrt{6}} \, \big(f_1^{(A)}+3 \, f_3^{(A)}\big)$ \\
$ A_{\bar{K} \Sigma_\mu, (\Xi)}^{(a_{\mu}^{\bar{K}} \Lambda),1}=-\frac{\sqrt{3}}{4\sqrt{2}}\, \big(f_1^{(A)}-f_3^{(A)}\big)$ \\
$ A_{\bar{K} \Sigma_\mu, (\Xi)}^{(a_{\mu}^{\bar{K}} \Sigma),1}=-\frac{1}{4\sqrt{6}}\, \big(f_1^{(A)}+3 \, f_3^{(A)}\big)$ \\
$ A_{\bar{K} \Sigma_\mu, (\Xi)}^{(a_{\mu}^{\eta} \Xi),1}=\frac{1}{4\sqrt{6}}\, \big(f_1^{(A)}-5 \, f_3^{(A)}\big)$ \\
$ A_{\eta \Xi_\mu, (\Xi)}^{(a_{\mu}^{\pi} \Xi),1}=\frac{1}{2\sqrt{6}}\, \big(f_1^{(A)}+f_3^{(A)}\big)$ \\
$ A_{\eta \Xi_\mu, (\Xi)}^{(a_{\mu}^{\bar{K}} \Lambda),1}=\frac{1}{4\sqrt{6}}\, \big(f_1^{(A)}-2 \, f_3^{(A)}\big)$ \\
$ A_{\eta \Xi_\mu, (\Xi)}^{(a_{\mu}^{\bar{K}} \Sigma),1}=\frac{1}{4\sqrt{6}}\, \big(f_1^{(A)}+4 \, f_3^{(A)}\big)$ \\
$ A_{\eta \Xi_\mu, (\Xi)}^{(a_{\mu}^{\eta} \Xi),1}=-\frac{1}{2\sqrt{6}} \, \big(f_1^{(A)}-2 \, f_3^{(A)}\big)$ \\
$ A_{K \Omega_\mu, (\Xi)}^{(a_{\mu}^{\pi} \Xi),1}=\frac{\sqrt{3}}{4} \, f_1^{(A)}$ \\
$ A_{K \Omega_\mu, (\Xi)}^{(a_{\mu}^{\bar{K}} \Sigma),1}=\frac{\sqrt{3}}{2} \, f_3^{(A)}$ \\
$ A_{K \Omega_\mu, (\Xi)}^{(a_{\mu}^{\eta} \Xi),1}=-\frac{1}{4\sqrt{3}}\, \big(f_1^{(A)}-2f_3^{(A)}\big)$ 
\\

$ A_{K \Xi_\mu, (\Sigma)}^{(a_{\mu}^{\eta} \Sigma),1}=-\frac{1}{12} \, \big(f_1^{(A)}+4 \, f_3^{(A)}\big)$ \\
$ A_{K \Xi_\mu, (\Sigma)}^{(a_{\mu}^{K} \Xi),1}=-\frac{1}{2\sqrt{6}}\, f_3^{(A)}$ }
\\ \hline

\end{tabular}
\caption{CGCs  $A_{QL,(\bar{B})}^{(iB),1}$ with $L\in[10]$. 
}
\label{Tab:Clebsches4point10n1}
\end{table}

\begin{table}[ht]
\centering \small
\renewcommand{\arraystretch}{1.0}
\begin{tabular}{| l  l |}\hline
\rule{0pt}{20ex}

\parbox[c]{7cm}{
$ A_{\pi \Delta_\mu, (N)}^{(a_{\mu}^{\pi} N),2}=\frac{1}{2\sqrt{6}}\, \big(-2 \, f_2^{(A)}+f_4^{(A)}\big)$ \\
$ A_{\pi \Delta_\mu, (N)}^{(a_{\mu}^{\eta} N),2}=\frac{1}{2\sqrt{6}} \, f_4^{(A)}$ \\
$ A_{\pi \Delta_\mu, (N)}^{(a_{\mu}^{K} \Lambda),2}=-\frac{1}{2\sqrt{6}}\, f_4^{(A)}$ \\
$ A_{\pi \Delta_\mu, (N)}^{(a_{\mu}^{K} \Sigma),2}=\frac{1}{2\sqrt{6}}\, \big(-2 \, f_2^{(A)}-f_4^{(A)}\big)$ \\
$ A_{K \Sigma_\mu, (N)}^{(a_{\mu}^{\pi} N),2}=\frac{1}{4\sqrt{6}}\, \big(-f_2^{(A)}+f_4^{(A)}\big)$ \\
$ A_{K \Sigma_\mu, (N)}^{(a_{\mu}^{\eta} N),2}=\frac{1}{4\sqrt{6}}\, \big(3 \, f_2^{(A)}+f_4^{(A)}\big)$ \\
$ A_{K \Sigma_\mu, (N)}^{(a_{\mu}^{K} \Lambda),2}=\frac{1}{4\sqrt{6}} \, \big(3 \, f_2^{(A)}-f_4^{(A)}\big)$ \\
$ A_{K \Sigma_\mu, (N)}^{(a_{\mu}^{K} \Sigma),2}=\frac{1}{4\sqrt{6}}\, (-f_2^{(A)}-f_4^{(A)}\big)$ }

&

\parbox[c]{7cm}{
$ A_{\pi \Sigma_\mu, (\Lambda)}^{(a_{\mu}^{\pi} \Sigma),2}=\frac{1}{\sqrt{6}} \, f_2^{(A)}$ \\
$ A_{\pi \Sigma_\mu, (\Lambda)}^{(a_{\mu}^{\bar{K}} N),2}=-\frac{1}{4} \, \big(f_2^{(A)}-f_4^{(A)}\big)$ \\
$ A_{\pi \Sigma_\mu, (\Lambda)}^{(a_{\mu}^{\eta} \Lambda),2}=0$ \\
$ A_{\pi \Sigma_\mu, (\Lambda)}^{(a_{\mu}^{K} \Xi),2}=\frac{1}{4}\, \big(f_2^{(A)}+f_4^{(A)}\big)$ \\
$ A_{K \Xi_\mu, (\Lambda)}^{(a_{\mu}^{\pi} \Sigma),2}=\frac{1}{4}\, f_2^{(A)}$ \\
$ A_{K \Xi_\mu, (\Lambda)}^{(a_{\mu}^{\bar{K}} N),2}=\frac{1}{2\sqrt{6}}\, f_4^{(A)}$ \\
$ A_{K \Xi_\mu, (\Lambda)}^{(a_{\mu}^{\eta} \Lambda),2}=-\frac{\sqrt{3}}{4}\, f_2^{(A)}$ \\
$ A_{K \Xi_\mu, (\Lambda)}^{(a_{\mu}^{K} \Xi),2}=\frac{1}{2\sqrt{6}} \, f_4^{(A)}$ }

\\ \hline

\rule{0pt}{40ex}

\parbox[c]{7cm}{
$ A_{\pi \Sigma_\mu, (\Sigma)}^{(a_{\mu}^{\pi} \Lambda),2}=\frac{1}{6}\big(3 \, f_2^{(A)}-f_4^{(A)}\big)$ \\
$ A_{\pi \Sigma_\mu, (\Sigma)}^{(a_{\mu}^{\pi} \Sigma),2}=\frac{1}{2\sqrt{6}}\, f_2^{(A)}$ \\
$ A_{\pi \Sigma_\mu, (\Sigma)}^{(a_{\mu}^{\bar{K}} N),2}=-\frac{1}{2\sqrt{6}}\, \big(f_2^{(A)}-f_4^{(A)}\big)$ \\
$ A_{\pi \Sigma_\mu, (\Sigma)}^{(a_{\mu}^{\eta} \Sigma),2}=\frac{1}{6}\, f_4^{(A)}$ \\
$ A_{\pi \Sigma_\mu, (\Sigma)}^{(a_{\mu}^{K} \Xi),2}=-\frac{1}{2\sqrt{6}} \, \big(-f_2^{(A)}+f_4^{(A)}\big)$ \\
$ A_{\bar{K} \Delta_\mu, (\Sigma)}^{(a_{\mu}^{\pi} \Lambda),2}=\frac{1}{6}\, f_4^{(A)}$ \\
$ A_{\bar{K} \Delta_\mu, (\Sigma)}^{(a_{\mu}^{\pi} \Sigma),2}=-\frac{1}{2\sqrt{6}} \, \big(f_2^{(A)}-2 \, f_4^{(A)}\big)$ \\
$ A_{\bar{K} \Delta_\mu, (\Sigma)}^{(a_{\mu}^{\bar{K}} N),2}=\frac{1}{\sqrt{6}}\, f_2^{(A)}$ \\
$ A_{\bar{K} \Delta_\mu, (\Sigma)}^{(a_{\mu}^{\eta} \Sigma),2}=-\frac{1}{6}\, \big(-3 \, f_2^{(A)}+f_4^{(A)}\big)$ \\
$ A_{\eta \Sigma_\mu, (\Sigma)}^{(a_{\mu}^{\pi} \Sigma),2}=-\frac{1}{6} \, f_4^{(A)}$ \\
$ A_{\eta \Sigma_\mu, (\Sigma)}^{(a_{\mu}^{\bar{K}} N),2}=\frac{1}{12} \, \big(-3 \, f_2^{(A)}-f_4^{(A)}\big)$ \\
$ A_{\eta \Sigma_\mu, (\Sigma)}^{(a_{\mu}^{K} \Xi),2}=\frac{1}{12}\, \big(-3 \, f_2^{(A)}+f_4^{(A)}\big)$ \\
$ A_{K \Xi_\mu, (\Sigma)}^{(a_{\mu}^{\pi} \Lambda),2}=-\frac{1}{12}\, \big(-3 \, f_2^{(A)}+2 \, f_4^{(A)}\big)$ \\
$ A_{K \Xi_\mu, (\Sigma)}^{(a_{\mu}^{\pi} \Sigma),2}=\frac{1}{2\sqrt{6}} \, f_2^{(A)}$ \\
$ A_{K \Xi_\mu, (\Sigma)}^{(a_{\mu}^{\bar{K}} N),2}=\frac{1}{2\sqrt{6}} \, f_4^{(A)}$ \\
$ A_{K \Xi_\mu, (\Sigma)}^{(a_{\mu}^{\eta} \Sigma),2}=-\frac{1}{12} \, \big(3 \, f_2^{(A)}-2 \, f_4^{(A)}\big)$ \\
$ A_{K \Xi_\mu, (\Sigma)}^{(a_{\mu}^{K} \Xi),2}=\frac{1}{2\sqrt{6}}\, \big(2 \, f_2^{(A)}-f_4^{(A)}\big)$ }

&

\parbox[c]{7cm}{
$ A_{\pi \Xi_\mu, (\Xi)}^{(a_{\mu}^{\pi} \Xi),2}=-\frac{1}{\sqrt{6}}\, f_2^{(A)}$ \\
$ A_{\pi \Xi_\mu, (\Xi)}^{(a_{\mu}^{\bar{K}} \Lambda),2}=\frac{1}{4\sqrt{6}}\, \big(3 \, f_2^{(A)}-2 \, f_4^{(A)}\big)$ \\
$ A_{\pi \Xi_\mu, (\Xi)}^{(a_{\mu}^{\bar{K}} \Sigma),2}=-\frac{1}{4\sqrt{6}} \, f_2^{(A)}$ \\
$ A_{\pi \Xi_\mu, (\Xi)}^{(a_{\mu}^{\eta} \Xi),2}=\frac{1}{2\sqrt{6}} \, f_4^{(A)}$ \\
$ A_{\bar{K} \Sigma_\mu, (\Xi)}^{(a_{\mu}^{\pi} \Xi),2}=-\frac{1}{4\sqrt{6}} \, \big(f_2^{(A)}-3 \, f_4^{(A)}\big)$ \\
$ A_{\bar{K} \Sigma_\mu, (\Xi)}^{(a_{\mu}^{\bar{K}} \Lambda),2}=-\frac{1}{4\sqrt{6}}\, \big(-3 \, f_2^{(A)}+f_4^{(A)}\big)$ \\
$ A_{\bar{K} \Sigma_\mu, (\Xi)}^{(a_{\mu}^{\bar{K}} \Sigma),2}=-\frac{1}{4\sqrt{6}}\, \big(-5 \, f_2^{(A)}+3 \, f_4^{(A)}\big)$ \\
$ A_{\bar{K} \Sigma_\mu, (\Xi)}^{(a_{\mu}^{\eta} \Xi),2}=\frac{1}{4\sqrt{6}}\, \big(-3 \, f_2^{(A)}+f_4^{(A)}\big)$ \\
$ A_{\eta \Xi_\mu, (\Xi)}^{(a_{\mu}^{\pi} \Xi),2}=-\frac{1}{2\sqrt{6}}\, f_4^{(A)}$ \\
$ A_{\eta \Xi_\mu, (\Xi)}^{(a_{\mu}^{\bar{K}} \Lambda),2}=-\frac{\sqrt{3}}{4\sqrt{2}}\, f_2^{(A)}$ \\
$ A_{\eta \Xi_\mu, (\Xi)}^{(a_{\mu}^{\bar{K}} \Sigma),2}=\frac{1}{4\sqrt{6}}\, \big(-3 \, f_2^{(A)}+2 \, f_4^{(A)}\big)$ \\
$ A_{\eta \Xi_\mu, (\Xi)}^{(a_{\mu}^{\eta} \Xi),2}=0$ \\
$ A_{K \Omega_\mu, (\Xi)}^{(a_{\mu}^{\pi} \Xi),2}=-\frac{\sqrt{3}}{4} \, f_2^{(A)}$ \\
$ A_{K \Omega_\mu, (\Xi)}^{(a_{\mu}^{\bar{K}} \Lambda),2}=-\frac{1}{2\sqrt{3}} \, f_4^{(A)}$ \\
$ A_{K \Omega_\mu, (\Xi)}^{(a_{\mu}^{\bar{K}} \Sigma),2}=0$ \\
$ A_{K \Omega_\mu, (\Xi)}^{(a_{\mu}^{\eta} \Xi),2}=-\frac{1}{4\sqrt{3}} \, \big(3 \, f_2^{(A)}-2 \, f_4^{(A)}\big)$ }
\\ \hline

\end{tabular}
\caption{CGCs $A_{QL,(\bar{B})}^{(iB),2}$ with $L\in[10]$
.}
\label{Tab:Clebsches4point10n2}
\end{table}

The 4-point vector couplings $A_{QL,(\bar{B})}^{(iB),3}$ are given by 
\begin{eqnarray}
 A_{QL\in [8],(\bar{B})}^{(iB),3}=\frac{1}{4}\, A_{QL\in [8],(\bar{B})}^{(iB),1} \;\,\text{ with }\;\, g_X^{(S)} \rightarrow g_X^{(V)}\,.
 \label{Eq:Clebsch-Vector-Couplings}
\end{eqnarray}
Furthermore, the parameter $n$ of the sum in Eq. \eqref{Eq:GA-all-contributions-of-diagrams} for $L\in [10]$ only leads to non-vanishing contributions up to $n=2$, so 
\begin{eqnarray}
 A_{QL\in [10],(\bar{B})}^{(iB),3} = A_{QL\in [10],(\bar{B})}^{(iB),4} = 0 \, .
\label{Eq:Clebsch-n-running-for-transitions} \\ \nonumber
 \end{eqnarray}
 
  For the bubble diagram  with internal baryon $R$ in Eq. \eqref{Eq:GA-all-contributions-of-diagrams} the 4-point vertices $A_{QR,(B)}^{(i\bar{B}),n}$ can be extracted from   $A_{QL,(\bar{B})}^{(iB),n}$ by 
\begin{eqnarray}
 &&A_{QR\in [8],(B)}^{(i\bar{B}),1}=-A_{QL\in [8],(\bar{B})}^{(\tilde{i}B),1}\,, 
 \hspace{0.8cm}
 A_{QR\in [8],(B)}^{(i\bar{B}),2}=A_{QL\in [8],(\bar{B})}^{(\tilde{i}B),2}\,,
 \nonumber\\ 
 &&A_{QR\in [8],(B)}^{(i\bar{B}),3}=A_{QL\in [8],(\bar{B})}^{(\tilde{i}B),3}\,,
 \hspace{1.07cm}
 A_{QR\in [8],(B)}^{(i\bar{B}),4}=-A_{QL\in [8],(\bar{B})}^{(\tilde{i}B),4}\,,
 \nonumber\\ 
 &&A_{QR\in [10],(B)}^{(i\bar{B}),1}= A_{QL\in [10],(\bar{B})}^{(\tilde{i}B),1}\,,
\hspace{0.8cm}
 A_{QR\in [10],(B)}^{(i\bar{B}),2}= A_{QL\in [10],(\bar{B})}^{(\tilde{i}B),2}\,,
 \nonumber\\
 &&A_{QR\in [10],(B)}^{(i\bar{B}),3} = A_{QR\in [10],(B)}^{(i\bar{B}),4} = 0 \,.
 \label{Eq:Clebsch-4-point-vertices-RvsL}
 \end{eqnarray}
Note that the axial-vector current has to be turned around, in other words, we have to replace $a_\mu^{\bar{K}}\leftrightarrow a_\mu^K$, but the contributions of $a_\mu^\pi$ and $a_\mu^\eta$ remain unchanged:
\begin{eqnarray}
 \tilde{i}=\begin{cases}
1 \hspace{1cm} \text{for} \hspace{1cm}i=1  \,,  \\        
6 \hspace{1cm} \text{for} \hspace{1cm}i=4  \,,   \\
4 \hspace{1cm} \text{for} \hspace{1cm}i=6  \,,  \\
8 \hspace{1cm} \text{for} \hspace{1cm}i=8  \,.  \\
           \end{cases}
\label{Eq:Clebsch-R-bubble-i-tilde}
\end{eqnarray}

\clearpage

\section{Recoupling Constants}
The recoupling constants of the bubble diagrams $\tilde{C}_i^{\bar{B}B}$ are shown in the following Table: 
\begin{table}[ht]
\centering
\renewcommand{\arraystretch}{1.0}
\begin{tabular}{|l l l l|}\hline

\rule{0pt}{13ex}
\parbox[c]{2.6cm}{
 $ \tilde{C}_{a_{\mu}^{\pi}}^{NN}= -\sqrt{\frac{2}{3}}$\\
 $ \tilde{C}_{a_{\mu}^{\pi}}^{\Sigma\Sigma}= \frac{1}{\sqrt{2}}$\\
 $ \tilde{C}_{a_{\mu}^{\pi}}^{\Xi\Xi}= \sqrt{\frac{2}{3}}$\\
 $ \tilde{C}_{a_{\mu}^{\pi}}^{\Lambda\Sigma}= -\frac{1}{\sqrt{3}}$\\
 $ \tilde{C}_{a_{\mu}^{\pi}}^{\Sigma\Lambda}= -1$}
 &
 
 \parbox[c]{2.6cm}{
 $ \tilde{C}_{a_{\mu}^{K}}^{\Sigma\Xi}= -1$\\
 $ \tilde{C}_{a_{\mu}^{K}}^{N\Sigma}= -\sqrt{\frac{2}{3}}$\\
 $ \tilde{C}_{a_{\mu}^{K}}^{N\Lambda}= -1$\\
 $ \tilde{C}_{a_{\mu}^{K}}^{\Lambda\Xi}= \frac{1}{\sqrt{2}}$}
 &
 
 \parbox[c]{2.6cm}{
 $ \tilde{C}_{a_{\mu}^{\bar{K}}}^{\Xi\Sigma}= \sqrt{\frac{2}{3}}$\\
 $ \tilde{C}_{a_{\mu}^{\bar{K}}}^{\Sigma N}= -1$\\
 $ \tilde{C}_{a_{\mu}^{\bar{K}}}^{\Lambda N}= -\frac{1}{\sqrt{2}}$\\
 $ \tilde{C}_{a_{\mu}^{\bar{K}}}^{\Xi \Lambda}= -1$}
 &

 \parbox[c]{2.1cm}{
 $ \tilde{C}_{a_{\mu}^{\eta}}^{NN}= -1$\\
 $ \tilde{C}_{a_{\mu}^{\eta}}^{\Sigma\Sigma}= -1$\\
 $ \tilde{C}_{a_{\mu}^{\eta}}^{\Xi\Xi}= -1$\\
 $ \tilde{C}_{a_{\mu}^{\eta}}^{\Lambda\Lambda}= -1$}
\\ \hline
\end{tabular}
\caption{Recoupling constants $\tilde{C}_i^{\bar{B}B}$ of the bubble diagram with internal baryon $L$.}
\label{Tab:Recouplingconstants-Bubble}
\end{table}
 
 The coefficients $\tilde{C}_{\bar{i}}^{B\bar{B}}$ in Eq. \eqref{Eq:Def-Recoupling-Constants}, connected to the bubble diagram with internal baryon $R$, are derived from $\tilde{C}_i^{\bar{B}B}$ by switching the incoming and outgoing external baryon. Note that the axial-vector current also needs to be turned around ($i\leftrightarrow\bar{i}$, see Eq. \eqref{Eq:Clebsch-R-bubble-i-tilde}). In the isospin basis this only affects the strangeness-changing current, implying $a_\mu^{K}\leftrightarrow a_\mu^{\bar{K}}$. 
 
 For the recoupling constants of the triangle diagrams we introduce a new notation for internal particles with the same isospin and strangeness: $\Xi_*$ includes the octet $\Xi$ and the decuplet $\Xi_{\mu}$, 
$\Sigma_*$ includes the octet $\Sigma$ and the decuplet $\Sigma_{\mu}$.
We find a simple structure for the recoupling constants with axial-vector current $a_\mu^\eta$:
\begin{eqnarray}
\tilde{C}_{\bar{B}B,LR}^{a_{\mu}^{\eta} Q}=1\,,
 \label{Eq:Recouplingsconstants-Triangle-Eta}
\end{eqnarray}
for all $Q$, $\bar{B}$, $B$, $L$, and $R$. 

The isospin structure of all other axial-vector currents complicates the treatment. We give all recoupling constants explicitly in the following Tables \ref{Tab:Recouplingconstants-Triangle-Pi} and \ref{Tab:Recouplingconstants-Triangle-K}. 
\begin{table}[ht]
\centering
\renewcommand{\arraystretch}{1.0}
\begin{tabular}{|l l l|}\hline

\rule{0pt}{22ex}
\parbox[c]{3cm}{
 $ \tilde{C}_{\Sigma\Sigma,\Sigma_*\Sigma_*}^{a_{\mu}^{\pi}\pi}=\frac{1}{2}$ \\
$ \tilde{C}_{\Sigma\Sigma,NN}^{a_{\mu}^{\pi}\bar{K}}=\frac{1}{\sqrt{2}}$ \\
$ \tilde{C}_{\Sigma\Sigma,N\Delta}^{a_{\mu}^{\pi}\bar{K}}=-\frac{1}{\sqrt{6}}$ \\
$ \tilde{C}_{\Sigma\Sigma,\Delta N}^{a_{\mu}^{\pi}\bar{K}}=-\frac{1}{\sqrt{6}}$ \\
$ \tilde{C}_{\Sigma\Sigma,\Xi_*\Xi_*}^{a_{\mu}^{\pi}K}=-\frac{1}{\sqrt{2}}$ \\
$ \tilde{C}_{\Sigma\Sigma,\Delta\Delta}^{a_{\mu}^{\pi}\bar{K}}=\frac{5}{2\sqrt{6}}$ \\
$ \tilde{C}_{\Sigma\Sigma,\Sigma_*\Sigma_*}^{a_{\mu}^{\pi}\eta}=1$ \\
$ \tilde{C}_{\Sigma\Sigma,\Lambda\Sigma_*}^{a_{\mu}^{\pi}\pi}=\frac{1}{\sqrt{2}}$ \\
$ \tilde{C}_{\Sigma\Sigma,\Sigma_*\Lambda}^{a_{\mu}^{\pi}\pi}=\frac{1}{\sqrt{2}}$ }

&

\parbox[c]{3cm}{
$ \tilde{C}_{NN,\Sigma_*\Sigma_*}^{a_{\mu}^{\pi}K}=\frac{2\sqrt{2}}{3}$ \\
$ \tilde{C}_{NN,NN}^{a_{\mu}^{\pi}\pi}=-\frac{1}{3}$ \\
$ \tilde{C}_{NN,NN}^{a_{\mu}^{\pi}\eta}=1$ \\
$ \tilde{C}_{NN,N\Delta}^{a_{\mu}^{\pi}\pi}=\frac{4}{3\sqrt{3}}$ \\
$ \tilde{C}_{NN,\Delta N}^{a_{\mu}^{\pi}\pi}=\frac{4}{3\sqrt{3}}$ \\
$ \tilde{C}_{NN,\Delta \Delta}^{a_{\mu}^{\pi}\pi}=\frac{5}{3\sqrt{3}}$ \\
$ \tilde{C}_{NN,\Sigma_*\Lambda}^{a_{\mu}^{\pi}K}=\frac{\sqrt{2}}{\sqrt{3}}$ \\
$ \tilde{C}_{NN,\Lambda\Sigma_*}^{a_{\mu}^{\pi}K}=\frac{\sqrt{2}}{\sqrt{3}} $}

&

\parbox[c]{3cm}{
$ \tilde{C}_{\Xi\Xi,\Sigma_*\Sigma_*}^{a_{\mu}^{\pi}\bar{K}}=-\frac{2\sqrt{2}}{3}$ \\
$ \tilde{C}_{\Xi\Xi,\Xi_*\Xi_*}^{a_{\mu}^{\pi}\pi}=-\frac{1}{3}$ \\
$ \tilde{C}_{\Xi\Xi,\Xi_*\Xi_*}^{a_{\mu}^{\pi}\eta}=1$ \\
$ \tilde{C}_{\Xi\Xi,\Sigma_*\Lambda}^{a_{\mu}^{\pi}\bar{K}}=-\frac{\sqrt{2}}{\sqrt{3}}$ \\
$ \tilde{C}_{\Xi\Xi,\Lambda\Sigma_*}^{a_{\mu}^{\pi}\bar{K}}=-\frac{\sqrt{2}}{\sqrt{3}} $}

\\ \hline

\rule{0pt}{15ex}
\parbox[c]{3cm}{
 $ \tilde{C}_{\Sigma\Lambda,\Sigma_*\Sigma_*}^{a_{\mu}^{\pi}\pi}=\sqrt{\frac{2}{3}}$ \\
  $ \tilde{C}_{\Sigma\Lambda,NN}^{a_{\mu}^{\pi}\bar{K}}=\frac{1}{\sqrt{2}}$ \\
   $ \tilde{C}_{\Sigma\Lambda,\Xi_*\Xi_*}^{a_{\mu}^{\pi}K}=-\frac{1}{\sqrt{2}}$ \\
    $ \tilde{C}_{\Sigma\Lambda,\Delta N}^{a_{\mu}^{\pi}\bar{K}}=\sqrt{\frac{2}{3}}$ \\
    $ \tilde{C}_{\Sigma\Lambda,\Lambda\Sigma_*}^{a_{\mu}^{\pi}\pi}=\frac{1}{\sqrt{3}}$ \\
      $ \tilde{C}_{\Sigma\Lambda,\Sigma_*\Lambda}^{a_{\mu}^{\pi}\eta}=1 $ }
      
&

\parbox[c]{3cm}{
 $ \tilde{C}_{\Lambda\Sigma,\Sigma_*\Sigma_*}^{a_{\mu}^{\pi}\pi}=\sqrt{\frac{2}{3}}$ \\
 $ \tilde{C}_{\Lambda\Sigma,NN}^{a_{\mu}^{\pi}\bar{K}}=\frac{1}{\sqrt{2}}$ \\
 $ \tilde{C}_{\Lambda\Sigma,\Xi_*\Xi_*}^{a_{\mu}^{\pi}K}=-\frac{1}{\sqrt{2}}$ \\
 $ \tilde{C}_{\Lambda\Sigma,N\Delta}^{a_{\mu}^{\pi}\bar{K}}=\sqrt{\frac{2}{3}}$ \\
     $ \tilde{C}_{\Lambda\Sigma,\Sigma_*\Lambda}^{a_{\mu}^{\pi}\pi}=\frac{1}{\sqrt{3}}$ \\
 $ \tilde{C}_{\Lambda\Sigma,\Lambda\Sigma_*}^{a_{\mu}^{\pi}\eta}=1$ }
 
&
\\ \hline
\end{tabular}
\caption{Recoupling Constants with axial-vector current $a_{\mu}^{\pi}$:
$\tilde{C}_{\bar{B}B,LR}^{a_{\mu}^{\pi} Q}$.}
\label{Tab:Recouplingconstants-Triangle-Pi}
\end{table}

\begin{table}[ht]
\centering
\renewcommand{\arraystretch}{1.0}
\begin{tabular}{|l l l l|}\hline
\rule{0pt}{18ex}

\parbox[c]{3.2cm}{
 $ \tilde{C}_{\Sigma\Xi,\Sigma_*\Xi_*}^{a_{\mu}^{K}\pi}=-\frac{\sqrt{2}}{\sqrt{3}}$ \\
  $ \tilde{C}_{\Sigma\Xi,\Sigma_*\Xi_*}^{a_{\mu}^{K}\eta}=1$ \\
    $ \tilde{C}_{\Sigma\Xi,N\Sigma_*}^{a_{\mu}^{K}\bar{K}}=\frac{1}{\sqrt{6}}$ \\
 $ \tilde{C}_{\Sigma\Xi,N\Lambda}^{a_{\mu}^{K}\bar{K}}=1$ \\
  $ \tilde{C}_{\Sigma\Xi,\Xi_*\Omega}^{a_{\mu}^{K}K}=1$ \\ 
  $ \tilde{C}_{\Sigma\Xi,\Lambda\Xi_*}^{a_{\mu}^{K}\pi}=-\sqrt{\frac{2}{3}}$ \\
   $ \tilde{C}_{\Sigma\Xi,\Delta\Sigma_*}^{a_{\mu}^{K}\bar{K}}=-2\sqrt{\frac{2}{3}}$ }
   
&
\parbox[c]{3cm}{
 $ \tilde{C}_{N\Sigma,\Sigma_*\Xi_*}^{a_{\mu}^{K}K}=-\frac{1}{\sqrt{6}}$ \\
  $ \tilde{C}_{N\Sigma,N\Sigma_*}^{a_{\mu}^{K}\pi}=-\frac{\sqrt{2}}{\sqrt{3}}$ \\
    $ \tilde{C}_{N\Sigma,N\Sigma_*}^{a_{\mu}^{K}\eta}=1$ \\
        $ \tilde{C}_{N\Sigma,N\Lambda}^{a_{\mu}^{K}\pi}=\sqrt{\frac{2}{3}}$ \\
   $ \tilde{C}_{N\Sigma,\Lambda\Xi_*}^{a_{\mu}^{K}K}=1$ \\
       $ \tilde{C}_{N\Sigma,\Delta\Sigma_*}^{a_{\mu}^{K}\pi}=\sqrt{\frac{2}{3}}$ }
       
&
\parbox[c]{3cm}{ 
$ \tilde{C}_{N\Lambda,\Sigma_*\Xi_*}^{a_{\mu}^{K}K}=-\frac{\sqrt{3}}{2}$ \\
 $ \tilde{C}_{N\Lambda,N\Sigma_*}^{a_{\mu}^{K}\pi}=\frac{1}{\sqrt{2}}$ \\
  $ \tilde{C}_{N\Lambda,N\Lambda}^{a_{\mu}^{K}\eta}=1$ \\
   $ \tilde{C}_{N\Lambda,\Lambda\Xi_*}^{a_{\mu}^{K}K}=-\frac{1}{\sqrt{2}}$ \\
    $ \tilde{C}_{N\Lambda,\Delta\Sigma_*}^{a_{\mu}^{K}\pi}=\sqrt{2}$ }
    
&
\parbox[c]{3cm}{
 $ \tilde{C}_{\Lambda\Xi,\Sigma_*\Xi_*}^{a_{\mu}^{K}\pi}=-\frac{1}{\sqrt{2}}$ \\
  $ \tilde{C}_{\Lambda\Xi,N\Sigma_*}^{a_{\mu}^{K}\bar{K}}=-\frac{\sqrt{3}}{2}$ \\
   $ \tilde{C}_{\Lambda\Xi,N\Lambda}^{a_{\mu}^{K}\bar{K}}=\frac{1}{\sqrt{2}}$ \\
    $ \tilde{C}_{\Lambda\Xi,\Xi_*\Omega}^{a_{\mu}^{K}K}=\frac{1}{\sqrt{2}}$ \\
     $ \tilde{C}_{\Lambda\Xi,\Lambda\Xi_*}^{a_{\mu}^{K}\eta}=1$ }
     
\\ \hline
\end{tabular}
\caption{Recoupling Constants with  axial-vector current $a_{\mu}^{K}$:
$\tilde{C}_{\bar{B}B,LR}^{a_{\mu}^{K} Q}$.}
\label{Tab:Recouplingconstants-Triangle-K}
\end{table}
The recoupling constants with axial-vector current $a_{\mu}^{\bar{K}}$ are derived from the recoupling constant $\tilde{C}_{\bar{B}B,LR}^{a_{\mu}^{K} Q}$ by
\begin{eqnarray}
 \tilde{C}_{\bar{B}B,LR}^{a_{\mu}^{\bar{K}} Q}=\tilde{C}_{B\bar{B},RL}^{a_{\mu}^{K} Q}\,.
  \label{Eq:Recouplingsconstants-Triangle-Kbar}
\end{eqnarray}

\newpage

\section{Kinematic Factors} 
\label{Chapter:Appendix-K-factors}

Pre-factors $K_{abc}^{(ABC)}$  in Eq. \eqref{Eq:PVR-unrenormalized-results-after-PVR} are connected to the integral involving the internal particles $A$, $B$, $C$ and originating from a diagram with internal particles $a$, $b$, $c$. Power-counting violating terms are subtracted: $K_{abc}^{(ABC)} \rightarrow \bar{K}_{abc}^{(ABC)}$. We expand $\bar{K}_{abc}^{(ABC)}$ in terms of the parameters $m_Q^2$, $t$, $\delta_L$, $\delta_R$, and $\delta_B$ \eqref{Eq:Power-Counting-Rules} in this Appendix. 
\begin{eqnarray}
&& \bar{K}_{Q}^{(Q)}=1\,,
 \nonumber\\
&& \bar{K}_{L\in[8] Q }^{(Q),n\in\{1,2,3,4,5\}}=\bar{K}_{QR\in[8] }^{(Q),n\in\{1,2,3,4,5\}}=O(Q^2)\,,
 \nonumber\\
&& \bar{K}_{L\in[8] Q R\in[8]}^{(Q),n=1}=-1+O(Q^2)\,,
 \nonumber\\
&& \bar{K}_{L\in[10] Q R\in[8]}^{(Q),n\in\{1,2\}}=
 \bar{K}_{L\in[8] Q R\in[10]}^{(Q),n\in\{1,2\}}=
 O(Q^2)\,,
 \nonumber\\
&& \bar{K}_{L\in[10] Q R\in[10]}^{(Q),1}=
O(Q^2)\,,
 \nonumber\\
&& \bar{K}_{L\in[10] Q }^{(Q),n\in\{1,2,3\}}=
\bar{K}_{Q R\in[10]}^{(Q),n\in\{1,2,3\}}=
  O(Q^2)\,,
\label{Eq:K-factors-Q}
 \\ \nonumber\\ \nonumber\\ \nonumber
&&\bar{K}_{L\in[8] Q }^{(LQ),n=1}=-m_Q^2 +O(Q^4)\,,
\\ \nonumber
&& \frac{1}{M_B}\bar{K}_{L\in[8] Q }^{(LQ),n=2}=\frac{2}{3}m_Q^2 +O(Q^4)\,,
\nonumber\\
&&\bar{K}_{L\in[8] Q }^{(LQ),n=3,5}=O(Q^4)\,,
\nonumber\\
&&\frac{1}{M_B}\bar{K}_{L\in[8] Q }^{(LQ),n=4}=\frac{4}{3}m_Q^2 +O(Q^4)\,,
\nonumber\\
&&\bar{K}_{L\in[8] Q R\in[8]}^{(LQ),1}=\frac{1}{2}((\delta_L+\delta_R)M_B-m_Q^2)+O(Q^4)\,,
\nonumber\\
&&\bar{K}_{L\in[8] Q R\in[10]}^{(LQ),1}=
-\frac{5}{48}\frac{\Delta}{M}\alpha_{11}t 
-\frac{5}{36}\alpha_{12}m_Q^2 
\nonumber\\
&&\hspace{1.55cm}
-\frac{5}{12}\alpha_{14}\delta_L M_B
-\frac{5}{12}\alpha_{13}\delta_R M_B 
+0 \,\alpha_{15}\delta_B M_B +O(Q^4)\,,
\nonumber\\
&&\frac{1}{M_B}\bar{K}_{L\in[8] Q R\in[10]}^{(LQ),2}=
\frac{1}{6}\frac{\Delta}{M}\alpha_{91}t 
-\frac{2}{3}\alpha_{92}m_Q^2 
\nonumber\\
&&\hspace{1.55cm}
+\frac{2}{3}\alpha_{94}\delta_L M_B
-\frac{2}{3}\alpha_{93}\delta_R M_B 
- \frac{3}{4}\,\frac{\Delta}{M}\alpha_{95}\delta_B M_B +O(Q^4)\,,
\nonumber\\
&&\bar{K}_{L\in[10] Q R\in[8]}^{(LQ),1}=
-\frac{5}{48}\frac{\Delta}{M}\alpha_{21}t 
+\frac{19}{36}\alpha_{22}m_Q^2 
\nonumber\\
&&\hspace{1.55cm}
-\frac{5}{12}\alpha_{23}\delta_L M_B
-\frac{5}{12}\alpha_{24}\delta_R M_B 
-\frac{5}{12}\frac{\Delta}{M}\alpha_{25}\delta_B M_B +O(Q^4)\,,
\nonumber\\
&&\frac{1}{M_B}\bar{K}_{L\in[10] Q R\in[8]}^{(LQ),2}=
\frac{1}{6}\frac{\Delta}{M}\alpha_{101}t 
+\frac{2}{3}\alpha_{102}m_Q^2 
\nonumber\\
&&\hspace{1.55cm}
+\frac{2}{3}\alpha_{103}\delta_L M_B
-\frac{2}{3}\alpha_{104}\delta_R M_B 
+ \frac{3}{4}\,\frac{\Delta}{M}\alpha_{105}\delta_B M_B +O(Q^4)\,,
\nonumber\\
&&\bar{K}_{L\in[10] Q R\in[10]}^{(LQ),1}=
-\frac{1}{36}\frac{\Delta}{M}\alpha_{31}t 
-\frac{1}{54}\alpha_{32}m_Q^2 
\nonumber\\
&&\hspace{1.55cm}
-\frac{1}{18}\alpha_{33}\delta_L M_B
+\frac{5}{6}\alpha_{34}\delta_R M_B 
-\frac{1}{18}\frac{\Delta}{M} \,\alpha_{35}\delta_B M_B +O(Q^4)\,,
\nonumber\\
&&\frac{1}{M_B}\bar{K}_{L\in[10] Q }^{(LQ),n=1}=
0\,\alpha_{41}t 
+\frac{20}{9}\alpha_{42}m_Q^2 
\nonumber\\
&&\hspace{1.55cm}
-\frac{40}{9}\frac{\Delta}{M}\alpha_{43}\delta_L M_B
-\frac{10}{3}\frac{\Delta^2}{M^2} \,\alpha_{45}\delta_B M_B +O(Q^4)\,,
\nonumber\\
&&\frac{1}{M_B}\bar{K}_{L\in[10] Q }^{(LQ),n=2}=
0\,\alpha_{51}t 
-\frac{4}{9}\alpha_{52}m_Q^2 
\nonumber\\
&&\hspace{1.55cm}
+\frac{8}{9}\frac{\Delta}{M}\alpha_{53}\delta_L M_B
+\frac{2}{3}\frac{\Delta^2}{M^2} \,\alpha_{55}\delta_B M_B +O(Q^4)\,,
\nonumber\\
&&\frac{1}{M_B}\bar{K}_{L\in[10] Q }^{(LQ),n=3}= - \frac{8}{9}\,\frac{\Delta^2}{M^2}\,\,\alpha_{111}\,t 
-\frac{8}{9}\,\frac{\Delta^2}{M^2} \,\alpha_{115}\,\delta_B M_B +O(Q^4)\,,
\label{Eq:K-factors-LQ}
\\ \nonumber\\ \nonumber\\ \nonumber
&& \bar{K}_{Q R\in[8]}^{(QR),n=1}=m_Q^2 +O(Q^4)\,, 
\nonumber\\
&& \frac{1}{M_B}\bar{K}_{Q R\in[8]}^{(QR),n=2}=\frac{2}{3}m_Q^2 +O(Q^4)\,,
\nonumber\\
&&\bar{K}_{Q R\in[8]}^{(QR),n=3,5}=O(Q^4)\,,
\nonumber\\
&&\frac{1}{M_B}\bar{K}_{Q R\in[8]}^{(QR),n=4}=-\frac{4}{3}m_Q^2 +O(Q^4)\,, 
\nonumber\\
&&\bar{K}_{L\in[8] Q R\in[8]}^{(QR),1}=\frac{1}{2}((\delta_L+\delta_R)M_B-m_Q^2)+O(Q^4)\,,
\nonumber\\
&&\bar{K}_{L\in[10] Q R\in[8]}^{(QR),1}=
-\frac{5}{48}\frac{\Delta}{M}\alpha_{11}t 
-\frac{5}{36}\alpha_{12}m_Q^2 
\nonumber\\
&&\hspace{1.55cm}
-\frac{5}{12}\alpha_{13}\delta_L M_B
-\frac{5}{12}\alpha_{14}\delta_R M_B 
+0 \,\alpha_{15}\delta_B M_B +O(Q^4)\,,
\nonumber\\
&&\frac{1}{M_B}\bar{K}_{L\in[10] Q R\in[8]}^{(QR),2}=
\frac{1}{6}\frac{\Delta}{M}\alpha_{91}t 
-\frac{2}{3}\alpha_{92}m_Q^2 
\nonumber\\
&&\hspace{1.55cm}
-\frac{2}{3}\alpha_{93}\delta_L M_B
+\frac{2}{3}\alpha_{94}\delta_R M_B 
+ \frac{3}{4}\,\frac{\Delta}{M}\alpha_{95}\delta_B M_B +O(Q^4)\,,
\nonumber\\
&&\bar{K}_{L\in[8] Q R\in[10]}^{(QR),1}=
-\frac{5}{48}\frac{\Delta}{M}\alpha_{21}t 
+\frac{19}{36}\alpha_{22}m_Q^2 
\nonumber\\
&&\hspace{1.55cm}
-\frac{5}{12}\alpha_{24}\delta_L M_B
-\frac{5}{12}\alpha_{23}\delta_R M_B
+\frac{5}{12}\frac{\Delta}{M}\alpha_{25}\delta_B M_B +O(Q^4)\,,
\nonumber\\
&&\frac{1}{M_B}\bar{K}_{L\in[8] Q R\in[10]}^{(QR),2}=
\frac{1}{6}\frac{\Delta}{M}\alpha_{101}t 
+\frac{2}{3}\alpha_{102}m_Q^2 
\nonumber\\
&&\hspace{1.55cm}
-\frac{2}{3}\alpha_{104}\delta_L M_B
+\frac{2}{3}\alpha_{103}\delta_R M_B 
- \frac{3}{4}\,\frac{\Delta}{M}\alpha_{105}\delta_B M_B +O(Q^4)\,,
\nonumber\\
&&\bar{K}_{L\in[10] Q R\in[10]}^{(QR),1}=
-\frac{1}{36}\frac{\Delta}{M}\alpha_{31}t 
-\frac{1}{54}\alpha_{32}m_Q^2 
\nonumber\\
&&\hspace{1.55cm}
+\frac{5}{6}\alpha_{34}\delta_L M_B 
-\frac{1}{18}\alpha_{33}\delta_R M_B
+\frac{1}{18}\frac{\Delta}{M} \,\alpha_{35}\delta_B M_B +O(Q^4)\,,
\nonumber\\
&&\frac{1}{M_B}\bar{K}_{Q R\in[10]}^{(QR),n=1}=
0\,\alpha_{41}t 
+\frac{20}{9}\alpha_{42}m_Q^2 
\nonumber\\
&&\hspace{1.55cm}
-\frac{40}{9}\frac{\Delta}{M}\alpha_{43}\delta_R M_B
+\frac{10}{3}\frac{\Delta^2}{M^2} \,\alpha_{45}\delta_B M_B +O(Q^4)\,,
\nonumber\\
&&\frac{1}{M_B}\bar{K}_{ Q R\in[10]}^{(QR),n=2}=
0\,\alpha_{51}t 
-\frac{4}{9}\alpha_{52}m_Q^2 
\nonumber\\
&&\hspace{1.55cm}
+\frac{8}{9}\frac{\Delta}{M}\alpha_{53}\delta_R M_B
-\frac{2}{3}\frac{\Delta^2}{M^2} \,\alpha_{55}\delta_B M_B +O(Q^4)\,,
\nonumber\\
&&\frac{1}{M_B}\bar{K}_{QR\in[10]  }^{(QR),n=3}=  \frac{8}{9}\,\frac{\Delta^2}{M^2}\,\,\alpha_{111}\,t 
-\frac{8}{9}\,\frac{\Delta^2}{M^2} \,\alpha_{115}\,\delta_B M_B +O(Q^4)\,,
\label{Eq:K-factors-QR}
\\ \nonumber\\ \nonumber\\ \nonumber
&&\bar{K}_{L\in[8] Q R\in[8]}^{(LQR),1}=\bar{K}_{L\in[8] Q R\in[8]}^{'(LQR),1}=O(Q^4)\,,
\nonumber\\
&&\frac{1}{M_B^2}\bar{K}_{L\in[10] Q R\in[8]}^{(LQR),1}=
\frac{5}{24}\frac{\Delta^2}{M^2}\alpha_{61}t 
-2\alpha_{62}m_Q^2
\nonumber\\
&&\hspace{1.55cm}
+\frac{5}{3}\frac{\Delta}{M}\alpha_{63}\delta_L M_B
+\frac{1}{3}\frac{\Delta}{M}\alpha_{64}\delta_R M_B 
+\frac{5}{6}\frac{\Delta^2}{M^2}\alpha_{65}\delta_B M_B +O(Q^4)\,,
\nonumber\\
&&\frac{1}{M_B^3}\bar{K}_{L\in[10] Q R\in[8]}^{(LQR),2}=
-\frac{1}{3}\frac{\Delta^2}{M^2}\alpha_{121}t 
+\frac{4}{3}\frac{\Delta}{M}\alpha_{122}m_Q^2
\nonumber\\
&&\hspace{1.55cm}
+\frac{8}{3}\alpha_{123}\delta_L M_B
-\frac{8}{3}\alpha_{124}\delta_R M_B 
-\frac{1}{6}\alpha_{125}\delta_B M_B +O(Q^4)\,,
\nonumber\\
&&\frac{1}{M_B^4}\bar{K}_{L\in[10] Q R\in[8]}^{'(LQR),1}=
0 \, \alpha_{71}t 
-\frac{2}{3}\frac{\Delta^2}{M^2}\alpha_{72}m_Q^2
\nonumber\\
&&\hspace{1.55cm}
+\frac{4}{3}\frac{\Delta}{M}\alpha_{73}\delta_L M_B
-\frac{4}{3}\frac{\Delta}{M}\alpha_{74}\delta_R M_B 
+\frac{2}{3}\frac{\Delta^2}{M^2}\alpha_{75}\delta_B M_B +O(Q^4)\,,
\nonumber\\
&&\frac{1}{M_B^2}\bar{K}_{L\in[8] Q R\in[10]}^{(LQR),1}=
\frac{5}{24}\frac{\Delta^2}{M^2}\alpha_{61}t 
-2\alpha_{62}m_Q^2
\nonumber\\
&&\hspace{1.55cm}
+\frac{1}{3}\frac{\Delta}{M}\alpha_{64}\delta_L M_B
+\frac{5}{3}\frac{\Delta}{M}\alpha_{63}\delta_R M_B 
-\frac{5}{6}\frac{\Delta^2}{M^2}\alpha_{65}\delta_B M_B +O(Q^4)\,,
\nonumber\\
&&\frac{1}{M_B^3}\bar{K}_{L\in[8] Q R\in[10]}^{(LQR),2}=
-\frac{1}{3}\frac{\Delta^2}{M^2}\alpha_{121}t 
+\frac{4}{3}\frac{\Delta}{M}\alpha_{122}m_Q^2
\nonumber\\
&&\hspace{1.55cm}
-\frac{8}{3}\alpha_{124}\delta_L M_B
+\frac{8}{3}\alpha_{123}\delta_R M_B 
+\frac{1}{6}\alpha_{125}\delta_B M_B +O(Q^4)\,,
\nonumber\\
&&\frac{1}{M_B^4}\bar{K}_{L\in[8] Q R\in[10]}^{'(LQR),1}=
0 \, \alpha_{71}t 
-\frac{2}{3}\frac{\Delta^2}{M^2}\alpha_{72}m_Q^2
\nonumber\\
&&\hspace{1.55cm}
-\frac{4}{3}\frac{\Delta}{M}\alpha_{74}\delta_L M_B
+\frac{4}{3}\frac{\Delta}{M}\alpha_{73}\delta_R M_B 
-\frac{2}{3}\frac{\Delta^2}{M^2}\alpha_{75}\delta_B M_B +O(Q^4)\,,
\nonumber\\
&&\frac{1}{M_B^5}\bar{K}_{L\in[10] Q R\in[8]}^{'(LQR),2}=
-\frac{2}{3}\frac{\Delta^2}{M^2}\alpha_{135}\delta_B M_B +O(Q^4)\,,
\nonumber\\
&&\frac{1}{M_B^5}\bar{K}_{L\in[8] Q R\in[10]}^{'(LQR),2}=
\frac{2}{3}\frac{\Delta^2}{M^2}\alpha_{135}\delta_B M_B +O(Q^4)\,,
\nonumber\\
&&\frac{1}{M_B^2}\bar{K}_{L\in[10] Q R\in[10]}^{(LQR),1}=
-\frac{7}{9}\frac{\Delta^2}{M^2}\alpha_{81}t 
+\frac{4}{3}\alpha_{82}m_Q^2
\nonumber\\
&&\hspace{1.55cm}
-\frac{4}{3}\frac{\Delta}{M}\alpha_{83}\delta_L M_B
-\frac{4}{3}\frac{\Delta}{M}\alpha_{84}\delta_R M_B 
+0\alpha_{85}\delta_B M_B +O(Q^4)\,,
\nonumber\\
&&\bar{K}_{L\in[10] Q R\in[10]}^{'(LQR),1}=O(Q^4)\,.
\label{Eq:K-factors-LQR}
\end{eqnarray}

The coefficients $\alpha_{ij}$, all functions of $\Delta/M$ only, are given in Eq. \eqref{Eq:alphas-explicit}. They are normalized to 1 in the limit $\Delta\rightarrow 0$. Power-counting violating contributions are subtracted, denoted by $\alpha_{0j}\rightarrow 0$ and $\alpha_{j0}\rightarrow 0$. All other factors $\alpha_{ij}$ with $i,j\neq 0$ fully enter our calculations, no expansion in $\Delta/M$ is applied.  
\begin{eqnarray}
 &&         r  = \Delta/M\,, \nonumber\\
 &&         \alpha_{01 } = (12 + 26\,r + 18\,r^2 + 6\,r^3 + r^4)/(12\,(1 + r)^2)\,,\nonumber\\  
 &&         \alpha_{02 } = (2 + r)^2\,(6 + 13\,r + 18\,r^2 + 12\,r^3 + 2\,r^4)/(24\,(1 + r)^4)\,,\nonumber\\
 &&         \alpha_{03 } = (2 + r)^3\,(5 + r)/(40\,(1 + r)^2)\,,\nonumber\\
 &&         \alpha_{04 } = -(-1 + r)\,(2 + r)^3/(8\,(1 + r)^2)\,, \nonumber\\
 &&         \alpha_{05 } = 0\,,\nonumber\\
 &&         \alpha_{10 } = (2 + r)^2\,(5 + 5\,r + r^2)/(20\,(1 + r)^2)\,,\nonumber\\
 &&         \alpha_{11 } = (2 + r)^2\,(5 + 5\,r + r^2)/(20\,(1 + r)^2)\,,\nonumber\\
 &&         \alpha_{12 } = -(-20 + 66\,r + 66\,r^2 + 9\,r^3)/(20\,(1 + r)^2)\,,\nonumber\\
 &&         \alpha_{13 } = (20 + 60\,r + 87\,r^2 + 65\,r^3 + 23\,r^4 + 3\,r^5)/(20\,(1 + r)^3)\,,\nonumber\\
 &&         \alpha_{14 } = (2 + r)^3\,(5 + 5\,r + r^2)/(40\,(1 + r)^2)\,,\nonumber\\
 &&         \alpha_{15 } = 0\,,\nonumber\\
 &&         \alpha_{20 } = (2 + r)^2\,(15 + 31\,r + 19\,r^2 + 4\,r^3)/(60\,(1 + r)^2)\,,\nonumber\\
 &&         \alpha_{21 } = (2 + r)^2\,(5 + 5\,r + r^2)/(20\,(1 + r)^2)\,,\nonumber\\
 &&         \alpha_{22 } = (76 + 170\,r + 170\,r^2 + 73\,r^3 + 12\,r^4)/(76\,(1 + r)^2)\,,\nonumber\\
 &&         \alpha_{23 } = (60 + 308\,r + 645\,r^2 + 707\,r^3 + 421\,r^4 + 129\,r^5 + 16\,r^6)/(60\,(1 + r)^3)\,,\nonumber\\
 &&         \alpha_{24 } = (2 + r)^3\,(5 + 5\,r + r^2)/(40\,(1 + r)^2)\,,\nonumber\\
 &&         \alpha_{25 } = (2 + r)^2\,(15 + 31\,r + 19\,r^2 + 4\,r^3)/(60\,(1 + r)^2)\,,\nonumber\\
 &&         \alpha_{30 } = (2 + r)^3\,(42 + 106\,r + 129\,r^2 + 72\,r^3 + 11\,r^4)/(336\,(1 + r)^4)\,,\nonumber\\
 &&         \alpha_{31 } = (2 + r)^3\,(6 + 114\,r + 163\,r^2 + 48\,r^3 + 5\,r^4)/(48\,(1 + r)^4)\,,\nonumber\\
 &&         \alpha_{32 } = (2 + r)\,(4 + 268\,r + 538\,r^2 + 449\,r^3 + 188\,r^4 + 29\,r^5)/(8\,(1 + r)^4)\,,\nonumber\\
 &&         \alpha_{33 } = -(2 + r)^2\,(-6 + 106\,r + 574\,r^2 + 949\,r^3 + 744\,r^4 + 291\,r^5 + 42\,r^6)/(24\,(1 + r)^5)\,, \nonumber\\
 &&         \alpha_{34 } = (2 + r)^2\,(90 + 234\,r + 306\,r^2 + 249\,r^3 + 128\,r^4 + 35\,r^5 + 2\,r^6)/(360\,(1 + r)^5)\,, \nonumber\\
 &&         \alpha_{35 } = -(2 + r)^3\,(-6 - 2\,r + 37\,r^2 + 72\,r^3 + 19\,r^4)/(48\,(1 + r)^4)\,,\nonumber\\
 &&         \alpha_{40 } = (2 + r)^4\,(5 + r)/(80\,(1 + r)^2)\,,\nonumber\\
 &&         \alpha_{41 } = 0\,,\nonumber\\
 &&         \alpha_{42 } = (2 + r)^2\,(20 + 24\,r + 19\,r^2 + 3\,r^3)/(80\,(1 + r)^2)\,,\nonumber\\
 &&         \alpha_{43 } = (2 + r)^3\,(20 + 36\,r + 29\,r^2 + 5\,r^3)/(160\,(1 + r)^3)\,,\nonumber\\
 &&         \alpha_{44 } = 0\,,\nonumber\\
 &&         \alpha_{45 } = (2 + r)^4\,(5 + r)/(80\,(1 + r)^2)\,,\nonumber\\
 &&         \alpha_{50 } = -(-1 + r)\,(2 + r)^4/(16\,(1 + r)^2)\,,\nonumber\\
 &&         \alpha_{51 } = 0\,,\nonumber\\
 &&         \alpha_{52 } = -(2 + r)^2\,(-4 + r^2 + 3\,r^3)/(16\,(1 + r)^2)\,,\nonumber\\
 &&         \alpha_{53 } = -(2 + r)^3\,(-4 + 5\,r^2 + 5\,r^3)/(32\,(1 + r)^3)\,,\nonumber\\
 &&         \alpha_{54 } = 0\,,\nonumber\\
 &&         \alpha_{55 } = -(-1 + r)\,(2 + r)^4/(16\,(1 + r)^2)\,,\nonumber\\
 &&         \alpha_{60 } = (2 + r)^3\,(5 + 5\,r + r^2)/(40\,(1 + r)^2)\,,\nonumber\\
 &&         \alpha_{61 } = (2 + r)^3\,(5 + 5\,r + r^2)/(40\,(1 + r)^2)\,,\nonumber\\
 &&         \alpha_{62 } = (2 + r)\,(48 + 88\,r + 76\,r^2 + 28\,r^3 + 5\,r^4)/(96\,(1 + r)^2)\,,\nonumber\\
 &&         \alpha_{63 } = (2 + r)^2\,(20 + 55\,r + 63\,r^2 + 31\,r^3 + 5\,r^4)/(80\,(1 + r)^3)\,,\nonumber\\
 &&         \alpha_{64 } = (2 + r)^2\,(8 + 28\,r + 33\,r^2 + 9\,r^3 + r^4)/(32\,(1 + r)^2)\,,\nonumber\\
 &&         \alpha_{65 } = (2 + r)^3\,(5 + 5\,r + r^2)/(40\,(1 + r)^2)\,,\nonumber\\
 &&         \alpha_{70 } = (2 + r)^3\,(1 + r + r^2)/(8\,(1 + r)^2)\,,\nonumber\\
 &&         \alpha_{71 } = 0\,,\nonumber\\
 &&         \alpha_{72 } = (2 + r)^3/(8\,(1 + r)^2)\,,\nonumber\\
 &&         \alpha_{73 } = (2 + r)^2\,(4 + 11\,r + 19\,r^2 + 15\,r^3 + 5\,r^4)/(16\,(1 + r)^3)\,,\nonumber\\
 &&         \alpha_{74 } = -(2 + r)^2\,(-8 - 4\,r - 3\,r^2 + 5\,r^3 + r^4)/(32\,(1 + r)^2)\,,\nonumber\\
 &&         \alpha_{75 } = (2 + r)^3\,(1 + r + r^2)/(8\,(1 + r)^2)\,,\nonumber\\
 &&         \alpha_{80 } = (2 + r)^4\,(6 + 6\,r + r^2)/(96\,(1 + r)^2)\,,\nonumber\\
 &&         \alpha_{81 } = (2 + r)^4\,(14 + 10\,r + 7\,r^2 + 8\,r^3 + r^4)/(224\,(1 + r)^4)\,,\nonumber\\
 &&         \alpha_{82 } = (2 + r)^2\,(12 + 24\,r + 25\,r^2 + 13\,r^3 + 2\,r^4)/(48\,(1 + r)^2)\,,\nonumber\\
 &&         \alpha_{83 } = (2 + r)^3\,(12 + 36\,r + 43\,r^2 + 21\,r^3 + 3\,r^4)/(96\,(1 + r)^3)\,,\nonumber\\
 &&         \alpha_{84 } = (2 + r)^3\,(12 + 36\,r + 43\,r^2 + 21\,r^3 + 3\,r^4)/(96\,(1 + r)^3)\,,\nonumber\\
 &&         \alpha_{85 } = 0\,,\nonumber\\
 &&         \alpha_{90 } = (2 + r)^2/(4\,(1 + r)^2)\,,\nonumber\\
 &&         \alpha_{91 } = (2 + r)^2/(4\,(1 + r)^2)\,,\nonumber\\
 &&         \alpha_{92 } = (4 + 3\,r + r^2)/(4\,(1 + r)^2)\,,\nonumber\\
 &&         \alpha_{93 } = (4 + 4\,r + 3\,r^2 + r^3)/(4\,(1 + r)^3)\,,\nonumber\\
 &&         \alpha_{94 } = (8 + 10\,r + 11\,r^2 + 6\,r^3 + r^4)/(8\,(1 + r)^2)\,,\nonumber\\
 &&         \alpha_{95 } = (2 + r)^2\,(9 + 2\,r)/(36\,(1 + r)^2)\,,\nonumber\\
 &&         \alpha_{100 } = (2 + r)^2/(4\,(1 + r)^2)\,,\nonumber\\
 &&         \alpha_{101 } = (2 + r)^2\,(3 + 16\,r + 16\,r^2 + 4\,r^3)/(12\,(1 + r)^2)\,,\nonumber\\
 &&         \alpha_{102 } = (4 + 3\,r + r^2)/(4\,(1 + r)^2)\,,\nonumber\\
 &&         \alpha_{103 } = (4 + 4\,r + 3\,r^2 + r^3)/(4\,(1 + r)^3)\,,\nonumber\\
 &&         \alpha_{104 } = (8 + 10\,r + 11\,r^2 + 6\,r^3 + r^4)/(8\,(1 + r)^2)\,,\nonumber\\
 &&         \alpha_{105 } = (2 + r)^2\,(27 + 38\,r + 32\,r^2 + 8\,r^3)/(108\,(1 + r)^2)\,,\nonumber\\
 &&         \alpha_{110 } = 0\,,\nonumber\\
 &&         \alpha_{111 } = (2 + r)^4/(16\,(1 + r)^2)\,,\nonumber\\
 &&         \alpha_{112 } = 0\,,\nonumber\\
 &&         \alpha_{113 } = 0\,,\nonumber\\
 &&         \alpha_{114 } = 0\,,\nonumber\\
 &&         \alpha_{115 } = (2 + r)^4/(16\,(1 + r)^2)\,,\nonumber\\
 &&         \alpha_{120 } = (2 + r)^3/(8\,(1 + r)^2)\,,\nonumber\\
 &&         \alpha_{121 } = (2 + r)^3/(8\,(1 + r)^2)\,,\nonumber\\
 &&         \alpha_{122 } = (2 + r)\,(4 + 3\,r + r^2)/(8\,(1 + r)^2)\,,\nonumber\\
 &&         \alpha_{123 } = (2 + r)^2\,(4 + 5\,r + 3\,r^2)/(16\,(1 + r)^3)\,,\nonumber\\
 &&         \alpha_{124 } = (2 + r)^2\,(8 + 3\,r + 4\,r^2 + r^3)/(32\,(1 + r)^2)\,,\nonumber\\
 &&         \alpha_{125 } = (2 + r)^3\,(1 + 2\,r)/(8\,(1 + r)^2)\,,\nonumber\\
 &&         \alpha_{130 } = 0\,,\nonumber\\
 &&         \alpha_{131 } = 0\,,\nonumber\\
 &&         \alpha_{132 } = 0\,,\nonumber\\
 &&         \alpha_{133 } = 0\,,\nonumber\\
 &&         \alpha_{134 } = 0\,,\nonumber\\
 &&         \alpha_{135 } = (2 + r)^3\,(1 + 2\,r)/(8\,(1 + r)^2)\,.
  \label{Eq:alphas-explicit}
\end{eqnarray}
Note that all $\alpha_{ij}$ are the same as in Ref. \cite{Lutz:2020dfi}. The only exception is $\alpha_{33}$ of Ref. \cite{Lutz:2020dfi}, which is given by
\begin{eqnarray}
-\frac{1}{14} \, \alpha_{33}+\frac{15}{14} \, \alpha_{34}\,. 
\end{eqnarray}

\clearpage
\bibliography{literature} 

\end{document}